\documentclass[aip,amsmath,amssymb,reprint]{revtex4-1}

\usepackage{graphicx}
\usepackage{svg}
\usepackage{dcolumn}
\usepackage{bm}
\usepackage[utf8]{inputenc}
\usepackage[T1]{fontenc}
\usepackage{etoolbox}
\usepackage[separate-uncertainty = true, multi-part-units = repeat, per-mode=symbol]{siunitx}
\usepackage{xcolor}
\usepackage{lipsum}
\usepackage[hidelinks]{hyperref}
\usepackage{float}
\usepackage{multirow}
\usepackage{ulem}
\usepackage{xcolor}
\usepackage{fancyhdr}
\draft 

\newcommand{\ColorVariable}[1]{\textcolor{black}{#1}}

\begin{document}

\title{
\texorpdfstring{
\textit{Ab initio} quantum scattering calculations and a new potential energy surface for the HCl($X^1\Sigma^+$)-O$_{2}$($X^3\Sigma^-_g$) system: collision-induced line-shape parameters for O$_{2}$-perturbed R(0) 0-0 line in H$^{35}$Cl
}{}
}

\author{Artur Olejnik}
\email[corresponding author (e-mail): ]{artur.olejnik.umk@gmail.com}
\author{Hubert Jóźwiak}
\author{Maciej Gancewski}
\affiliation{Institute of Physics, Faculty of Physics, Astronomy and Informatics, Nicolaus Copernicus University in Toruń, Grudziądzka 5, 87-100 Toruń, Poland}

\author{Ernesto Quintas-S{\'a}nchez}
\author{Richard Dawes}
\affiliation{Department of Chemistry,
Missouri University of Science and Technology,
Rolla, MO 65409-0010,
USA}

\author{Piotr Wcisło}
\email[e-mail: ]{piotr.wcislo@umk.pl}
\affiliation{Institute of Physics, Faculty of Physics, Astronomy and Informatics, Nicolaus Copernicus University in Toruń, Grudziądzka 5, 87-100 Toruń, Poland}

\date{\today}

\begin{abstract}
The remote sensing of abundance and properties of HCl -- the main atmospheric reservoir of Cl atoms which directly participate in ozone depletion -- are important for monitoring the partitioning of chlorine between "ozone-depleting" and "reservoir" species. 
Such remote studies require knowledge of the shapes of molecular resonances of HCl, which are perturbed by collisions with the molecules of the surrounding air. In this work, we report the first fully quantum calculations of collisional perturbations of the shape of a pure rotational line in H$^{35}$Cl perturbed by an air-relevant molecule (as the first model system we choose the R(0) line in HCl perturbed by O$_2$). The calculations are performed on our new highly-accurate HCl($X^1\Sigma^+$)-O$_2$($X^3\Sigma^-_g$) potential energy surface. In addition to pressure broadening and shift, we determine also their speed dependencies and the complex Dicke parameter. This gives important input to the community discussion on the physical meaning of the complex Dicke parameter and its relevance for atmospheric spectra (previously, the complex Dicke parameter for such systems was mainly determined from phenomenological fits to experimental spectra and the physical meaning of its value in that context is questionable). We also calculate the temperature dependence of the line-shape parameters and obtain agreement with the available experimental data. We estimate the total combined uncertainties of our calculations at 2\% relative RMSE residuals in the simulated line shape at 296~K. This result constitutes an important step towards computational population of spectroscopic databases with accurate \textit{ab initio} line-shape parameters for molecular systems of terrestrial atmospheric importance.
\end{abstract}

\pacs{}

\maketitle 

\thispagestyle{fancy}
\fancyhf{}
\fancyhead[L]{Copyright (2023) A. Olejnik, H. J\'o\'zwiak, M. Gancewski, E. Q. S{\'a}nchez, R. Dawes and P. Wcis{\l}o.\\This article is distributed under a Creative Commons Attribution (CC-BY NC ND) License.}

\section{\label{sec:introduction}Introduction}
The ozone layer protecting the Earth from harmful ultraviolet radiation is destroyed in chemical reactions induced in the stratosphere mainly by chlorofluorocarbons (CFCs), whose heightened atmospheric presence is of anthropogenic origin.~\cite{zander1992, solomon1999} The discovery of the Antarctic ozone hole~\cite{farman1985} stimulated and accelerated the ratification, in 1987, of the Montreal Protocol -- an international agreement to reduce the production and emission of various ozone-depleting substances. As a direct impact of this treaty, now signed by all the United Nations members,~\cite{polvani2023} the concentration of the ozone-depleting compounds in our planet's atmosphere has been reduced~\cite{newman2009, mader2010} and the ozone layer over the Antarctica has started to recover,~\cite{solomon2016} which elevated the role of the Montreal Protocol as one of the most successful ongoing global human ecosystem initiatives.~\cite{WMO2022}

Among the ozone-depleting species originating from the breakdown of CFCs, atomic fluorine F and chlorine Cl are the most abundant.~\cite{solomon1999} These halogens react with other molecular constituents of Earth's atmosphere, forming acids and nitrates that result in the formation of stratospheric reservoirs of F and Cl. In the case of fluorine, the abstraction of hydrogen allows HF molecules to be formed, which due to their resilience to further photo-chemical breakdown keep the total amount of ozone-depleting FO molecules at negligible levels.~\cite{solomon1999} Thus, although fluorine itself has a relatively low impact on the ozone layer, the concentration of stratospheric HF may reflect the rate and distribution of the emission of surface-released fluorine-bearing gases.~\cite{zander1992} On the other hand, the main stratospheric reservoirs of chlorine are HCl and ClONO$_2$ which are broken-up by photolysis and temperature-induced processes, thus feeding the ClO$_x$ catalytic cycle: ozone depletion by Cl (Cl + O$_3$ $\rightarrow$ ClO + O$_2$) and the subsequent provision of chlorine atoms (ClO + O $\rightarrow$ Cl + O$_2$).~\cite{zander1992, solomon1999, nassar2006} As most of the stratospheric chlorine resides in the hydrochloric acid reservoir,~\cite{solomon1999} the monitoring of HCl abundance using spectroscopic and remote-sensing techniques is necessary for the determination of the chlorine-partitioning between the "reservoir" HCl and "active" Cl and ClO that directly participate in the ozone-depleting processes.~\cite{solomon1999, zander1992, nassar2006} 

Beyond the Earth's atmosphere, HCl spectra have been observed in various astrophysical conditions. Traces of hydrochloric acid were detected in the Venusian~\cite{connes1967} and stellar~\cite{ridgway1984} atmospheres. HCl absorption lines were also detected in one of the largest molecular gas clouds in our Galaxy, Sagittarius B2, against its dust continuum.~\cite{zmuidzinas1995} This observation pointed to the validity of the prediction of Jura~\cite{jura1974} and Dalgarno \textit{et al.}~\cite{dalgarno1974} that chlorine reservoirs in the cosmic molecular gas clouds should be deposited mostly in HCl, and that for very dense clouds the cooling due to HCl can be surprisingly effective, comparable in its effect to CO molecules. 
Due to relatively low pressures typical for the interstellar medium, the shape of the HCl emission spectra can be well described without considerations involving collisional processes. However, as far as terrestrial-atmospheric applications are concerned, the shapes and positions of the lines in the required spectra are perturbed by collisions with other molecules, whose presence cannot be neglected. 
Various molecular transitions in such foreign gas-perturbed HCl that contribute to its atmospheric absorption spectrum have been studied extensively in the past, mostly in experiment~\cite{jones1964, deLucia1971, pourcin1973, houdeau1980, sergent-rozey1986, pine1987, park1991, klaus1998, zu2003, drouin2004, morino2005, hurtmans2009, li2015} and less in theoretical works.~\cite{fitz1975, tran2014, tran2017} 

HCl is a simple system of great interest in the spectroscopy of small molecules owing in part to its large rotational constant and permanent dipole moment. The spectral lines of HCl are therefore intense and relatively well separated, which allows for accurate 
probing of the beyond-Voigt line-shape effects, in particular the Dicke narrowing~\cite{dicke1953}
especially pronounced in the rotation-vibration spectra of HCl.\cite{barret2005, ramachandra1987}
The temperature- and pressure-dependent line-shape perturbations examined in experimental and theoretical works mentioned above were studied for HCl interacting with N$_2$~\cite{pine1987, park1991, sergent-rozey1986, hurtmans2009, houdeau1980, zu2003, drouin2004, morino2005} and O$_2$~\cite{pine1987, park1991, houdeau1980, zu2003, drouin2004, morino2005} -- the two most abundant constituents of the Earth's atmosphere -- as well as with the noble gases: He,~\cite{rank1963, fitz1975, hurtmans2009} Ar~\cite{rank1963, sergent-rozey1986, ramachandra1987, hurtmans2009, fitz1975, tran2014, tran2017, morino2005}, Kr~\cite{rank1963} and Xe,~\cite{rank1963, fitz1975, hurtmans2009} with CF$_4$~\cite{rank1963}, and with the hydrogen isotopologues: H$_2$ and D$_2$.~\cite{houdeau1980}. The self-perturbation effects on the line-shape in the pure HCl mixtures were also studied.~\cite{pine1985, benedict1956, barret2005}

As far as the laboratory studies are concerned, a well-known problem with handling HCl is its highly corrosive and adsorptive nature.~\cite{sergent-rozey1986, zu2003, morino2005, li2015, tran2014} Due to sticking of the individual molecules on the inner walls of the gas cell, the collision-induced line-shape parameters determined from independent measurements may differ depending on the experimental approach and the measurement conditions (involving, e.g., the type of absorption cell surface, flowing or static measurement approach).~\cite{zu2003} 
Theoretical calculations of the collision-induced line-shape parameters of HCl, performed \textit{ab initio}, can thus provide valuable guidance in the interpretation and possible refinement of the measurements' results.

Previous theoretical studies of the line-shape parameters of the HCl lines, that go beyond fitting experimental spectra with phenomenological line-shape profiles, concern only the case of atom-perturbed HCl lines. For example, in an paper by Fitz and Marcus~\cite{fitz1975} the half-widths and shifts of the HCl lines due to collisions with Ar and He are calculated using a semi-classical approach developed by Marcus\cite{Marcus_1970} and Miller\cite{Miller_1970} (see Ref.\cite{Fitz_1973} and references therein). In recent papers, Tran and Domenech~\cite{tran2014}, and Tran \textit{et al.}~\cite{tran2017} simulated the Ar-perturbed spectra of HCl directly using classical molecular dynamics simulations, utilizing a requantization procedure for determining the angular momentum quantum numbers corresponding to classical (continuous) angular frequencies. The relevant line-shape parameters, such as the Lorentzian half-width and the Dicke parameter, were subsequently extracted from the simulated spectrum by fitting it with a beyond-Voigt line-shape model.

It turns out, however, that no systematic efforts have been undertaken to determine the line-shape parameters for the HCl spectra perturbed by collisions with diatomic molecules starting directly from the first principles of quantum mechanics. In this work, we report the first fully quantum calculations of collisional perturbations of the shape of a pure rotational line in HCl perturbed by an air-relevant molecule (as the first model system we choose the R(0) line in HCl perturbed by O$_2$).

We report a new \textit{ab initio} potential energy surface (PES) for the HCl($X^1\Sigma^+$)-O$_{2}$($X^3\Sigma^-_g$) system, and we use it to determine the $S$-matrices required for the computation of the generalized spectroscopic cross sections by solving numerically the relevant close-coupling equations employing the scattering boundary conditions. By averaging the cross sections over the collisional kinetic energies, we obtain the collisional line-shape parameters: the pressure broadening and shift, the speed-dependence of broadening and shift, and the real and imaginary parts of the Dicke-narrowing parameter. The parameters are determined for temperatures ranging from 150~K to 360~K. This work constitutes an important step towards populating line-by-line spectroscopic databases, such as HITRAN~\cite{hitran2022} and GEISA,~\cite{geisa2021} with accurate \textit{ab initio} line-shape parameters. This is a very promising direction since the theoretical calculations can cover temperature and spectral ranges that are very challenging to access experimentally. Furthermore, the \textit{ab initio} calculations give direct access to all the beyond-Voigt line-shape parameters, while experimentally it is often difficult to disentangle contributions of different line-shape effects from the fits to measured spectra. 
This gives important input to the community discussion on the physical meaning of the complex Dicke parameter and its relevance for atmospheric spectra. Previous fully \textit{ab initio} calculations of the complex Dicke parameter
were done for simpler systems of atom-perturbed\cite{thibault2017,Jozwiak_2018,Martinez_2018,thibault_2020,Kowzan_2020a,Kowzan_2020b,słowiński2020,Stankiewicz_2020,Serov_2021,słowiński2022,stolarczyk2023} and self-perturbed hydrogen molecule.\cite{wcisło2018,Lamperti_2023} Recently, the complex Dicke parameter was calculated for some N$_2$-perturbed CO lines.\cite{Paredes_Roibas_2021} In the case of molecular systems relevant to the Earth's atmosphere, the complex Dicke parameter has been mainly determined from the phenomenological fits to experimental spectra, and in these cases the physical interpretation of its value is questionable. In particular, in this work, we show that the value of the real part of the complex Dicke parameter that was obtained from the fit to experimental spectra\cite{morino2005} (for the case of the line considered in this article) is almost three times larger than its best estimate. 

This paper is organized as follows. In Sec.~\ref{sec:Potential energy surface} we report the \textit{ab initio} PES for the HCl($X^1\Sigma^+$)-O$_{2}$($X^3\Sigma^-_g$) system used in the line-shape calculations. We discuss the computational methodology behind this PES, as well as its topography and basic characteristics, and we further decompose the PES into angle-dependent and radial parts. The latter are used in the quantum scattering calculations, the details of which are described in Sec.~\ref{sec:Quantum scattering calculations}. There, we discuss our methodology of setting up and numerically solving the close-coupling equations whose solutions allow us to obtain the relevant S-matrix elements from the appropriate scattering boundary conditions, and to determine the generalized spectroscopic cross sections (GSXS). We report the results of our calculations and we discuss the collision simulation parameters as well as the method of constructing the expansion basis for the close-coupling wave functions used by us in the computations. Section~\ref{sec:Collision-induced line-shape parameters} contains the details and the results of the calculations of line-shape parameters for the R(0) 0-0 ground-vibrational transition in the H$^{35}$Cl isotopomer perturbed by O$_2$, and the comparison of our results with the available experimental data. Finally, we conclude the present investigation in Sec.~\ref{sec:Conclusion} where we summarize our results and put them in perspective of future work.

\section{\label{sec:Potential energy surface}Potential energy surface}

\begin{figure}[t]
 \includegraphics[width=0.45\textwidth]{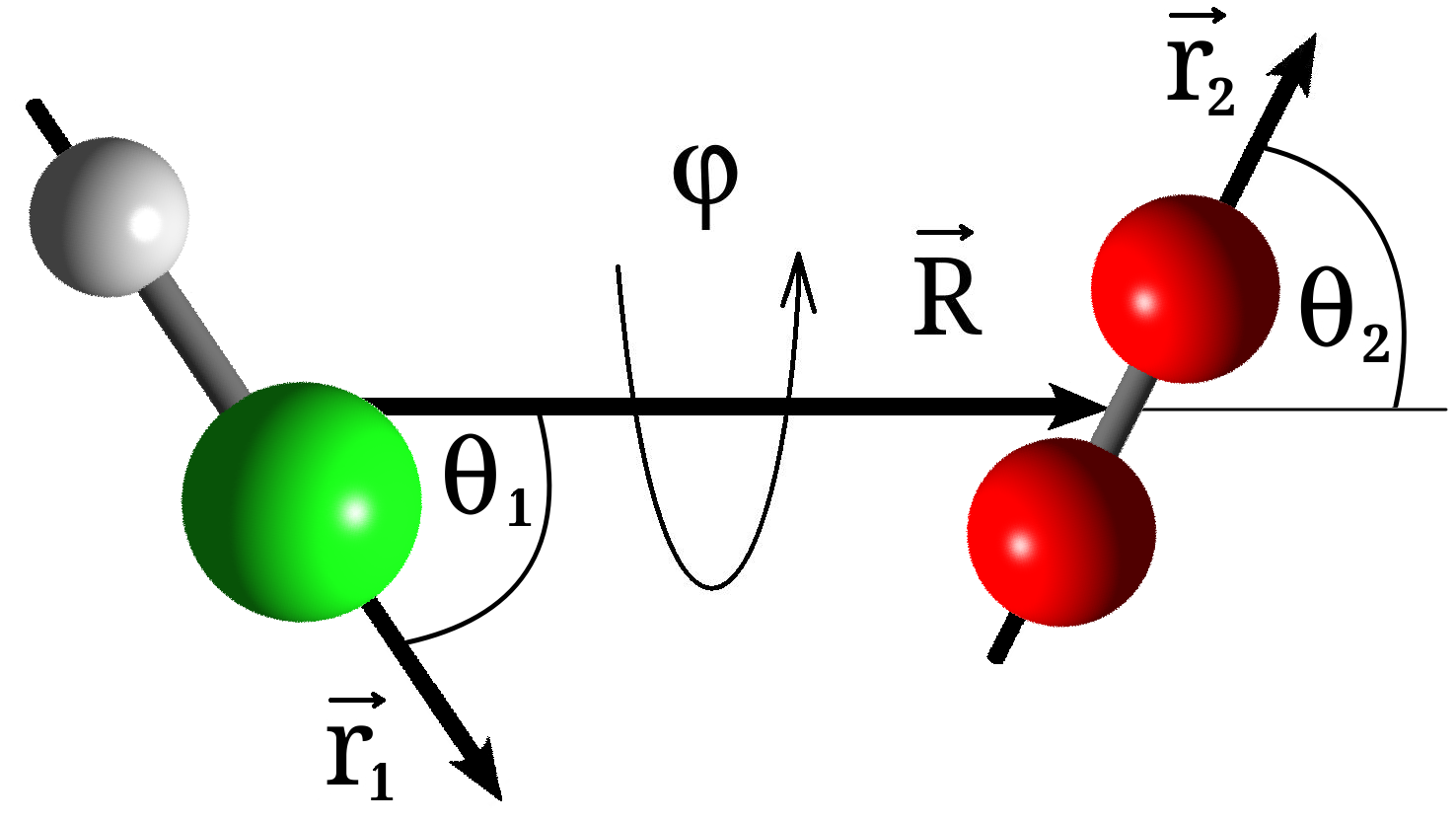}
 \caption{Coordinates used to describe the HCl--O$_2$ interaction. 
  See the text for details.}
  \label{SR}
\end{figure}

As depicted in Figure~\ref{SR}, the coordinates used to represent the four-dimensional (4D) HCl--O$_2$ PES are the Jacobi coordinates $R$, $\theta_1$, $\theta_2$, and $\varphi$.
$\vec{R}$ is the vector between the centers of mass of the two fragments, and $\vec{r}_1$ and $\vec{r}_2$ are vectors aligned with each molecule.
Coordinate $R$ is the length of vector $\vec{R}$, while coordinates $\theta_1$ and $\theta_2$ represent (respectively) the angles between $\vec{R}$ and the vectors $\vec{r}_1$ and $\vec{r}_2$. 
The fourth coordinate is the dihedral (out of plane) torsional angle, labeled $\varphi$, which is the angle between the vectors $\vec{R}\times\vec{r}_1$ and $\vec{R}\times\vec{r}_2$. 
Notice that for $\theta_1=0^{\circ}$ the HCl molecule aligns with the Cl atom pointing to the center-of-mass of the O$_2$ molecule.

\subsection{\label{pes.esc} Electronic structure calculations}

For construction of the PES, both monomers were held rigid. The bond distances for HCl and O$_2$ were fixed at their ground rovibrational state vibrationally-averaged bond distances, $r_{HCl}=1.28387$~{\AA} and $r_{OO}=1.20752$~{\AA} respectively, consistent with their experimental rotational constants.

The final high-level PES was computed using explicitly-correlated unrestricted coupled-cluster theory~\cite{knizia2009simplified} (UCCSD(T)-F12b). The complete basis set limit was estimated by extrapolating calculations at the UCCSD(T)-F12b/VTZ-F12 and UCCSD(T)-F12b/VQZ-F12 levels,~\cite{peterson2008systematically} using the $l^{-3}$ formula.\cite{feller2006sources} All \textit{ab initio} calculations were performed using the Molpro electronic structure code package.\cite{Werner2012-molpro} To ensure convergence to the desired triplet electron configuration, the restricted open-shell Hartree--Fock (roHF) reference was obtained by first using \textsc{molpro}'s CASSCF (multi) algorithm with the occupation of the desired configuration specified, followed by a single iteration of the roHF SCF algorithm to prepare the orbitals for the UCCSD(T)-F12b procedure. To avoid placing expensive high-level data in energetically inaccessible regions, a lower-level guide surface was first constructed. This was done using similarly computed data, but at the reduced UCCSD(T)-F12b/VDZ-F12 level of theory.
Exploiting the system's symmetry, energies were only computed in the reduced angular range: $0<\theta_1<180^{\circ}$, $0<\theta_2<90^{\circ}$, and $0<\varphi<180^{\circ}$.

\subsection{\label{pes.ar} Analytical representation and characterization of the PES}

As we have done in the past for other vdW dimers composed of linear monomers,\cite{dumouchel2023collisional,zadrożny2022,denis2022state,ajili2022theoretical} the PES analytical representation was constructed using an automated interpolating moving least squares (IMLS) methodology, freely available as a software package under the name \textsc{autosurf}.\cite{quintas2018autosurf} As usual,\cite{Dawes2018,majumder2016automated} a local fit was expanded about each data point, and the final potential is obtained as the normalized weighted sum of the local fits. This interpolative approach can accommodate arbitrary energy-surface topographies, being particularly advantageous in systems with large anisotropy. The fitting basis and most other aspects of the IMLS procedure  were similar to those for other previously treated systems that have been described in detail elsewhere.\cite{dawes2010nitrous,majumder2016automated,quintas2018autosurf}

\begin{figure*}[t]
	\includegraphics[width=0.8\textwidth]{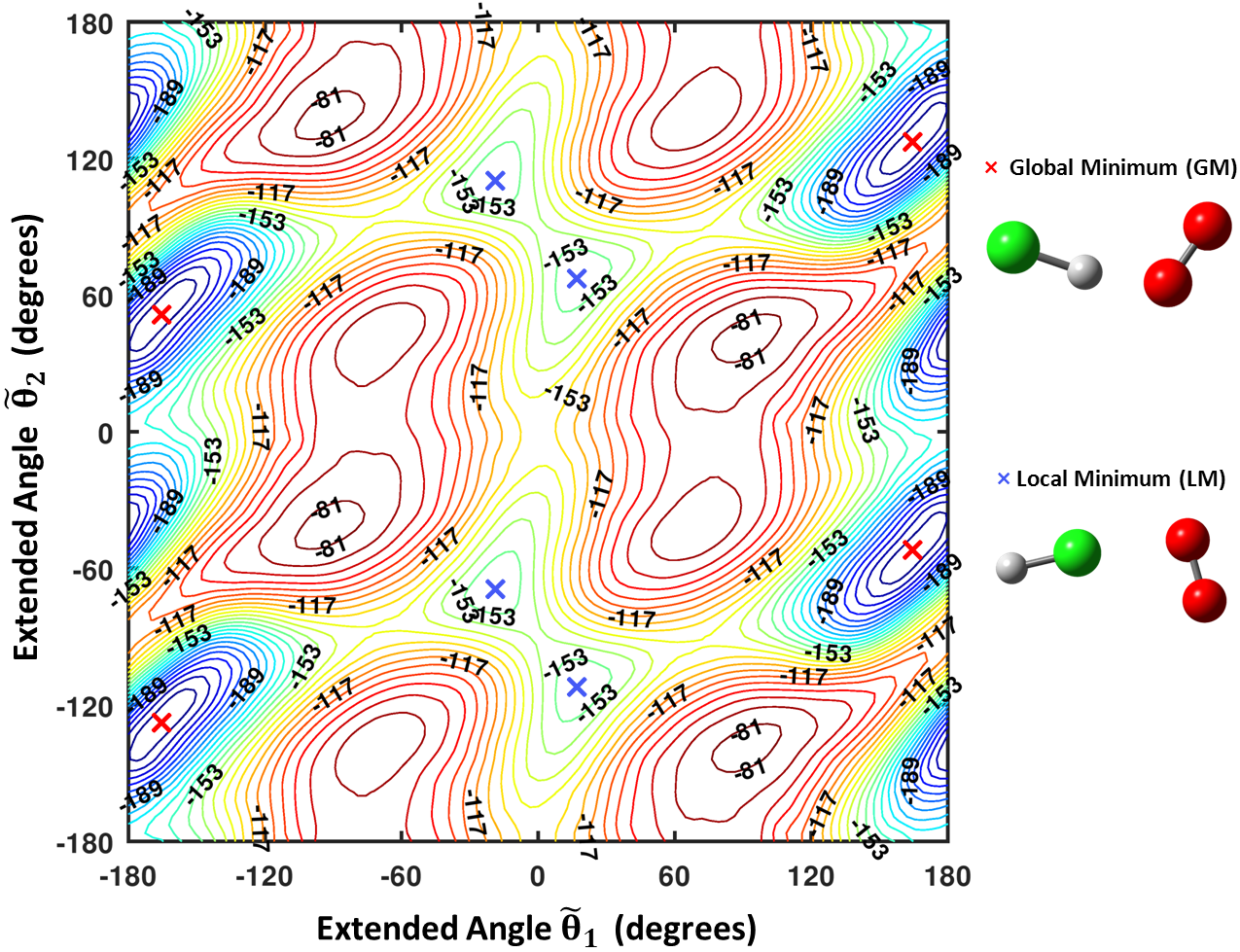}
	\caption{$R$-optimized contour plot of the PES as a function of the extended angles $\tilde{\theta}_1$ and $\tilde{\theta}_2$ for planar configurations ($\varphi=0^{\circ}$). For each pair of angles, the energy (given in cm$^{-1}$) is optimized with respect to the center-of-mass distance $R$.}
	\label{Ropt}
\end{figure*}

The shortest intermonomer center-of-mass distance considered is $R = 2.3$~{\AA}. The short-range part of the PES is restricted by excluding regions with repulsive energies above a maximum of  $8$~kcal/mol ($\sim2\,800$~cm$^{-1}$) relative to the separated monomers asymptote. The \textit{ab initio} data coverage in the fitted PES extends to $R=20$~{\AA}, while the zero of energy is set at infinite center-of-mass separation between the monomers.

For the high-level PES, the global estimated root-mean-squared fitting error tolerance was set to $0.64$~cm$^{-1}$, and the total number of automatically generated symmetry-unique points needed to reach that target was $2\,452$ (the final estimated error is $0.09$~cm$^{-1}$ for energies below the asymptote). To guide the placement of high-level data, a lower-level guide surface was constructed using $2\,000$ symmetry-unique points, distributed using a Sobol sequence~\citep{sobol1976uniformly} biased to sample the short range region more densely. The analytical representation of the PES is available from the authors upon request.

\begin{table}[b]
\caption{\label{table1} Geometric parameters and  potential energy for stable structures in the PES. Energies are given relative to the asymptote. Units are angstr\"{o}ms, degrees, and cm$^{-1}$.}
	\centering
	\begin{tabular}{lrr}
	               & GM        & LM         \\
	\hline
	$R$           & $3.952$   & $3.541$    \\
	$\theta_1$ & $162.1$   & $18.1$     \\
	$\theta_2$ & $124.2$    & $68.0$    \\
	$\varphi$        & $\pi$     & $\pi$      \\
	$V$           & $-223.47$ & $-157.52$  \\
	\end{tabular}
\end{table} 

Figure~\ref{Ropt} shows a 2D representation of the PES (denoted $R$-optimized) as a function of the extended angles $\tilde{\theta}_1$ and $\tilde{\theta}_2$ for planar configurations.
The extended-angle coordinates have been described in detail elsewhere.\cite{Dawes2013}
For planar geometries ($\varphi=0$ for quadrants II and IV, and $\varphi=\pi$ for quadrants I and III), the plot describes the complete ranges of  $\tilde{\theta}_1$ and $\tilde{\theta}_2$, relaxing the intermonomer distance coordinate $r_0$ for each pair of angles. This type of plot provides unique insight into the isomers in the system, since for many systems---those (such as this one) without non-planar minima---the plot will include all isomers and any planar isomerization paths between them, making it easy to visualize planar motions during which $\varphi$ changes from $0$ to $\pi$.
There are two equivalent local minima (LM), and two equivalent global minima (GM) in the PES, each duplicated and appearing twice in the extended angles plot, which shows four wells of each type. 

The geometric parameters of the two isomeric planar structures are given in Table \ref{table1} and their images are provided with Figure \ref{Ropt}. Both types of minima can be described as skewed-T-shaped. The GM, with a well-depth of $223.47$ cm$^{-1}$, finds the H-atom of HCl pointed nearly directly towards the O$_2$ molecule ($\theta_1$=162.1$^{\circ}$), while the O$_2$ molecule is rotated significantly from side-on ($\theta_2$=124.2$^{\circ}$), bringing one of the O-atoms closer to the H-atom. The LM, with a well-depth of $157.52$ cm$^{-1}$, is similar, but with the Cl-atom of HCl pointing toward the O$_2$ molecule, which is also rotated to bring an O-atom closer to the Cl-atom, but in the case of LM, less rotation from side-on is needed to maximize the interaction. Beginning at one of the LM, a conrotatory motion of the two monomers provides a very facile path to the other, symmetry-equivalent LM. A somewhat larger barrier separates LM and GM along a disrotatory path. 

\begin{figure}[t]
 \includegraphics[width=0.45\textwidth]{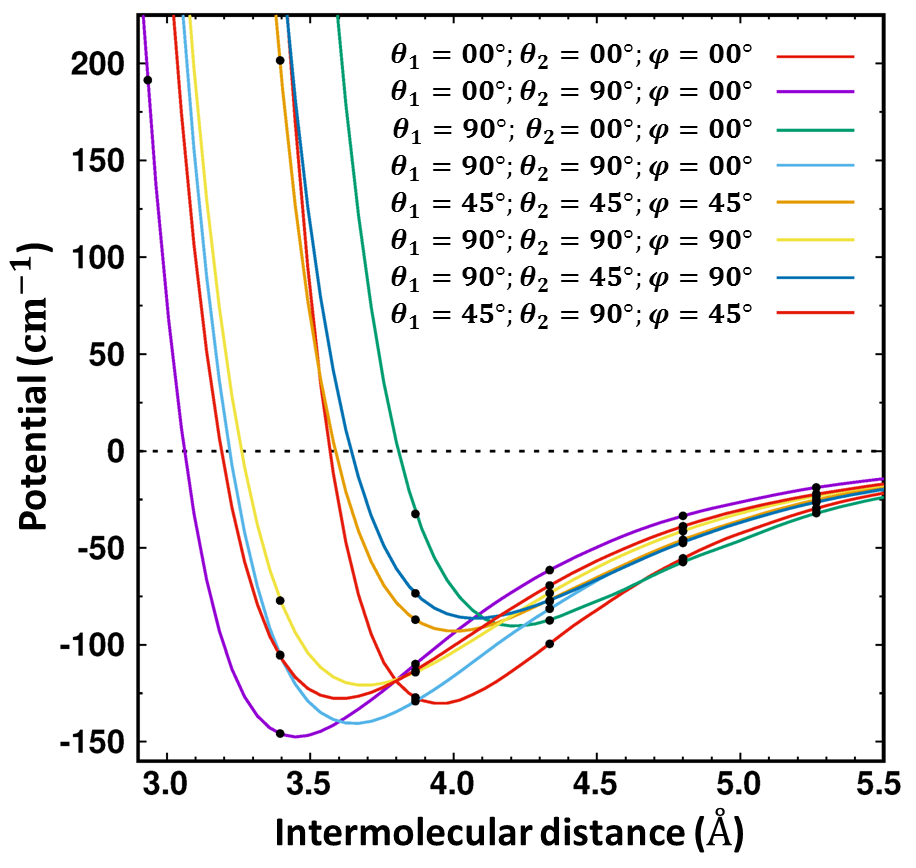}
 \caption{Various radial cuts defined by different orientational poses of the monomers. In all cases,
energies are in cm$^{-1}$, lines represent the fitted PES, and points represent \textit{ab initio} calculations (not used in the fit).}
 \label{R-cut}
\end{figure}

Figure~\ref{R-cut} shows radial cuts through the PES for eight different angular poses. The points represent \textit{ab initio} data (not included in the fit), while the lines plot the fitted PES. This helps one appreciate the anisotropy of the PES as reflected in the different values of $R$ at which the onset of a steep repulsive wall begins, as well as the accuracy of the fit as the lines pass through the data points.

\subsection{\label{pes.dec} Decomposition of the PES}

\begin{figure}[t]
    \includegraphics[width=\linewidth]{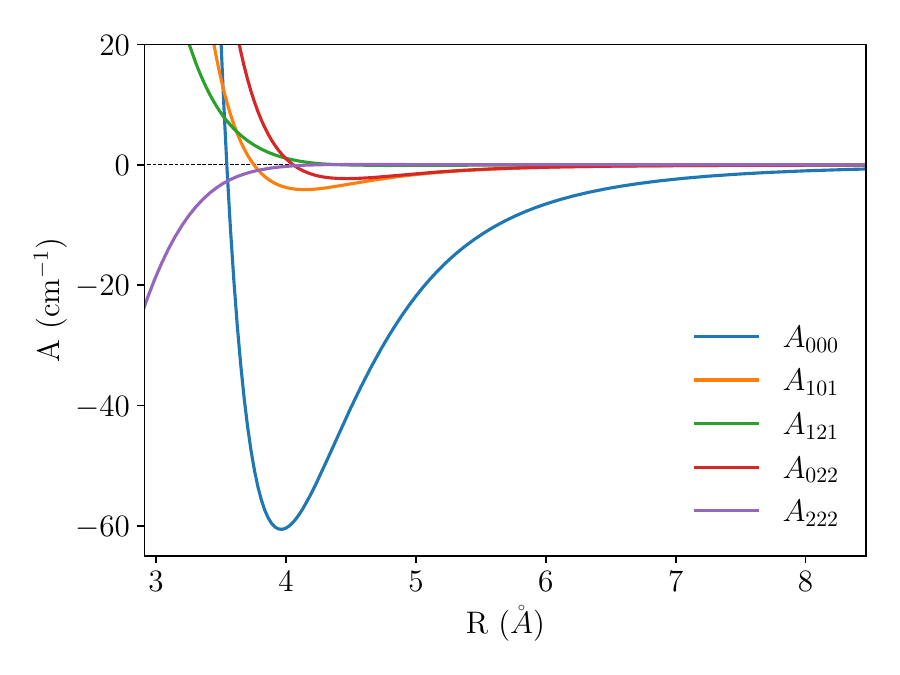}
    \caption{The comparison between the isotropic  radial term (blue line) and largest anisotropic contributions. Note that the radial terms are shown divided by the normalization factor $8\pi^2 / (2L+1)$.}
    \label{fig:radialterms}
\end{figure}

The four-dimensional PES is prepared for scattering calculations by projecting over the set of bispherical harmonics 
\begin{align}
    V(R, \theta_1, \theta_2, \varphi) =\sum_{L_1,L_2,L} A_{L_1 L_2 L}(R) I_{L_1 L_2 L}(\theta_1, \theta_2, \varphi) \text{,} \label{eq:potential}
\end{align}
where $L_1$ and $L_2$ are non-negative quantum numbers, and $L$ satisfies the triangle inequality $|L_1 - L_2| \leq L \leq L_1 + L_2$. Additionally, $L_2$ can take only even values due to the symmetry of the O$_2$ molecule and $L_1 + L_2 + L$ must be an even integer.~\cite{jóźwiak2021} The bispherical harmonics are defined as~\cite{green1975}
\begin{align}
    I_{L_1 L_2 L}(&\theta_1, \theta_2, \varphi=\varphi_1-\varphi_2) ={} \sqrt{\frac{2L+1}{4\pi}} \sum_{m} 
    C_{m -m 0}^{L_1 L_2 L} \notag \\
    &\times Y_{L_1 m}(\theta_1, \varphi_1) Y_{L_2 {-m}}(\theta_2, \varphi_2) \text{,} \label{eq:bishperics}
\end{align}
where $Y_{L_i m}(\theta_i, \varphi_i)$ denotes the usual spherical harmonics for each molecule and $C_{m -m 0}^{L_1 L_2 L}$ is the Clebsch-Gordan coefficient. $A_{L_1 L_2 L}(R)$ is the radial term, which can be obtained by integrating Eq.~\eqref{eq:potential} over the angular distribution,~\cite{jóźwiak2021, gancewski2021}
\begin{align}
    A_{L_1 L_2 L}(R) &= \frac{8\pi^2}{2L+1} \int_0^{2\pi} d\varphi \int_0^{\pi} \sin \theta_1 d\theta_1 \int_0^{\pi} \sin \theta_2 d\theta_2 \nonumber \\
    &\times V(R, \theta_1, \theta_2, \varphi) I_{L_1 L_2 L}(\theta_1, \theta_2, \varphi) \label{eq:radialterm} \text{.}
\end{align}
The PES  was prepared as the set of 271 radial terms with the maximum $(L_1, L_2, L) = (10, 10, 20)$. The calculations were performed on the radial grid of 1001 points, from 2.6 to 28.6~\AA. Figure~\ref{fig:radialterms} shows the comparison between the dominating isotropic radial term $A_{000}$ and several anisotropic contributions. The radial coefficients allow to express the PES as the potential matrix, whose elements are used in the close-coupling equations, see Eq.~(21) in Ref.~\citenum{gancewski2021}.

\section{\label{sec:Quantum scattering calculations}Quantum scattering calculations}

We perform quantum scattering calculations in the body-fixed (BF) frame of reference, following the formalism introduced by Launay~\cite{Launay_1977} and extensively elaborated upon by Alexander and DePristo.~\cite{Alexander_1977} The scattering wavefunction is expanded in a complete set of eigenfunctions of $\hat{\mathbf{N}}_{12}^2$, $|\hat{N}_{12_{Z'}}|$, $\hat{\mathbf{J}}^{2}$, $\hat{J}_{Z}$, $\hat{\Pi}$
\begin{align}
    \begin{split}
    \label{eq:BF_basis}
    &|(N_{1}N_{2})N_{12}\Bar{\Omega} J M \epsilon\rangle = \frac{1}{\sqrt{2(1+\epsilon p\delta_{\Omega,0})}} \\
    &\times \Bigl(|(N_{1}N_{2})N_{12} {\Bar{\Omega}}  J M \rangle \Bigr. +\epsilon p |(N_{1}N_{2})N_{12} -{\Bar{\Omega}} J M \rangle \Bigr),    
    \end{split}
\end{align}
where
\begin{align}
    \begin{split}
        |(N_{1}N_{2})N_{12} {\Omega} J M \rangle = |(N_{1}N_{2})N_{12} {\Omega} \rangle |JM\Omega\rangle
    \end{split}
\end{align}
and $\hat{\mathbf{J}}$ is the total angular momentum, $\hat{J}_{Z}$ is its projection on the space-fixed $Z$-axis and $\hat{\Pi}$ is the parity operator. $\hat{\mathbf{N}}_{12}$ is an angular momentum resulting from the coupling of the rotational angular momenta of HCl ($\hat{\mathbf{N}}_{1}$) and O$_{2}$ ($\hat{\mathbf{N}}_{2}$), and $\hat{N}_{12_{Z'}}$ is the projection of $\hat{\mathbf{N}}_{12}$ on the body-fixed (intermolecular) $Z'$ axis with the direction determined by the centers of mass of the molecules). Here, $\Bar{\Omega} = |{\Omega}|$ denotes an absolute value of a given eigenvalue of $\hat{N}_{12_{Z'}}$, $\epsilon$ is the spatial parity of the system, ${p=(-1)^{N_{1}+N_{2}+J-N_{12}}}$, and $\Bar{\Omega} \in  \langle 0, \mathrm{min}(N_{12}, J)\rangle$. $J$ is the total angular momentum quantum number associated with the eigenvalue of $\hat{\mathbf{J}}^2$. $M$ denotes the eigenvalue of $\hat{J}_Z$. The eigenfunctions of $\hat{\mathbf{J}}^{2}$, $\hat{J}_{Z}$, $\hat{\mathbf{N}}_{12}$, and $\hat{N}_{12_{Z'}}$ are products of the eigenfunctions of $\hat{\mathbf{N}}_{12}$, and $\hat{N}_{12_{Z'}}$ ($|(N_{1}N_{2})N_{12} {\Omega} \rangle$) and symmetric-top eigenfunctions
\begin{equation}
   \langle \Theta, \Phi, \chi  |JM\Omega\rangle = \sqrt{\frac{2J+1}{4\pi}} D^{J\,*}_{M \Omega} (\Theta, \Phi, 0),
\end{equation}
defined using Wigner rotation matrix, $D$, and Euler angles $(\Theta, \Phi, \chi)$ describing a transformation between the space-fixed (SF) and body-fixed frames. Note that the basis vectors commonly used in calculations performed in the SF framed are related to basis vectors defined in Eq.~\eqref{eq:BF_basis} through a transformation
\begin{align}
    \begin{split}
        &|(N_{1}N_{2})N_{12}L J M \epsilon\rangle  =(-1)^{-N_{12}+L} \sum_{\Bar{\Omega}}(-1)^{\Bar{\Omega}} \\
        &\times \sqrt{\frac{2(2L+1)}{(1+\epsilon p\delta_{\Omega,0})}} 
        \begin{pmatrix}
            N_{12}& L &J\\
            \Bar{\Omega} & 0 & -\Bar{\Omega}
        \end{pmatrix}
        |(N_{1}N_{2})N_{12}\Bar{\Omega} J M \epsilon \rangle .
    \end{split}
\end{align}
$L$ is the quantum number labeling the eigenvalues of the $\hat{\mathbf{L}}^{2}$ operator for the squared orbital angular momentum of relative motion of HCl and O$_{2}$.

The scattering wavefunction of the system is expanded in the basis set introduced in Eq.~\eqref{eq:BF_basis}
\begin{align}
    \begin{split}
        |\Psi \rangle&=\sum_{\substack{J,M,\epsilon, N_{1},\\N_{2}, N_{12}, \Bar{\Omega}}} |(N_{1}N_{2})N_{12}\Bar{\Omega} J M \epsilon \rangle  \frac{f^{JM\epsilon}_{N_{1}, N_{2}, N_{12}, \Bar{\Omega}}(R)}{R} .
    \end{split}
\end{align}
Substituting this expansion to the Schr\"{o}dinger equation for the scattering system involving two rigid diatomic molecules in $^{1}\Sigma$ electronic states leads to a set of coupled equations on coefficients $f^{J\epsilon}_{\gamma}(R)$
\begin{align}
\frac{d^2}{dR^2} f^{J\epsilon}_{\gamma}(R)
= \sum_{\gamma'} W^{J\epsilon}_{\gamma,\gamma'}(R)f^{J\epsilon}_{\gamma'}(R) ,\label{eq:close-coupling}
\end{align}
where we introduced a shorthand notation for ${\gamma=\{N_1,N_2,N_{12},\Bar{\Omega}\}}$. Since $\hat{\mathbf{J}}^{2}$, $\hat{J}_{Z}$, and $\hat{\Pi}$ commute with the scattering Hamiltonian, $J$, $M$ and $\epsilon$ are conserved during the collision, and coupled equations can be solved for each $J$, $M$, $\epsilon$ block independently. Additionally, in the absence of external fields, the equations are independent of $M$, and this index is suppressed in Eq.~\eqref{eq:close-coupling}. The coupling matrix
\begin{align}
    \begin{split}
    \label{eq:coupling_matrix}
        W^{J\epsilon}_{\gamma,\gamma'}(R) = 2\mu V^{J\epsilon}_{\gamma,\gamma'}(R)+ \frac{1}{R^{2}}  \hat{\mathbf{L}}^{2\,J\epsilon}_{\gamma,\gamma'} - \delta_{\gamma,\gamma'}k_{\gamma}^{2}
    \end{split}
\end{align}
involves contributions from the HCl-O$_{2}$ potential energy surface, $V^{J\epsilon}_{\gamma,\gamma'}(R)$, the relative motion of O$_{2}$ with respect to HCl, quantified by the square relative angular momentum operator, $\hat{\mathbf{L}}^{2\,J\epsilon}_{\gamma,\gamma'}$, and the relative kinetic energy of the colliding pair, expressed using the wavevector $k_{\gamma} = \sqrt{2\mu(E-E_{N_{1}}-E_{N_{2}})}$. Here, $E$ is the total energy of the scattering system, and $E_{N_{i}}$ denotes the internal (rotational) energy of the $i$-th molecule. We assume that both molecules are in their ground vibrational states.

\begin{table}[h!]
\caption{\label{tab:Tab1}H$^{35}$Cl and O$_2$ rotational energy levels in cm$^{-1}$ used in our calculations of the generalized spectroscopic cross sections.}
\begin{ruledtabular}
\begin{tabular}{crcr}
\multicolumn{2}{c}{HCl}&\multicolumn{2}{c}{O$_2$}\\
$N_1$ & \multicolumn{1}{c}{$E_{N_1}$} & $N_2$ & \multicolumn{1}{c}{$E_{N_2}$} \\
\hline
0 & 0.000000 & 1 & 2.875330 \\
1 & 20.878402 & 3 & 17.251397 \\
2 & 62.622535 & 5 & 43.125880 \\
3 & 125.207074 & 7 & 80.494597 \\
4 & 208.594056 & 9 & 129.351507 \\
5 & 312.732923 & 11 & 189.688714 \\
6 & 437.560568 & 13 & 261.496460 \\
7 & 583.001390 & 15 & 344.763133 \\
\end{tabular}
\end{ruledtabular}
\end{table}

Table~\ref{tab:Tab1} gathers values of rotational levels of both monomers used in scattering calculations. The energy levels of the H$^{35}$Cl molecule were calculated by solving the nuclear Schr\"{o}dinger equation in the Born-Oppenheimer approximation for an isolated HCl molecule using the finite basis discrete variable representation method~\cite{Lill_1982} and the potential energy curve reported by Coxon and Hajigeorgiou.~\cite{coxon2015} We neglect the hyperfine structure of rotational levels in HCl.~\cite{Cazzoli_2004}
\ColorVariable{The purely-rotational energies of the O$_2$ monomer were calculated using
\begin{align}
\begin{split}
E_{N_2} &= B N_2(N_2+1) - D\left[N_2(N_2+1)\right]^2 \\&+ H \left[N_2(N_2+1)\right]^3\label{eq:Ej2}  ,
\end{split}
\end{align}
where $B$ is the rotational constant, and $D$ and $H$ denote the first- and second-order centrifugal distortion corrections to the energy of a rigid rotor. Here, we neglect the fine-structure splitting in ground-electronic O$_2$ originating mostly from spin-spin interactions, since it is much lower than the collision energies considered in this work. For the rotational constants in Eq.~\eqref{eq:Ej2} we use the values reported by Hajigeorgiou.~\cite{photos2013}: $B=\SI{1.437674521}{\cm^{-1}}$, $D=\SI{4.839482e-6}{\cm^{-1}}$ and $H=-\SI{4.220376e-14}{\cm^{-1}}$ (only the first few digits matters for our quantum-scattering calculations, but for consistency with previous literature we quote here the full accuracy).
}

The coupling matrix~(Eq.~\eqref{eq:coupling_matrix}) has (for given $J$ and $\epsilon$) almost a block-diagonal structure: the contribution from the HCl-O$_{2}$ PES is diagonal with respect to $\bar{\Omega}$, but ${\bar{\Omega}'=\bar{\Omega}\pm 1}$ blocks are coupled by the $\hat{\mathbf{L}}^{2}$ operator (see Fig.~3 in Ref.~\citenum{Rabitz_1975} and Appendix~2 in Ref.~\citenum{Launay_1977}). The sparse structure of the matrix  enables a significant reduction in computational time and memory requirements.

The coupled equations are solved \ColorVariable{in BF frame} using renormalized Numerov's algorithm~\cite{Johnson_1978} implemented in the BIGOS code.~\cite{BIGOS} \ColorVariable{$f^{J\epsilon}_{\gamma}(R)$} 
is transformed to the SF frame of reference at a sufficiently large value of $R$. By imposing boundary conditions on the scattering wave functions, we can derive the scattering $S$-matrix elements,~\cite{Launay_1977} which are necessary to obtain the complex generalized spectroscopic cross-sections, $\sigma_{\lambda}^q$. The generalized spectroscopic cross-sections describe how the shape of molecular resonance is perturbed by collisions. For the isolated spectral transition $N_{i} \rightarrow N_{f}$ of rank $q$ in a diatomic molecule perturbed by collisions with a diatomic perturber in the state $N_{2}$, the generalized cross-sections are given as~\cite{monchick1986diatomic, schafer1992line}
\begin{widetext}
 \begin{align}
     \sigma_{\lambda}^q &(N_{i}, N_{f}, N_{2}, E_{\mathrm{kin}}) = \frac{\pi}{k^2} \sum_{N_{2}'} \sum_{L, L', \bar{L}, \bar{L}'} \sum_{J, N_{12}, N_{12}'} \sum_{\bar{J}, \bar{N}_{12}, \bar{N}_{12}'} i^{L-L'-\bar{L}+\bar{L}'}
     (-1)^{\lambda + N_2 - N_2' + L - L' - \bar{L} + \bar{L}'}
     [J] [\bar{J}] 
     \nonumber \\
     &\times \sqrt{[L][L'][\bar{L}][\bar{L}'][N_{12}][N_{12}'][\bar{N}_{12}][\bar{N}_{12}']}
     \begin{pmatrix}
        L       & \bar{L}  & \lambda \\
        0       & 0        & 0
     \end{pmatrix} \begin{pmatrix}
        L'      & \bar{L}' & \lambda \\
        0       & 0        & 0
     \end{pmatrix}
     \begin{Bmatrix}
        q       & N_{12}'       & \bar{j}_{12}' \\
        N_2'    & N_f     & N_i
     \end{Bmatrix} \begin{Bmatrix}
        q       & \bar{N}_{12}  & N_{12} \\
        N_2     & N_i      & N_f
     \end{Bmatrix}\begin{bmatrix}
        N_{12}       & N_{12}'       & \bar{L}   & \bar{L}' \\
        \bar{N}_{12} & L        & \bar{N}_{12}'  & L'      \\
        q       & \bar{J}      & J       & \lambda
     \end{bmatrix} 
     \nonumber \\
     &\times\bigg(
        \delta_{N_2 N_{12} \bar{N}_{12} L \bar{L}, N_2' N_{12}' \bar{N}_{12}' L' \bar{L}'} - \langle (N_i N_2) N_{12} L | S^J(E_{T_{i}}) | (N_i N_2') N_{12}' L' \rangle
        \langle (N_f N_2) \bar{N}_{12} \bar{L} | S^{\bar{J}}(E_{T_{f}}) | (N_f N_2') \bar{N}_{12}' \bar{L}' \rangle^{*}
     \bigg), \label{eq:crosssection}
 \end{align}
\end{widetext}
 where $[x] = 2x + 1$ and $\delta_{a \ldots z, a' \ldots z'}= \delta_{aa'}\ldots \delta_{zz'}$, and the quantities $(\ldots)$, $\{\ldots\}$ and $[\ldots]$ are Wigner 3-$j$, 6-$j$ and 12-$j$ symbols, respectively.~\cite{Yutsis} For $\lambda=0$, the real and imaginary parts of this cross-section describe pressure broadening and pressure shift of the spectral line, respectively.~\cite{Ben_Reuven_1966a,Ben_Reuven_1966b} For $\lambda=1$, the complex cross-section describes the collisional perturbation of the translational motion of the molecule which undergoes a spectral transition and is crucial in a proper description of the Dicke effect~\cite{Corey_1984,monchick1986diatomic} and other effects related to the velocity-changing collisions such as a reduction of inhomogeneous collisional broadening and asymmetry.~\cite{wcisło2015} The angular momenta are coupled in the following order
\begin{align*}
\begin{split}
    \hat{\mathbf{N}}_{i} + \hat{\mathbf{N}}_{2} = \hat{\mathbf{N}}_{12},&\,\,\hat{\mathbf{N}}_{f} + \hat{\mathbf{N}}_{2} = \hat{\bar{\mathbf{N}}}_{12},\\
    \hat{\mathbf{N}}_{i} + \hat{\mathbf{N}}_{2}' = \hat{\mathbf{N}}_{12}',&\,\,\hat{\mathbf{N}}_{f} + \hat{\mathbf{N}}_{2}' = \hat{\bar{\mathbf{N}}}_{12}',\\
    \hat{\mathbf{N}}_{12} + \hat{\mathbf{L}} = \hat{\mathbf{J}},&\,\,\hat{\bar{\mathbf{N}}}_{12} + \hat{\bar{\mathbf{L}}} = \hat{\bar{\mathbf{J}}},\\
    \hat{\mathbf{N}}_{12}' + \hat{\mathbf{L}}' = \hat{\mathbf{J}},&\,\,\hat{\bar{\mathbf{N}}}_{12}' + \hat{\bar{\mathbf{L}}}' = \hat{\bar{\mathbf{J}}},
\end{split}
\end{align*}
where the non-primed and primed symbols correspond to pre- and post-collisional operators (or quantum numbers). Note that Eq.~\eqref{eq:crosssection} implies that two distinct scattering $S$-matrices are needed to obtain $\sigma^{q}_{\lambda}$ for given kinetic energy, $E_{\rm{kin}}$. The first matrix is computed at the total energy ${E_{T_{i}} = E_{\rm{kin}}+E_{N_{i}}+E_{N_{2}}}$, and the second matrix is calculated at ${E_{T_{f}} = E_{\rm{kin}}+E_{N_{f}}+E_{N_{2}}}$. For the R(0) line in HCl considered in this work, $N_{i}=0$, $N_{f}=1$ and $q=1$.

\begin{figure*}
\includegraphics[width=\linewidth]{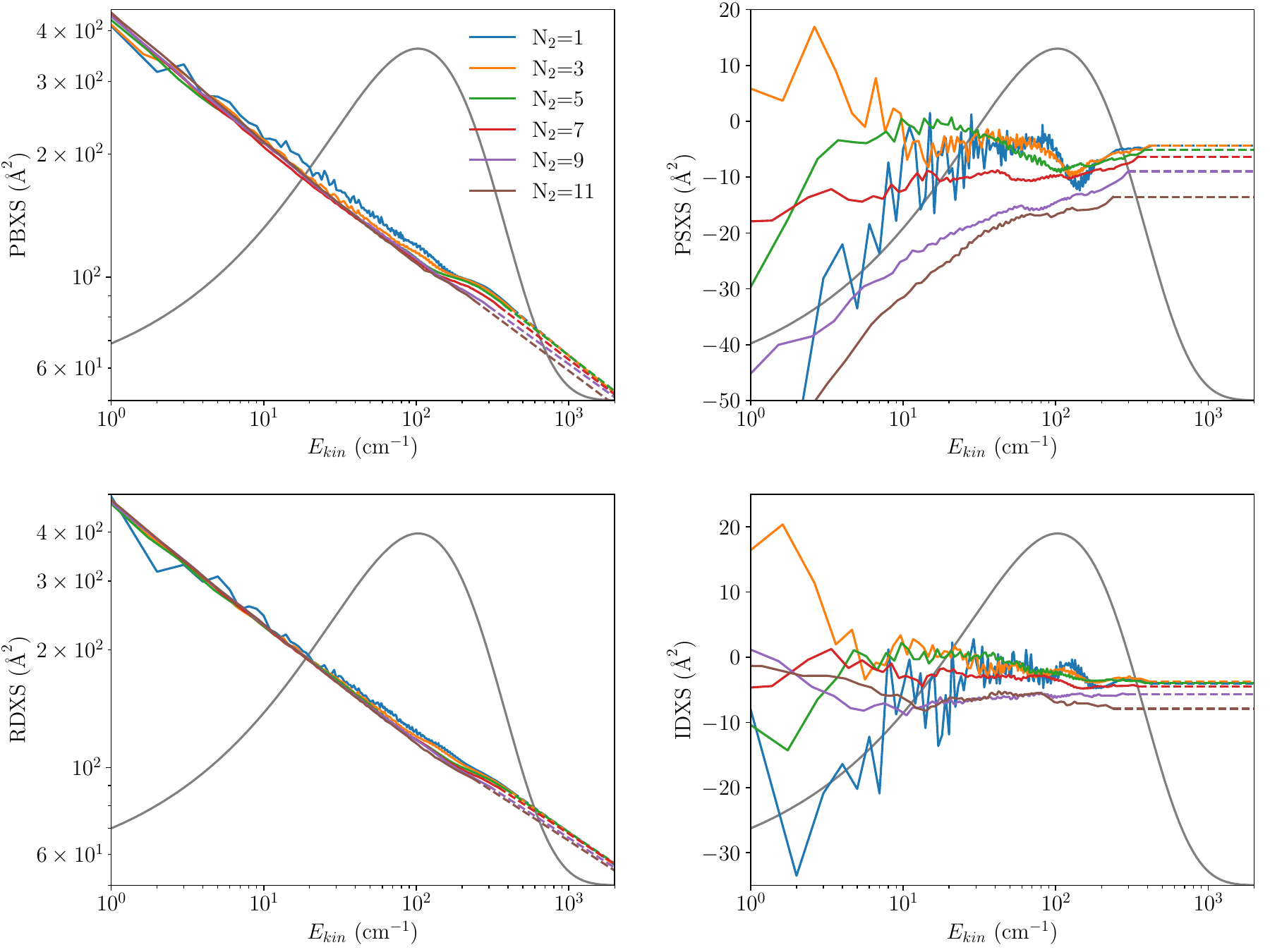}
\caption{\label{fig:xs} Generalized spectroscopic cross sections as a function of relative kinetic energy, $E_{\mathrm{kin}}$, for the first six rotational states ($N_2=1$ to 11) of the perturber molecule (i.e., oxygen molecule). The left and right upper panels correspond to pressure broadening and shift ($\gamma_0$ and $\delta_0$), and the left and right lower panels correspond to the real and imaginary parts of the complex Dicke parameter ($\widetilde{\nu}_{opt}^r$ and $\widetilde{\nu}_{opt}^i$). The solid lines are the results of our \textit{ab initio} calculations, and the dashed lines are the extrapolations (see text for details). As a reference, we plot the Boltzmann $E_{\mathrm{kin}}$ distribution at $T=296$~K as a grey curve (note that for the Boltzmann distribution in all the panels, the vertical axis is linear and has an arbitrary unit).}
\end{figure*}

We made a \ColorVariable{huge} 
effort to ensure convergence of $\sigma^{q}_{\lambda}$ with respect to the rotational basis set, range of $R$ values used to solve the coupled equations, the step of the Numerov propagator, and the number of partial waves (or, equivalently, total angular momenta $J$ and $\bar{J}$) contributing to the sums in Eq.~\eqref{eq:crosssection}. Ultimately, the coupled equations are solved from $R_{\rm{min}} = 5$~$a_{0}$ to $R_{\rm{max}} = 45$~$a_{0}$. The number of steps per half-de Broglie wavelength of the scattering system varied between 15 and 30, depending on the total energy and the total angular momentum to ensure unitary of S-matrices. Additionally, the number of steps was corrected for the depth of the isotropic part of the PES (approximately~107.4~cm$^{-1}$). The sums over $J$ (and $\bar{J}$) were truncated after 4 consecutive $J$ (and $\bar{J}$) blocks contributed less than $10^{-4}$~\AA$^{2}$ to the largest elastic state-to-state cross-section at a given total energy.

In all calculations, the basis set involved all energetically accessible (open) rotational levels of the colliding pair, and a certain number of energetically inaccessible (closed) levels to ensure convergence of the generalized cross-sections at the level of 1\%. In particular, for kinetic energies lower than $250$~cm$^{-1}$ the basis involved 36 energy levels (up to $N_{1_{\rm{max}}}=6, N_{2_{\rm{max}}}=7$). For higher kinetic energies, the size of the basis was gradually increased, reaching 47 rotational levels of the colliding pair at $E_{\mathrm{kin}} = 425$~cm$^{-1}$.

\begin{table}[ht]
   \caption{\label{tab:Tab2} Coefficients of the power-law fits to the PBXS and RDXS (see the corresponding dashed lines in Fig.~\ref{fig:xs}). The coefficients are reported for every rotational state of the perturbed from $N_2=1$ to $11$.}
\begin{ruledtabular}
\begin{tabular}{ccccc}
    \multirow{2}{*}{$N_2$} & \multicolumn{2}{c}{PBXS} & \multicolumn{2}{c}{RDXS} \\
    & $A$ (\AA$^2$) & $b$ & $A$ (\AA$^2$) & $b$ \\
    \hline
    1 & 131.39 & 0.310 & 129.44 & 0.277 \\
    3 & 129.34 & 0.303 & 128.58 & 0.275 \\
    5 & 125.31 & 0.289 & 125.97 & 0.265 \\
    7 & 120.87 & 0.283 & 122.36 & 0.257 \\
    9 & 113.85 & 0.269 & 118.43 & 0.253 \\
    11 & 111.82 & 0.280 & 116.60 & 0.254 \\
\end{tabular}
\end{ruledtabular}
\end{table}

Figure \ref{fig:xs} shows the calculated dependencies of pressure broadening (PBXS), Re$\left[\sigma^1_0(E_{\mathrm{kin}})\right]$, pressure shift (PSXS), Im$\left[\sigma^1_0(E_{\mathrm{kin}})\right]$, and Dicke cross sections (RDXS for Re$\left[\sigma^1_1(E_{\mathrm{kin}})\right]$ and IDXS for Im$\left[\sigma^1_1(E_{\mathrm{kin}})\right]$) on the collsion kinetic energy for various rotational states of the perturber, $N_2$. The solid lines come from our \textit{ab initio} calculations and the dashed lines are extrapolations. The real (PBXS and RDXS) and imaginary (PSXS and IDXS) parts of the generalized cross sections as a function of kinetic energy exhibit very different behavior, therefore we discuss them separately in the two following paragraphs. 

The  PBXS and RDXS as a function of kinetic energy exhibit a power-law behavior. We use this property to extrapolate the \textit{ab initio} cross sections to higher energies,
\begin{align}
    \text{Re} \left[ \sigma^1_{\lambda}(E_{\mathrm{kin}}) \right] = A \left( \frac{E_0}{E_{\mathrm{kin}}} \right)^b \text{,}\label{eq:powerlaw}
\end{align}
where $E_0 = 100$~cm$^{-1}$ is a reference kinetic energy. The fit coefficients, $A$ and $b$, are determined for each rotational level and are given in Table~\ref{tab:Tab2}. The PBXS and RDXS become almost state-independent for larger $N_2$ (such behavior of the PBXS has already been observed for other systems, such as CO-O$_2$,~\cite{zadrożny2022} CO-N$_2$,~\cite{jóźwiak2021} O$_2$–N$_2$,~\cite{gancewski2021}  N$_2$–H$_2$~\cite{gomez2011} and N$_2$–N$_2$~\cite{thibault2011}). We use this property to extrapolate the PBXS and RDXS for $N_2 > 11$ (for $N_2 > 11$, we assume  the same value as for $N_2 = 11$). We discuss the estimations of the uncertainties caused by the extrapolations in Sec.~\ref{sec:Collision-induced line-shape parameters}.

The imaginary parts of the cross sections,
i.e. the PSXS and IDXS, reveal very different features from their real counterparts. First, they are much more demanding in terms of converging the quantum-scattering calculations (i.e., they require a much larger basis, larger integration range, etc., compared to the PBXS and RDXS). Secondly, their dependencies on the kinetic energy and $N_2$ are much less predictable and, hence, more difficult to extrapolate, see the right upper and lower panels in Fig.~\ref{fig:xs}. Due to a lack of more accurate information, we extrapolate PSXS and IDXS to higher energies simply with horizontal lines (see Fig.~\ref{fig:xs}) and assume the same value as for $N_2 = 11$ for $N_2 > 11$. On one hand, these crude extrapolations for PSXS and IDXS inherently induce much larger relative uncertainties compared to the PBXS and RDXS. On the other hand, the magnitudes of PSXS and IDXS are much smaller than PBXS, hence these large relative uncertainties of PSXS and IDXS have small impact on the collision-induced shapes of the HCl line. We provide a detailed discussion of the uncertainties in the next section.

\section{\label{sec:Collision-induced line-shape parameters}Collision-induced line-shape parameters}

The speed-dependent spectroscopic parameters, pressure broadening and shift, $\gamma(v)$ and $\delta(v)$, can be obtained from the generalized spectroscopic cross section, Eq.~\eqref{eq:crosssection}, from the following integral \cite{wcisło2018, stolarczyk2020}
\begin{align}
    \gamma(v) &- i\delta(v) ={} \frac{1}{2\pi c} \frac{1}{k_BT} \frac{2v^2_p}{\sqrt{\pi}v} e^{-\frac{v^2}{v^2_{p}}} \sum_{N_2} p_{N_2}\notag \\
    & \times  \int_0^\infty \text{d}x \,x^2 e^{-x^2} \sinh \left( \frac{2vx}{v_p}  \right) \notag \\
    & \times \sigma_0^1(N_i=0, N_f=1, N_2; E_{\mathrm{kin}}=\mu x^2v^2_p/2) \text{,} \label{eq:parametersv}
\end{align}
where $v$ is a given speed of the active molecule, $v_p=\sqrt{2k_BT/m_2}$ is the most probable perturber speed, $v_r$ is the relative absorber-perturber speed and $x=v_r/v_p$. $N_i$ and $N_f$ denote, respectively, the initial and final rotational quantum numbers of the active molecule. $k_B$ is the Boltzmann constant and $c$ is the speed of light in vacuum. The statistical weight $p_{N_2}$ denotes the population of the $N_2$-th rotational state of the perturbing molecule at given temperature, $T$,
\begin{align}
    p_{N_2}(T) = \frac{1}{Z(T)} (2N_2+1) e^{-E_{N_2}/(k_BT)} \text{,} \label{eq:pN2}
\end{align}
where $E_{N_2}$ is the energy of that rotational state, and
\begin{align}
    Z(T) = \sum_{N_2}(2N_2+1) e^{-E_{N_2}/(k_BT)} \label{eq:ZT}
\end{align}
is the partition function. The summation up to $N_2=39$ covers over 99\% of the perturber population at 296~K. The total nuclear spin of $^{16}$O$_2$ molecule is zero, hence the nuclear spin statistic does not influence Eqs.~\eqref{eq:pN2} and \eqref{eq:ZT}.

\begin{figure*}[ht]
\includegraphics[width=\linewidth]{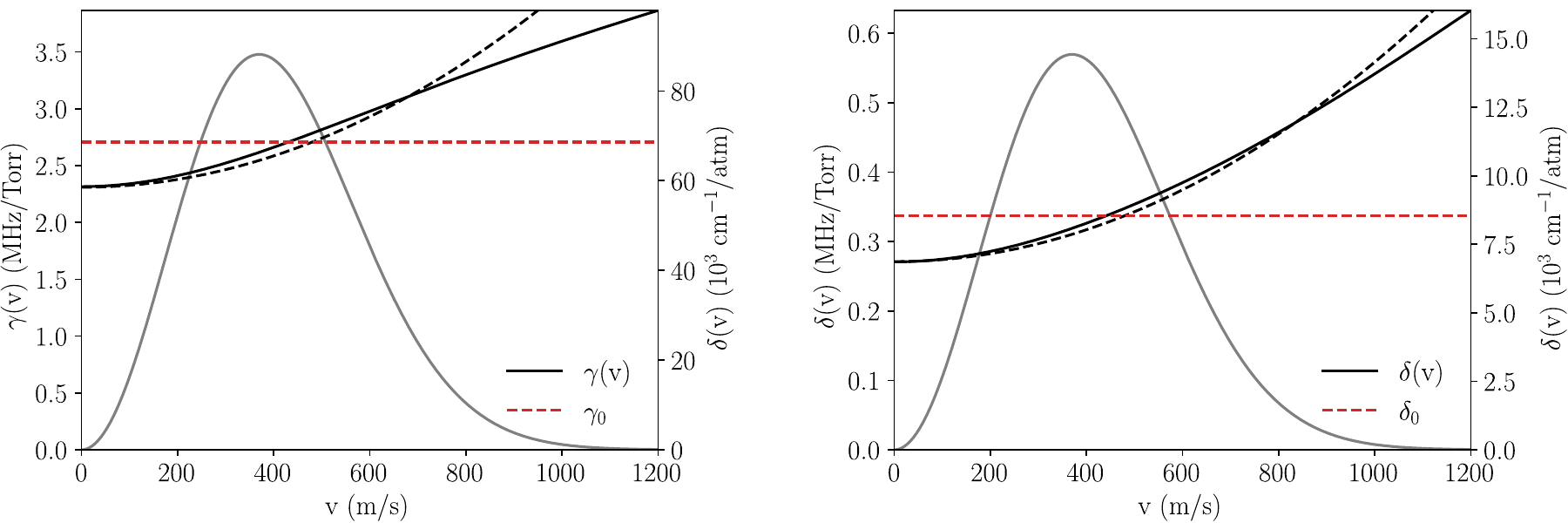}
\caption{\label{fig:parametersv}\textit{Ab initio} speed-dependent pressure broadening and shift, $\gamma(v)$ and $\delta(v)$, for the O$_2$-perturbed 0-0 R(0) line in the H$^{35}$Cl molecule at \SI{296}{\K}, their quadratic approximation (the dashed curve) and the corresponding speed-averaged parameters $\gamma_0$ and $\delta_0$ (the dashed red line). As a reference, the corresponding Maxwellian speed distribution of the perturber (gray solid line) at \SI{296}{\K} is shown in arbitrary units.}
\end{figure*}

The full speed dependencies of $\gamma$ and $\delta$ are often too complex and inconvenient from the perspective of spectroscopic applications (including populating the HITRAN database~\cite{wcisło2021,Stankiewicz_2021}). Therefore, we approximate the $\gamma(v)$ and $\delta(v)$ functions on simple quadratic functions~\cite{rohart1994, stolarczyk2020}
\begin{align}
    \gamma(v) + i \delta(v) \approx \gamma_0 + i \delta_0 + (\gamma_2 + i \delta_2) \left(\frac{v^2}{v_m^2}-\frac{3}{2}\right) \text{,} \label{gammavdeltav}
\end{align}
where $v_m$ is the most probable speed of the active molecules, $\gamma_0$ and $\delta_0$ are the speed-averaged pressure broadening and shift, respectively. They are obtained\cite{stolarczyk2020} at given gas temperature, $T$, as
\begin{align}
    \gamma_0(T) - i\delta_0(T) ={} \frac{1}{2\pi c} \frac{\langle v_r \rangle}{k_BT} \sum_{N_2} p_{N_2}\int_0^{\infty}\text{d}z \,z e^{-z} \notag \\
    \times \sigma_0^1 (N_i, N_f, N_2; E_{\mathrm{kin}}=zk_BT)\text{,} \label{eq:gamma0delta0}
\end{align}
where $\langle v_r \rangle = \sqrt{ 8k_BT/\pi\mu}$ is the mean relative speed of the colliding molecules. $\mu$ is their reduced mass. $z$ is a dimensionless kinetic energy of a collision, $z=E_{\mathrm{kin}}/(k_BT)$. Note that $\gamma_0$ and $\delta_0$ can also be obtained by averaging $\gamma(v)$ and $\delta(v)$ from Eq.~\eqref{eq:parametersv} over the Maxwell-Boltzmann distribution of the absorber, $v$, at a given temperature, $T$. To calculate $\gamma_2$ and $\delta_2$ we assume that the slopes of the quadratic approximation and the actual speed dependencies are equal at $v=v_m$,~\cite{wcisło2018, stolarczyk2020}
\begin{align}
    \frac{\text{d}}{\text{d}v} (\gamma(v) + i \delta(v))|_{v=v_m} = \frac{2}{v_m}(\gamma_2 + i \delta_2) \text{.} \label{eq:gammadeltacondition}
\end{align}
The derivative can be done analytically, which leads to the direct formulas for $\gamma_2$ and $\delta_2$\cite{wcisło2021, stolarczyk2020}
\begin{align}
    \gamma_2(T) &- i\delta_2(T) ={} \frac{1}{2\pi c} \frac{1}{k_BT} \frac{v_p}{\sqrt{\pi}} e^{-y^2} \sum_{N_2} p_{N_2} \int_0^\infty \text{d}x \notag \\
    &\times \bigg[ 2x\cosh (2xy) - \left(\frac{1}{y}+2y\right) \sinh(2xy)\bigg] \notag \\
    &\times x^2 e^{-x^2} \sigma_0^1(N_i, N_f, N_2; E_{\mathrm{kin}}=\mu x^2v^2_p/2) \text{,} \label{eq:gamma2delta2}
\end{align}
where $y=v_m/v_p$.  
In Figure~\ref{fig:parametersv}, the full speed-dependence of our \textit{ab initio} $\gamma(v)$ and $\delta(v)$ at 296~K (black solid lines) is compared with the corresponding quadratic approximation (black dashed lines), Eq.~\eqref{gammavdeltav}, and the speed-averaged values (red dashed lines), Eq.~\eqref{eq:gamma0delta0}. 

In some experiments conducted in the high-pressure regime, where the contribution from the Gaussian component is no longer significant, the spectra are fitted with simple Lorentz profiles. However, the speed-dependent of $\gamma$ and $\delta$ may be non negligible even at high pressure. Therefore, we calculate a weighted sum of Lorentz profiles (WSLPs) for each frequency $\nu$ taking average over the Maxwell-Boltzmann distribution $f_m(\vec{v})$ of the active molecule's velocity,~\cite{pickett1980}
\begin{align}
    I_{\text{WSLP}}(\nu-\nu'_0) &= \frac{1}{\pi} \int \text{d}^3 \vec{v}\, f_m(\vec{v}) \notag \\
    &\times \frac{\gamma(v)}  {\gamma^2(v) + (\nu - \nu'_0 - \delta(v))^2} \text{,} \label{eq:wslp}
\end{align}
where $\nu'_0$ is the unperturbed line position without the pressure shift effect. By fitting a Lorentzian profile to the simulated WSLP, we obtain the effective width of the line, which is smaller than $\gamma_0$ and can be compared with experimental broadening obtained from Lorentzian fits to the measured lines. Note that the effective width of WSLP is smaller than $\gamma_0$ because the speed-dependence of $\gamma$ dominates over the speed-dependence of $\delta$ (otherwise the effective width of WSLP would be larger).

The rate of the velocity-changing collisions is quantified by the complex Dicke parameter, $\widetilde{\nu}_{opt}$, which is calculated as~\cite{thibault2017, wcisło2018}
\begin{align}
    \text{Re} \left[ \widetilde{\nu}_{opt}(T) \right] - i\text{Im}\left[ \widetilde{\nu}_{opt}(T) \right] ={} \frac{1}{2\pi c} \frac{M_2\langle v_r \rangle}{k_BT} \sum_{N_2} p_{N_2} \notag \\
    \times \int_0^{\infty}\text{d}z \,z e^{-z} 
    \bigg[ \frac{2}{3}z\sigma_1^1 (N_i, N_f, N_2; E_{\mathrm{kin}}=zk_BT) \notag \\
    -\sigma_0^1 (N_i, N_f, N_2; E_{\mathrm{kin}}=zk_BT) \bigg] \text{,}
\end{align}
where $M_2=m_2/(m_1+m_2)$. $m_{1/2}$ is the mass of the active/perturbing molecule.

The speed-averaged coefficients were calculated for the given temperature range of 150--360~K and fitted with the power-law,
\begin{align}
    \gamma_0(T) = \gamma_0(T_0) \left(\frac{T_0}{T}\right)^{n_{\gamma_0}} \text{,} \label{eq:gammafit}
\end{align}
where $T_0 = 296$~K is a reference temperature and $n_{\gamma_0}$ is the fitted temperature exponent. Equation~\eqref{eq:gammafit} is written for the case of the $\gamma_0$ parameter but we use the same power-law function for other line-shape parameters. The power-law coefficients are given in Table~\ref{tab:Tab3}, where we compare them with Ref.~\citenum{drouin2004}.

\begin{figure*}
\includegraphics[width=\linewidth]{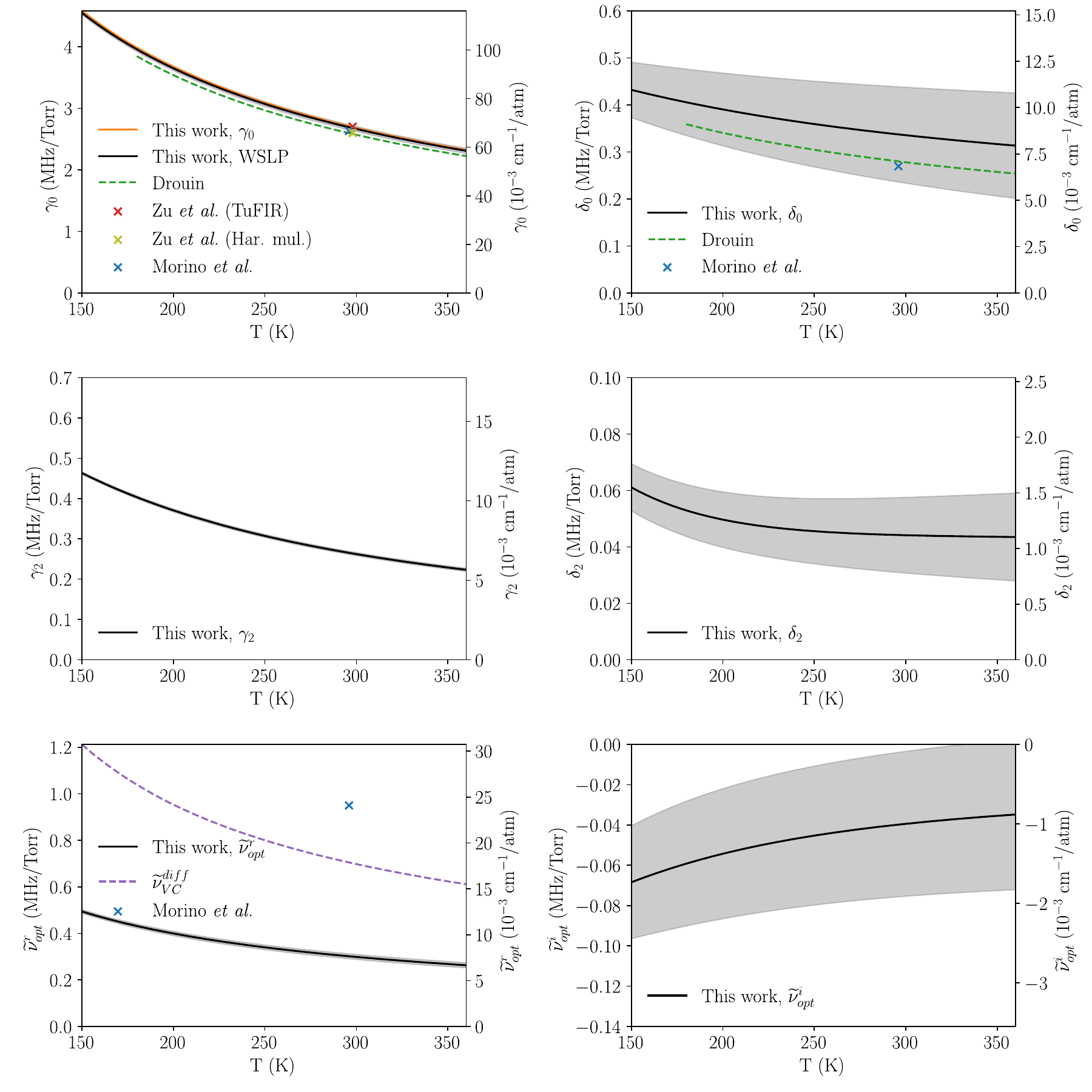}
\caption{\label{fig:parametersT}Temperature dependencies of the $\gamma_0$, $\delta_0$, $\gamma_2$, $\delta_2$ and $\widetilde{\nu}_{opt}$ line-shape parameters for the 0-0 R(0) line in H$^{35}$Cl perturbed by O$_2$. Our \textit{ab initio} results are given by black solid lines (with the effective width of the line, Eq.~\eqref{eq:wslp}, denoted with orange solid line). The shaded areas correspond to the estimated numerical uncertainty of the determined line-shape parameters. The experimental results~\cite{drouin2004, morino2005, zu2003} are marked with the green dashed lines and "x" symbols.}
\end{figure*}

In Figure~\ref{fig:parametersT} we show a comparison of the calculated line-shape parameters with experimental data. Our estimations of their total combined relative uncertainties at 296~K are $1.5$~\% for $\gamma_0$ and $\gamma_2$, $3.5$~\% for $\widetilde{\nu}_{opt}^r$, 30~\% for $\delta_0$ and $\delta_2$, and $90$~\% for $\widetilde{\nu}_{opt}^i$. The contributions from the quantum-scattering calculations are almost negligible, i.e., 1~\% level for imaginary parts and much smaller uncertainty for the real parts of the cross sections. The main contributions to the total uncertainties at 296~K originate from the cross-section extrapolations to higher kinetic energies and higher perturber's rotational states, see the dashed lines in Fig.~\ref{fig:xs}. Our estimated uncertainties are depicted in Fig.~\ref{fig:parametersT} as gray shadows - they decrease with decreasing temperature because the extrapolations to higher temperatures and higher $N_2$ play a smaller role at lower temperatures. The much larger uncertainty for the shift and imaginary part of the complex Dicke parameter is because the dependence of the corresponding cross sections (see the right panel in Fig.~\ref{fig:xs}) on kinetic energy and $N_2$ is more difficult to predict, and hence, to extrapolate.

\begin{table*}
   \caption{\label{tab:Tab3}Comparison between our \textit{ab initio} line-shape parameters and the available experimental data for the 0-0 R(0) line in H$^{35}$Cl perturbed by O$_2$. All line-shape parameters are expressed in units of MHz/Torr. $n$ is a dimensionless fitted temperature exponent, see Eq.~\eqref{eq:gammafit}. The numbers in parentheses correspond, in the case of our theoretical results, to the estimated numerical uncertainty, and in the case of experimental data, to one standard deviation.}
\begin{ruledtabular}
\begin{tabular}{lcllllll}
       & T~(K) & $\gamma_0$ & $\delta_0$ & $\gamma_2$ & $\delta_2$ & $\nu^r_{\text{opt}}$ & $\nu^i_{\text{opt}}$ \\
    \hline
    This work & 296 & 2.707(41) & 0.34(11) & 0.2656(40) & 0.044(14) & 0.302(11) & -0.040(36) \\
    This work, WSLP & 296
    & 2.691(41) & & & & & \\
    This work & 298 & 2.693(41) & 0.34(11) & 0.2636(40) & 0.044(14) & 0.300(11) & -0.040(36) \\
    This work, WSLP & 298
    & 2.676(41) & & & & & \\
    \hline
    Zu \textit{et al.} (TuFIR\footnote{A high resolution tunable far infrared spectrometer})~\cite{zu2003} & 298 & 2.70(5) & & & & & \\
    Zu \textit{et al.} (Har. mul.\footnote{A harmonic multiplier radiation source})~\cite{zu2003} & 298 & 2.60 & & & & & \\
    Drouin~\cite{drouin2004} & 296 & 2.595(5) & 0.280(1) & & & & \\
    Morino \textit{et al.}~\cite{morino2005} & 296 & 2.632(15) & 0.2697(37) & & & \ColorVariable{0.950(33)} & \\
    \hline
    \hline
      & T~(K) & $n_{\gamma_0}$ & $n_{\delta_0}$ & $n_{\gamma_2}$ & $n_{\delta_2}$ & $n_{\nu^r_{\text{opt}}}$ & $n_{\nu^i_{\text{opt}}}$ \\
    \hline
    This work & 296 & 0.776(21) & 0.37(20) & 0.837(22) & 0.36(18) & 0.721(45) & 0.79(74) \\
    This work, WSLP & 296 & 0.774(21) & & & & & \\
    Drouin~\cite{drouin2004} & 296 & 0.79(1) & 0.50(3) & & & & \\
\end{tabular}
\end{ruledtabular}
\end{table*}

However, the outcome of the uncertainty analysis, in the context of accurate spectra simulations and spectroscopic data analysis, should not be the uncertainties of the particular line-shape parameters, but how they propagate on the simulated line shapes. It was shown in Ref.~\citenum{słowiński2022} that a 1.5~\% uncertainty in $\gamma_0$ results in relative RMSE of profile residuals of 0.7~\% (see Eq.~(A.4) in Ref.~\citenum{słowiński2022}). The 30~\% uncertainty of the $\delta_0$ parameter may seem to be the most worrying in this context. However, to quantify its impact on the line shape one should not consider the relative $\delta_0$ uncertainty, but relate the absolute uncertainty of $\delta_0$ to the width of the line (note that $\gamma_0$ is much larger than $\delta_0$), see Eq.~(A.7) in Ref.~\citenum{słowiński2022}. It turns out that 30~\% uncertainty in $\delta_0$ results in only 1.7~\% line-shape residuals. The four other line-shape parameters ($\gamma_2$, $\delta_2$, $\widetilde{\nu}_{opt}^r$ and $\widetilde{\nu}_{opt}^i$) have small impacts on the line-shape. Assuming statistical independence of the uncertainties of the six line-shape parameters, we estimate that the total combined uncertainties of our calculations and extrapolations correspond to 2~\% relative RMSE residuals in the line-shape at 296~K.

The calculated line-shape parameters and the available experimental data are in a good agreement across the investigated temperature range (see Fig.~\ref{fig:parametersT}). In the case of the $\delta_0$ parameter, all the experimental points lie within one combined standard uncertainty from our \textit{ab initio} values. Despite that, at the first glance, the uncertainties of $\delta_0$ may seem large, we recall that they should be related to the magnitude of $\gamma_0$ (rather than $\delta_0$) to actually reflect its impact on the final collision-perturbed shape of the line.~\cite{słowiński2022}

The comparison between the experimentally determined real part of the Dicke parameter, $\widetilde{\nu}^r_{opt}$, with its \textit{ab initio} value  gives a unique chance to give an important input to the long-lasting discussion on physical meaning of the complex Dicke parameter. For the molecular systems relevant to the Earth's atmosphere, the complex Dicke parameter was determined only from phenomenological fits to experimental spectra and the physical meaning of its value is questionable. An notable exception is Ref.\citenum{Paredes_Roibas_2021}, where the authors provided values of the complex Dicke parameter for the N$_{2}$-perturbed 1-0 Q(0) line in CO (see Appendix A therein).  However, this particular line is not observed in the absorption spectra of CO. It is accessible only through Raman spectroscopy, which limits its direct relevance to atmospheric studies. In this work, we show that the value of real part of the complex Dicke parameter obtained from the fit to experimental spectra~\cite{morino2005} for the case of the line considered in this article is almost three times larger than its actual value, see Fig.~\ref{fig:parametersT}. This means that the line-shape fits performed in Ref.~\citenum{morino2005} well reproduced the line shapes in a geometrical sense but the underlying line-shape parameters converged to unphysical values. This is caused by strong numerical correlations between the line-shape parameters. The problem of retriving unphysical values of the fitted line-shape parameters is known in the literature and it is especially pronounced for fits based on the beyond-Voigt profiles. For instance, it was reported for the case of molecular hydrogen, see Table 2 and Fig. 3 in Ref.~\citenum{wcisło2016} or Table 1 and Fig. 4 in Ref.~\citenum{wcisło2018}.

Before, the complex Dicke parameter was calculated from the first principles only for simpler systems such as atom-perturbed hydrogen molecule~\cite{słowiński2020, słowiński2022, stolarczyk2023} or self-perturbed hydrogen molecule~\cite{wcisło2018} (the exception is the recent work for CO-N$_2$\cite{jóźwiak2021}). For the case of molecular hydrogen, however, the contribution from inelastic collisions is so small that $\widetilde{\nu}^r_{opt}$ is very close to the frequency of the velocity-changing collisions calculated from the diffusion coefficient, see Appendix B in Ref.~\citenum{thibault2017}. From this perspective the molecular hydrogen case is trivial because the simple kinetic considerations give almost the same results as the full quantum calculations of the generalized spectroscopic cross sections. In this work, we probe a very different regime where $\widetilde{\nu}^r_{opt}$ is over two times smaller than the frequency of the velocity changing collisions for HCl, $\widetilde{\nu}_{VC}^{diff}$ (see the dashed purple line in Fig.~\ref{fig:parametersT}). Comprehensive details regarding the determination of $\widetilde{\nu}_{VC}^{diff}$ are available in the Appendix~\ref{sec:diffusion}. In this case, $\widetilde{\nu}^r_{opt}$ really holds information on very nontrivial effect of correlation between a velocity-changing collision and perturbation of optical coherence associated with the considered transition.

Experimental values for $\gamma_2$, $\delta_2$ and the imaginary part of the Dicke parameter, $\widetilde{\nu}^i_{\text{opt}}$, are currently unavailable. A detailed summary of the results, which contains the room temperature coefficients, fitting parameters and their uncertainties, can be found in Table \ref{tab:Tab3}. The parameters are connected with their pressure-dependent counterparts as $\Gamma_0=\gamma_0 p$, $\Delta_0=\delta_0 p$, $\Gamma_2=\gamma_2 p$, $\Delta_2=\delta_2 p$ and $\nu_{opt}=\widetilde{\nu}_{opt} p$.

\section{\label{sec:Conclusion}Conclusion}
We presented the results of the first-ever fully quantum calculations of the line-shape parameters of HCl molecular resonances perturbed by a diatomic molecule. In particular, we investigated the effect of the atmospheric O$_2$-induced perturbations on the shape of the pure rotational R(0) transition in HCl. Using the newly calculated \textit{ab initio} HCl($X^1\Sigma^+$)-O$_2$($X^3\Sigma^-_g$) potential energy surface, we performed quantum scattering calculations using the full close-coupling approach, without making additional approximations. We used the determined S-matrix elements to compute the generalized spectroscopic cross sections and we analyzed their structure using the discussed convergence and extrapolation criteria. By averaging the cross sections, we determined the temperature-dependence of the pressure broadening and shift, as well as the complex Dicke parameter for this line. This is the first \textit{ab initio} calculation of the complex Dicke parameter done for the HCl molecule. We included the speed-dependence of the calculated line-shape parameters in the analysis, and obtained agreement with the available experimental data. 

This study constitutes a major step towards accurate theoretical determination of the line-shape parameters for molecular systems of atmospheric interest, perturbed by interactions with other air molecules. Such investigations are important for studying the intermolecular interactions between atmospheric species, as well as the interplay of  different collisional effects and the discrimination between the line-shape models used to recover the atmospheric information from the remote-sensed molecular spectra. These results are relevant for accurate modeling of the atmospheric spectra of HCl -- the main reservoir of ozone-depleting chlorine -- and for populating the spectroscopic databases (e.g., HITRAN, GEISA) with line-shape parameters determined using the first principles of quantum mechanics. Furthermore, such theoretical investigations provide an independent test (as well as guidance) for the methodologies of analysing the collisional data.

\section{\label{sec:supplementary}Supplementary material}
See the supplementary material associated with this article for the tabulated generalized spectroscopic cross sections used in the line-shape calculations.

\begin{acknowledgments}
A.O. and H.J. are supported by the National Science Centre in Poland through Project No. 2018/31/B/ST2/00720. M.G. is supported by the National Science Centre in Poland through Project No. 2021/43/O/ST2/0021. R.D. and E.Q.S. are supported by the U.S. Department of Energy (Award DE-SC0019740). P.W. is supported by the National Science Centre in Poland through Project No. 2022/46/E/ST2/00282. \ColorVariable{For the purpose of Open Access, the author has applied a CC-BY public copyright licence to any Author Accepted Manuscript (AAM) version arising from this submission.} We gratefully acknowledge Polish high-performance computing infrastructure PLGrid (HPC Centers: ACK Cyfronet AGH, CI TASK) for providing computer facilities and support within computational grant no. PLG/2023/016409. Calculations have been carried out using resources provided by Wroclaw Centre for Networking and Supercomputing (http://wcss.pl), grant no. 546. The research is a part of the program of the National Laboratory FAMO in Toruń, Poland.
\end{acknowledgments}

\section*{Author Declarations}
\subsection*{Conflict of interest}
The authors have no conflicts to disclose.
    
\section*{Data availability}
The data that support the findings of this study are available from the corresponding author upon reasonable request.

\appendix

\section{\label{sec:diffusion}Frequency of the velocity changing collisions}

In Fig.~\ref{fig:parametersT} we compare the real part of the Dicke parameter, $\widetilde{\nu}^r_{opt}$, with the frequency of the velocity changing collisions for HCl, $\widetilde{\nu}_{VC}^{diff}$. To determine this value, we employ the principle that $\widetilde{\nu}_{VC}^{diff}$ is related with the mass diffusion coefficient, $D$, via $\nu_{VC}^{diff} = v_m^2 / (2D)$, see Appendix A in Ref.~\citenum{wcisło2016}. We use a corresponding pressure-independent quantity as $\widetilde{\nu}_{VC}^{diff} = \nu_{VC}^{diff} / p$, where $p$ is pressure. The ﬁrst-order approximation of the diffusion coefficient can be calculated as~\cite{hirschfelder1954}
\begin{align}
    D = \frac{3}{16} \frac{(k_{B}T)^{2}}{\mu p} \frac{1}{\Omega^{(1,1)} (T)},
\end{align}
where $\Omega^{(1,1)} (T)$ is the collision integral introduced by Chapman and Cowling.~\cite{Chapman_Cowling} Here, we determine $\widetilde{\nu}_{VC}^{diff}$ by fitting the isotropic term of Eq.~\eqref{eq:potential} to the Lennard-Jones (LJ) potential. We use a standard form of the LJ potential
\begin{equation}
    V(R)=4\epsilon\Biggl[\Bigl(\frac{\sigma}{R}\Bigr)^{12}-\Bigl(\frac{\sigma}{R}\Bigr)^{6}\Biggr],
\end{equation}
where $\epsilon$ is the depth of the potential well, and $\sigma$ is the value of $R$ for which $V(R)=0$. The fitted LJ parameters (${\epsilon = 57.1\pm0.4}$~cm$^{-1}$ and $\sigma= 6.69\pm0.01\,a_{0}$), along with tabulated values of the reduced collision integrals $\Omega^{(1,1)\,*}$ (see Tab.~I-M and Chapter 8.4 in Ref.~\citenum{hirschfelder1954}), are then used to calculate $\Omega^{(1,1)}$.


\bibliography{bibliography}

\begin{thebibliography}{98}%
\makeatletter
\providecommand \@ifxundefined [1]{%
 \@ifx{#1\undefined}
}%
\providecommand \@ifnum [1]{%
 \ifnum #1\expandafter \@firstoftwo
 \else \expandafter \@secondoftwo
 \fi
}%
\providecommand \@ifx [1]{%
 \ifx #1\expandafter \@firstoftwo
 \else \expandafter \@secondoftwo
 \fi
}%
\providecommand \natexlab [1]{#1}%
\providecommand \enquote  [1]{``#1''}%
\providecommand \bibnamefont  [1]{#1}%
\providecommand \bibfnamefont [1]{#1}%
\providecommand \citenamefont [1]{#1}%
\providecommand \href@noop [0]{\@secondoftwo}%
\providecommand \href [0]{\begingroup \@sanitize@url \@href}%
\providecommand \@href[1]{\@@startlink{#1}\@@href}%
\providecommand \@@href[1]{\endgroup#1\@@endlink}%
\providecommand \@sanitize@url [0]{\catcode `\\12\catcode `\$12\catcode
  `\&12\catcode `\#12\catcode `\^12\catcode `\_12\catcode `\%12\relax}%
\providecommand \@@startlink[1]{}%
\providecommand \@@endlink[0]{}%
\providecommand \url  [0]{\begingroup\@sanitize@url \@url }%
\providecommand \@url [1]{\endgroup\@href {#1}{\urlprefix }}%
\providecommand \urlprefix  [0]{URL }%
\providecommand \Eprint [0]{\href }%
\providecommand \doibase [0]{http://dx.doi.org/}%
\providecommand \selectlanguage [0]{\@gobble}%
\providecommand \bibinfo  [0]{\@secondoftwo}%
\providecommand \bibfield  [0]{\@secondoftwo}%
\providecommand \translation [1]{[#1]}%
\providecommand \BibitemOpen [0]{}%
\providecommand \bibitemStop [0]{}%
\providecommand \bibitemNoStop [0]{.\EOS\space}%
\providecommand \EOS [0]{\spacefactor3000\relax}%
\providecommand \BibitemShut  [1]{\csname bibitem#1\endcsname}%
\let\auto@bib@innerbib\@empty
\bibitem [{\citenamefont {Zander}\ \emph {et~al.}(1992)\citenamefont {Zander},
  \citenamefont {Gunson}, \citenamefont {Farmer}, \citenamefont {Rinsland},
  \citenamefont {Irion},\ and\ \citenamefont {Mahieu}}]{zander1992}%
  \BibitemOpen
  \bibfield  {author} {\bibinfo {author} {\bibfnamefont {R.}~\bibnamefont
  {Zander}}, \bibinfo {author} {\bibfnamefont {M.~R.}\ \bibnamefont {Gunson}},
  \bibinfo {author} {\bibfnamefont {C.~B.}\ \bibnamefont {Farmer}}, \bibinfo
  {author} {\bibfnamefont {C.~P.}\ \bibnamefont {Rinsland}}, \bibinfo {author}
  {\bibfnamefont {F.~W.}\ \bibnamefont {Irion}}, \ and\ \bibinfo {author}
  {\bibfnamefont {E.}~\bibnamefont {Mahieu}},\ }\href {\doibase
  10.1007/BF00053758} {\bibfield  {journal} {\bibinfo  {journal} {J. Atmos.
  Chem.}\ }\textbf {\bibinfo {volume} {15}},\ \bibinfo {pages} {171} (\bibinfo
  {year} {1992})}\BibitemShut {NoStop}%
\bibitem [{\citenamefont {Solomon}(1999)}]{solomon1999}%
  \BibitemOpen
  \bibfield  {author} {\bibinfo {author} {\bibfnamefont {S.}~\bibnamefont
  {Solomon}},\ }\href {\doibase 10.1029/1999RG900008} {\bibfield  {journal}
  {\bibinfo  {journal} {Rev. Geophys.}\ }\textbf {\bibinfo {volume} {37}},\
  \bibinfo {pages} {275} (\bibinfo {year} {1999})}\BibitemShut {NoStop}%
\bibitem [{\citenamefont {Farman}\ \emph {et~al.}(1985)\citenamefont {Farman},
  \citenamefont {Gardiner},\ and\ \citenamefont {Shanklin}}]{farman1985}%
  \BibitemOpen
  \bibfield  {author} {\bibinfo {author} {\bibfnamefont {J.~C.}\ \bibnamefont
  {Farman}}, \bibinfo {author} {\bibfnamefont {B.~G.}\ \bibnamefont
  {Gardiner}}, \ and\ \bibinfo {author} {\bibfnamefont {J.~D.}\ \bibnamefont
  {Shanklin}},\ }\href {\doibase 10.1038/315207a0} {\bibfield  {journal}
  {\bibinfo  {journal} {Nature}\ }\textbf {\bibinfo {volume} {315}},\ \bibinfo
  {pages} {207–210} (\bibinfo {year} {1985})}\BibitemShut {NoStop}%
\bibitem [{\citenamefont {Polvani}\ \emph {et~al.}(2023)\citenamefont
  {Polvani}, \citenamefont {Keeble}, \citenamefont {Banerjee}, \citenamefont
  {Checa-Garcia}, \citenamefont {Chiodo}, \citenamefont {Rieder},\ and\
  \citenamefont {Rosenlof}}]{polvani2023}%
  \BibitemOpen
  \bibfield  {author} {\bibinfo {author} {\bibfnamefont {L.~M.}\ \bibnamefont
  {Polvani}}, \bibinfo {author} {\bibfnamefont {J.}~\bibnamefont {Keeble}},
  \bibinfo {author} {\bibfnamefont {A.}~\bibnamefont {Banerjee}}, \bibinfo
  {author} {\bibfnamefont {R.}~\bibnamefont {Checa-Garcia}}, \bibinfo {author}
  {\bibfnamefont {G.}~\bibnamefont {Chiodo}}, \bibinfo {author} {\bibfnamefont
  {H.~E.}\ \bibnamefont {Rieder}}, \ and\ \bibinfo {author} {\bibfnamefont
  {K.~H.}\ \bibnamefont {Rosenlof}},\ }\href {\doibase
  10.1038/s41467-023-37134-3} {\bibfield  {journal} {\bibinfo  {journal} {Nat.
  Commun.}\ }\textbf {\bibinfo {volume} {14}},\ \bibinfo {pages} {1608}
  (\bibinfo {year} {2023})}\BibitemShut {NoStop}%
\bibitem [{\citenamefont {Newman}\ \emph {et~al.}(2009)\citenamefont {Newman},
  \citenamefont {Oman}, \citenamefont {Douglass}, \citenamefont {Fleming},
  \citenamefont {Frith}, \citenamefont {Hurwitz}, \citenamefont {Kawa},
  \citenamefont {Jackman}, \citenamefont {Krotkov}, \citenamefont {Nash},
  \citenamefont {Nielsen}, \citenamefont {Pawson}, \citenamefont {Stolarski},\
  and\ \citenamefont {Velders}}]{newman2009}%
  \BibitemOpen
  \bibfield  {author} {\bibinfo {author} {\bibfnamefont {P.~A.}\ \bibnamefont
  {Newman}}, \bibinfo {author} {\bibfnamefont {L.~D.}\ \bibnamefont {Oman}},
  \bibinfo {author} {\bibfnamefont {A.~R.}\ \bibnamefont {Douglass}}, \bibinfo
  {author} {\bibfnamefont {E.~L.}\ \bibnamefont {Fleming}}, \bibinfo {author}
  {\bibfnamefont {S.~M.}\ \bibnamefont {Frith}}, \bibinfo {author}
  {\bibfnamefont {M.~M.}\ \bibnamefont {Hurwitz}}, \bibinfo {author}
  {\bibfnamefont {S.~R.}\ \bibnamefont {Kawa}}, \bibinfo {author}
  {\bibfnamefont {C.~H.}\ \bibnamefont {Jackman}}, \bibinfo {author}
  {\bibfnamefont {N.~A.}\ \bibnamefont {Krotkov}}, \bibinfo {author}
  {\bibfnamefont {E.~R.}\ \bibnamefont {Nash}}, \bibinfo {author}
  {\bibfnamefont {J.~E.}\ \bibnamefont {Nielsen}}, \bibinfo {author}
  {\bibfnamefont {S.}~\bibnamefont {Pawson}}, \bibinfo {author} {\bibfnamefont
  {R.~S.}\ \bibnamefont {Stolarski}}, \ and\ \bibinfo {author} {\bibfnamefont
  {G.~J.~M.}\ \bibnamefont {Velders}},\ }\href {\doibase
  10.5194/acp-9-2113-2009} {\bibfield  {journal} {\bibinfo  {journal} {Atmos.
  Chem. Phys.}\ }\textbf {\bibinfo {volume} {9}},\ \bibinfo {pages}
  {2113–2128} (\bibinfo {year} {2009})}\BibitemShut {NoStop}%
\bibitem [{\citenamefont {M{\"a}der}\ \emph {et~al.}(2010)\citenamefont
  {M{\"a}der}, \citenamefont {Staehelin}, \citenamefont {Peter}, \citenamefont
  {Brunner}, \citenamefont {Rieder},\ and\ \citenamefont {Stahel}}]{mader2010}%
  \BibitemOpen
  \bibfield  {author} {\bibinfo {author} {\bibfnamefont {J.~A.}\ \bibnamefont
  {M{\"a}der}}, \bibinfo {author} {\bibfnamefont {J.}~\bibnamefont
  {Staehelin}}, \bibinfo {author} {\bibfnamefont {T.}~\bibnamefont {Peter}},
  \bibinfo {author} {\bibfnamefont {D.}~\bibnamefont {Brunner}}, \bibinfo
  {author} {\bibfnamefont {H.~E.}\ \bibnamefont {Rieder}}, \ and\ \bibinfo
  {author} {\bibfnamefont {W.~A.}\ \bibnamefont {Stahel}},\ }\href {\doibase
  10.5194/acp-10-12161-2010} {\bibfield  {journal} {\bibinfo  {journal} {Atmos.
  Chem. Phys.}\ }\textbf {\bibinfo {volume} {10}},\ \bibinfo {pages}
  {12161–12171} (\bibinfo {year} {2010})}\BibitemShut {NoStop}%
\bibitem [{\citenamefont {Solomon}\ \emph {et~al.}(2016)\citenamefont
  {Solomon}, \citenamefont {Ivy}, \citenamefont {Kinnison}, \citenamefont
  {Mills}, \citenamefont {R.~R.~Neely},\ and\ \citenamefont
  {Schmidt}}]{solomon2016}%
  \BibitemOpen
  \bibfield  {author} {\bibinfo {author} {\bibfnamefont {S.}~\bibnamefont
  {Solomon}}, \bibinfo {author} {\bibfnamefont {D.~J.}\ \bibnamefont {Ivy}},
  \bibinfo {author} {\bibfnamefont {D.}~\bibnamefont {Kinnison}}, \bibinfo
  {author} {\bibfnamefont {M.~J.}\ \bibnamefont {Mills}}, \bibinfo {author}
  {\bibfnamefont {I.}~\bibnamefont {R.~R.~Neely}}, \ and\ \bibinfo {author}
  {\bibfnamefont {A.}~\bibnamefont {Schmidt}},\ }\href {\doibase
  10.1126/science.aae0061} {\bibfield  {journal} {\bibinfo  {journal}
  {Science}\ }\textbf {\bibinfo {volume} {353}},\ \bibinfo {pages} {269}
  (\bibinfo {year} {2016})}\BibitemShut {NoStop}%
\bibitem [{\citenamefont {{World Meteorological Organization (WMO),
  \textit{Scientific Assessment of Ozone Depletion: 2022}, GAW Report No. 278,
  509 pp., WMO, Geneva, 2022}}()}]{WMO2022}%
  \BibitemOpen
  \bibfield  {author} {\bibinfo {author} {\bibnamefont {{World Meteorological
  Organization (WMO), \textit{Scientific Assessment of Ozone Depletion: 2022},
  GAW Report No. 278, 509 pp., WMO, Geneva, 2022}}},\ }\href@noop {} {\
  }\BibitemShut {NoStop}%
\bibitem [{\citenamefont {Nassar}\ \emph {et~al.}(2006)\citenamefont {Nassar},
  \citenamefont {Bernath}, \citenamefont {Boone}, \citenamefont {Clerbaux},
  \citenamefont {Coheur}, \citenamefont {Dufour}, \citenamefont {Froidevaux},
  \citenamefont {Mahieu}, \citenamefont {McConnell}, \citenamefont {McLeod},
  \citenamefont {Murtagh}, \citenamefont {Rinsland}, \citenamefont {Semeniuk},
  \citenamefont {Skelton}, \citenamefont {Walker},\ and\ \citenamefont
  {Zander}}]{nassar2006}%
  \BibitemOpen
  \bibfield  {author} {\bibinfo {author} {\bibfnamefont {R.}~\bibnamefont
  {Nassar}}, \bibinfo {author} {\bibfnamefont {P.~F.}\ \bibnamefont {Bernath}},
  \bibinfo {author} {\bibfnamefont {C.~D.}\ \bibnamefont {Boone}}, \bibinfo
  {author} {\bibfnamefont {C.}~\bibnamefont {Clerbaux}}, \bibinfo {author}
  {\bibfnamefont {P.~F.}\ \bibnamefont {Coheur}}, \bibinfo {author}
  {\bibfnamefont {G.}~\bibnamefont {Dufour}}, \bibinfo {author} {\bibfnamefont
  {L.}~\bibnamefont {Froidevaux}}, \bibinfo {author} {\bibfnamefont
  {E.}~\bibnamefont {Mahieu}}, \bibinfo {author} {\bibfnamefont {J.~C.}\
  \bibnamefont {McConnell}}, \bibinfo {author} {\bibfnamefont {S.~D.}\
  \bibnamefont {McLeod}}, \bibinfo {author} {\bibfnamefont {D.~P.}\
  \bibnamefont {Murtagh}}, \bibinfo {author} {\bibfnamefont {C.~P.}\
  \bibnamefont {Rinsland}}, \bibinfo {author} {\bibfnamefont {K.}~\bibnamefont
  {Semeniuk}}, \bibinfo {author} {\bibfnamefont {R.}~\bibnamefont {Skelton}},
  \bibinfo {author} {\bibfnamefont {K.~A.}\ \bibnamefont {Walker}}, \ and\
  \bibinfo {author} {\bibfnamefont {R.}~\bibnamefont {Zander}},\ }\href
  {\doibase 10.1029/2006JD007073} {\bibfield  {journal} {\bibinfo  {journal}
  {J. Geophys. Res. Atmos.}\ }\textbf {\bibinfo {volume} {111}},\ \bibinfo
  {pages} {D22312} (\bibinfo {year} {2006})}\BibitemShut {NoStop}%
\bibitem [{\citenamefont {Connes}\ \emph {et~al.}(1967)\citenamefont {Connes},
  \citenamefont {Connes}, \citenamefont {Benedict},\ and\ \citenamefont
  {Kaplan}}]{connes1967}%
  \BibitemOpen
  \bibfield  {author} {\bibinfo {author} {\bibfnamefont {P.}~\bibnamefont
  {Connes}}, \bibinfo {author} {\bibfnamefont {J.}~\bibnamefont {Connes}},
  \bibinfo {author} {\bibfnamefont {W.~S.}\ \bibnamefont {Benedict}}, \ and\
  \bibinfo {author} {\bibfnamefont {L.~D.}\ \bibnamefont {Kaplan}},\ }\href
  {\doibase 10.1086/149124} {\bibfield  {journal} {\bibinfo  {journal} {ApJ}\
  }\textbf {\bibinfo {volume} {147}},\ \bibinfo {pages} {1230} (\bibinfo {year}
  {1967})}\BibitemShut {NoStop}%
\bibitem [{\citenamefont {Ridgway}\ \emph {et~al.}(1984)\citenamefont
  {Ridgway}, \citenamefont {Carbon}, \citenamefont {Hall},\ and\ \citenamefont
  {Jewell}}]{ridgway1984}%
  \BibitemOpen
  \bibfield  {author} {\bibinfo {author} {\bibfnamefont {S.~T.}\ \bibnamefont
  {Ridgway}}, \bibinfo {author} {\bibfnamefont {D.~F.}\ \bibnamefont {Carbon}},
  \bibinfo {author} {\bibfnamefont {D.~N.~B.}\ \bibnamefont {Hall}}, \ and\
  \bibinfo {author} {\bibfnamefont {J.}~\bibnamefont {Jewell}},\ }\href
  {\doibase 10.1086/190925} {\bibfield  {journal} {\bibinfo  {journal} {ApJ
  Suppl. Ser.}\ }\textbf {\bibinfo {volume} {54}},\ \bibinfo {pages} {177}
  (\bibinfo {year} {1984})}\BibitemShut {NoStop}%
\bibitem [{\citenamefont {Zmuidzinas}\ \emph {et~al.}(1995)\citenamefont
  {Zmuidzinas}, \citenamefont {Blake}, \citenamefont {Carlstrom}, \citenamefont
  {Keene},\ and\ \citenamefont {Miller}}]{zmuidzinas1995}%
  \BibitemOpen
  \bibfield  {author} {\bibinfo {author} {\bibfnamefont {J.}~\bibnamefont
  {Zmuidzinas}}, \bibinfo {author} {\bibfnamefont {G.~A.}\ \bibnamefont
  {Blake}}, \bibinfo {author} {\bibfnamefont {J.}~\bibnamefont {Carlstrom}},
  \bibinfo {author} {\bibfnamefont {J.}~\bibnamefont {Keene}}, \ and\ \bibinfo
  {author} {\bibfnamefont {D.}~\bibnamefont {Miller}},\ }\href {\doibase
  10.1086/309570} {\bibfield  {journal} {\bibinfo  {journal} {ApJ}\ }\textbf
  {\bibinfo {volume} {447}},\ \bibinfo {pages} {L125} (\bibinfo {year}
  {1995})}\BibitemShut {NoStop}%
\bibitem [{\citenamefont {Jura}(1974)}]{jura1974}%
  \BibitemOpen
  \bibfield  {author} {\bibinfo {author} {\bibfnamefont {M.}~\bibnamefont
  {Jura}},\ }\href {\doibase 10.1086/181497} {\bibfield  {journal} {\bibinfo
  {journal} {ApJ}\ }\textbf {\bibinfo {volume} {190}},\ \bibinfo {pages} {L33}
  (\bibinfo {year} {1974})}\BibitemShut {NoStop}%
\bibitem [{\citenamefont {Dalgarno}\ \emph {et~al.}(1974)\citenamefont
  {Dalgarno}, \citenamefont {de~Jong}, \citenamefont {Oppenheimer},\ and\
  \citenamefont {Black}}]{dalgarno1974}%
  \BibitemOpen
  \bibfield  {author} {\bibinfo {author} {\bibfnamefont {A.}~\bibnamefont
  {Dalgarno}}, \bibinfo {author} {\bibfnamefont {T.}~\bibnamefont {de~Jong}},
  \bibinfo {author} {\bibfnamefont {M.}~\bibnamefont {Oppenheimer}}, \ and\
  \bibinfo {author} {\bibfnamefont {J.~H.}\ \bibnamefont {Black}},\ }\href
  {\doibase 10.1086/181584} {\bibfield  {journal} {\bibinfo  {journal} {ApJ}\
  }\textbf {\bibinfo {volume} {192}},\ \bibinfo {pages} {L37} (\bibinfo {year}
  {1974})}\BibitemShut {NoStop}%
\bibitem [{\citenamefont {Jones}\ and\ \citenamefont
  {Gordy}(1964)}]{jones1964}%
  \BibitemOpen
  \bibfield  {author} {\bibinfo {author} {\bibfnamefont {G.}~\bibnamefont
  {Jones}}\ and\ \bibinfo {author} {\bibfnamefont {W.}~\bibnamefont {Gordy}},\
  }\href {\doibase 10.1103/PhysRev.136.A1229} {\bibfield  {journal} {\bibinfo
  {journal} {Phys. Rev.}\ }\textbf {\bibinfo {volume} {136}},\ \bibinfo {pages}
  {A1229} (\bibinfo {year} {1964})}\BibitemShut {NoStop}%
\bibitem [{\citenamefont {Lucia}\ \emph {et~al.}(1971)\citenamefont {Lucia},
  \citenamefont {Helminger},\ and\ \citenamefont {Gordy}}]{deLucia1971}%
  \BibitemOpen
  \bibfield  {author} {\bibinfo {author} {\bibfnamefont {F.~C.~D.}\
  \bibnamefont {Lucia}}, \bibinfo {author} {\bibfnamefont {P.}~\bibnamefont
  {Helminger}}, \ and\ \bibinfo {author} {\bibfnamefont {W.}~\bibnamefont
  {Gordy}},\ }\href {\doibase 10.1103/PhysRevA.3.1849} {\bibfield  {journal}
  {\bibinfo  {journal} {Phys. Rev. A.}\ }\textbf {\bibinfo {volume} {3}},\
  \bibinfo {pages} {1849} (\bibinfo {year} {1971})}\BibitemShut {NoStop}%
\bibitem [{\citenamefont {Pourcin}\ and\ \citenamefont
  {Romanetti}(1973)}]{pourcin1973}%
  \BibitemOpen
  \bibfield  {author} {\bibinfo {author} {\bibfnamefont {J.}~\bibnamefont
  {Pourcin}}\ and\ \bibinfo {author} {\bibfnamefont {R.}~\bibnamefont
  {Romanetti}},\ }\href {\doibase 10.1016/0020-0891(73)90025-0} {\bibfield
  {journal} {\bibinfo  {journal} {Infrared Phys.}\ }\textbf {\bibinfo {volume}
  {13}},\ \bibinfo {pages} {161} (\bibinfo {year} {1973})}\BibitemShut
  {NoStop}%
\bibitem [{\citenamefont {Houdeau}\ \emph {et~al.}(1980)\citenamefont
  {Houdeau}, \citenamefont {Larvor},\ and\ \citenamefont
  {Haeusler}}]{houdeau1980}%
  \BibitemOpen
  \bibfield  {author} {\bibinfo {author} {\bibfnamefont {J.-P.}\ \bibnamefont
  {Houdeau}}, \bibinfo {author} {\bibfnamefont {M.}~\bibnamefont {Larvor}}, \
  and\ \bibinfo {author} {\bibfnamefont {C.}~\bibnamefont {Haeusler}},\ }\href
  {\doibase 10.1139/P80-046} {\bibfield  {journal} {\bibinfo  {journal} {Can.
  J. Phys.}\ }\textbf {\bibinfo {volume} {58}},\ \bibinfo {pages} {318}
  (\bibinfo {year} {1980})}\BibitemShut {NoStop}%
\bibitem [{\citenamefont {Sergent-Rozey}\ \emph {et~al.}(1986)\citenamefont
  {Sergent-Rozey}, \citenamefont {Lacome},\ and\ \citenamefont
  {Levy}}]{sergent-rozey1986}%
  \BibitemOpen
  \bibfield  {author} {\bibinfo {author} {\bibfnamefont {M.}~\bibnamefont
  {Sergent-Rozey}}, \bibinfo {author} {\bibfnamefont {N.}~\bibnamefont
  {Lacome}}, \ and\ \bibinfo {author} {\bibfnamefont {A.}~\bibnamefont
  {Levy}},\ }\href {\doibase 10.1016/0022-2852(86)90013-5} {\bibfield
  {journal} {\bibinfo  {journal} {J. Mol. Spectrosc.}\ }\textbf {\bibinfo
  {volume} {120}},\ \bibinfo {pages} {403} (\bibinfo {year}
  {1986})}\BibitemShut {NoStop}%
\bibitem [{\citenamefont {Pine}\ and\ \citenamefont {Looney}(1987)}]{pine1987}%
  \BibitemOpen
  \bibfield  {author} {\bibinfo {author} {\bibfnamefont {A.~S.}\ \bibnamefont
  {Pine}}\ and\ \bibinfo {author} {\bibfnamefont {J.~P.}\ \bibnamefont
  {Looney}},\ }\href {\doibase 10.101/0022-2852(87)90217-7} {\bibfield
  {journal} {\bibinfo  {journal} {J. Mol. Spectrosc.}\ }\textbf {\bibinfo
  {volume} {122}},\ \bibinfo {pages} {41} (\bibinfo {year} {1987})}\BibitemShut
  {NoStop}%
\bibitem [{\citenamefont {Park}\ \emph {et~al.}(1991)\citenamefont {Park},
  \citenamefont {Chance}, \citenamefont {Nolt}, \citenamefont {Radostitz},
  \citenamefont {Vanek}, \citenamefont {Jennings},\ and\ \citenamefont
  {Evenson}}]{park1991}%
  \BibitemOpen
  \bibfield  {author} {\bibinfo {author} {\bibfnamefont {K.}~\bibnamefont
  {Park}}, \bibinfo {author} {\bibfnamefont {K.~V.}\ \bibnamefont {Chance}},
  \bibinfo {author} {\bibfnamefont {I.~G.}\ \bibnamefont {Nolt}}, \bibinfo
  {author} {\bibfnamefont {J.~V.}\ \bibnamefont {Radostitz}}, \bibinfo {author}
  {\bibfnamefont {M.~D.}\ \bibnamefont {Vanek}}, \bibinfo {author}
  {\bibfnamefont {D.~A.}\ \bibnamefont {Jennings}}, \ and\ \bibinfo {author}
  {\bibfnamefont {K.~M.}\ \bibnamefont {Evenson}},\ }\href {\doibase
  10.1016/0022-2852(91)90075-L} {\bibfield  {journal} {\bibinfo  {journal} {J.
  Mol. Spectrosc.}\ }\textbf {\bibinfo {volume} {147}},\ \bibinfo {pages} {521}
  (\bibinfo {year} {1991})}\BibitemShut {NoStop}%
\bibitem [{\citenamefont {Klaus}\ \emph {et~al.}(1998)\citenamefont {Klaus},
  \citenamefont {Belov},\ and\ \citenamefont {Winnewisser}}]{klaus1998}%
  \BibitemOpen
  \bibfield  {author} {\bibinfo {author} {\bibfnamefont {T.}~\bibnamefont
  {Klaus}}, \bibinfo {author} {\bibfnamefont {S.~P.}\ \bibnamefont {Belov}}, \
  and\ \bibinfo {author} {\bibfnamefont {G.}~\bibnamefont {Winnewisser}},\
  }\href {\doibase 10.1006/jmsp.1997.7465} {\bibfield  {journal} {\bibinfo
  {journal} {J. Mol. Spectrosc.}\ }\textbf {\bibinfo {volume} {187}},\ \bibinfo
  {pages} {109} (\bibinfo {year} {1998})}\BibitemShut {NoStop}%
\bibitem [{\citenamefont {Zu}\ \emph {et~al.}(2003)\citenamefont {Zu},
  \citenamefont {Hamilton}, \citenamefont {Chance},\ and\ \citenamefont
  {Davies}}]{zu2003}%
  \BibitemOpen
  \bibfield  {author} {\bibinfo {author} {\bibfnamefont {L.}~\bibnamefont
  {Zu}}, \bibinfo {author} {\bibfnamefont {P.~A.}\ \bibnamefont {Hamilton}},
  \bibinfo {author} {\bibfnamefont {K.~V.}\ \bibnamefont {Chance}}, \ and\
  \bibinfo {author} {\bibfnamefont {P.~B.}\ \bibnamefont {Davies}},\ }\href
  {\doibase 10.1016/S0022-2852(03)00096-1} {\bibfield  {journal} {\bibinfo
  {journal} {J. Mol. Spectrosc.}\ }\textbf {\bibinfo {volume} {220}},\ \bibinfo
  {pages} {107–112} (\bibinfo {year} {2003})}\BibitemShut {NoStop}%
\bibitem [{\citenamefont {Drouin}(2004)}]{drouin2004}%
  \BibitemOpen
  \bibfield  {author} {\bibinfo {author} {\bibfnamefont {B.~J.}\ \bibnamefont
  {Drouin}},\ }\href {\doibase 10.1016/S0022-4073(02)00360-6} {\bibfield
  {journal} {\bibinfo  {journal} {J. Quant. Spectrosc. Radiat. Transf.}\
  }\textbf {\bibinfo {volume} {83}},\ \bibinfo {pages} {321} (\bibinfo {year}
  {2004})}\BibitemShut {NoStop}%
\bibitem [{\citenamefont {Morino}\ and\ \citenamefont
  {Yamada}(2005)}]{morino2005}%
  \BibitemOpen
  \bibfield  {author} {\bibinfo {author} {\bibfnamefont {I.}~\bibnamefont
  {Morino}}\ and\ \bibinfo {author} {\bibfnamefont {K.~M.~T.}\ \bibnamefont
  {Yamada}},\ }\href {\doibase 10.1016/j.jms.2005.06.003} {\bibfield  {journal}
  {\bibinfo  {journal} {J. Mol. Spectrosc.}\ }\textbf {\bibinfo {volume}
  {233}},\ \bibinfo {pages} {77} (\bibinfo {year} {2005})}\BibitemShut
  {NoStop}%
\bibitem [{\citenamefont {Hurtmans}\ \emph {et~al.}(2009)\citenamefont
  {Hurtmans}, \citenamefont {Henry}, \citenamefont {Valentin},\ and\
  \citenamefont {Boulet}}]{hurtmans2009}%
  \BibitemOpen
  \bibfield  {author} {\bibinfo {author} {\bibfnamefont {D.}~\bibnamefont
  {Hurtmans}}, \bibinfo {author} {\bibfnamefont {A.}~\bibnamefont {Henry}},
  \bibinfo {author} {\bibfnamefont {A.}~\bibnamefont {Valentin}}, \ and\
  \bibinfo {author} {\bibfnamefont {C.}~\bibnamefont {Boulet}},\ }\href
  {\doibase 10.1016/j.jms.2009.01.015} {\bibfield  {journal} {\bibinfo
  {journal} {J. Mol. Spectrosc.}\ }\textbf {\bibinfo {volume} {254}},\ \bibinfo
  {pages} {126–136} (\bibinfo {year} {2009})}\BibitemShut {NoStop}%
\bibitem [{\citenamefont {Li}\ \emph {et~al.}(2015)\citenamefont {Li},
  \citenamefont {Serdyukov}, \citenamefont {Gisi}, \citenamefont {Werhahn},\
  and\ \citenamefont {Ebert}}]{li2015}%
  \BibitemOpen
  \bibfield  {author} {\bibinfo {author} {\bibfnamefont {G.}~\bibnamefont
  {Li}}, \bibinfo {author} {\bibfnamefont {A.}~\bibnamefont {Serdyukov}},
  \bibinfo {author} {\bibfnamefont {M.}~\bibnamefont {Gisi}}, \bibinfo {author}
  {\bibfnamefont {O.}~\bibnamefont {Werhahn}}, \ and\ \bibinfo {author}
  {\bibfnamefont {V.}~\bibnamefont {Ebert}},\ }\href {\doibase
  10.1016/j.jqsrt.2015.06.021} {\bibfield  {journal} {\bibinfo  {journal} {J.
  Quant. Spectrosc. Radiat. Transf.}\ }\textbf {\bibinfo {volume} {165}},\
  \bibinfo {pages} {76} (\bibinfo {year} {2015})}\BibitemShut {NoStop}%
\bibitem [{\citenamefont {Fitz}\ and\ \citenamefont {Marcus}(1975)}]{fitz1975}%
  \BibitemOpen
  \bibfield  {author} {\bibinfo {author} {\bibfnamefont {D.~E.}\ \bibnamefont
  {Fitz}}\ and\ \bibinfo {author} {\bibfnamefont {R.~A.}\ \bibnamefont
  {Marcus}},\ }\href {\doibase 10.1063/1.430930} {\bibfield  {journal}
  {\bibinfo  {journal} {J. Chem. Phys.}\ }\textbf {\bibinfo {volume} {62}},\
  \bibinfo {pages} {3788} (\bibinfo {year} {1975})}\BibitemShut {NoStop}%
\bibitem [{\citenamefont {Tran}\ and\ \citenamefont
  {Domenech}(2014)}]{tran2014}%
  \BibitemOpen
  \bibfield  {author} {\bibinfo {author} {\bibfnamefont {H.}~\bibnamefont
  {Tran}}\ and\ \bibinfo {author} {\bibfnamefont {J.-L.}\ \bibnamefont
  {Domenech}},\ }\href {\doibase 10.1063/1.4892590} {\bibfield  {journal}
  {\bibinfo  {journal} {J. Chem. Phys.}\ }\textbf {\bibinfo {volume} {141}},\
  \bibinfo {pages} {064313} (\bibinfo {year} {2014})}\BibitemShut {NoStop}%
\bibitem [{\citenamefont {Tran}\ \emph {et~al.}(2017)\citenamefont {Tran},
  \citenamefont {Hartmann}, \citenamefont {Li},\ and\ \citenamefont
  {Ebert}}]{tran2017}%
  \BibitemOpen
  \bibfield  {author} {\bibinfo {author} {\bibfnamefont {H.}~\bibnamefont
  {Tran}}, \bibinfo {author} {\bibfnamefont {J.-M.}\ \bibnamefont {Hartmann}},
  \bibinfo {author} {\bibfnamefont {G.}~\bibnamefont {Li}}, \ and\ \bibinfo
  {author} {\bibfnamefont {V.}~\bibnamefont {Ebert}},\ }\href {\doibase
  10.1088/1742-6596/810/1/012039} {\bibfield  {journal} {\bibinfo  {journal}
  {J. Phys.: Conf. Ser.}\ }\textbf {\bibinfo {volume} {810}},\ \bibinfo {pages}
  {012039} (\bibinfo {year} {2017})}\BibitemShut {NoStop}%
\bibitem [{\citenamefont {Dicke}(1953)}]{dicke1953}%
  \BibitemOpen
  \bibfield  {author} {\bibinfo {author} {\bibfnamefont {R.~H.}\ \bibnamefont
  {Dicke}},\ }\href {\doibase 10.1103/PhysRev.89.472} {\bibfield  {journal}
  {\bibinfo  {journal} {Phys. Rev.}\ }\textbf {\bibinfo {volume} {89}},\
  \bibinfo {pages} {472} (\bibinfo {year} {1953})}\BibitemShut {NoStop}%
\bibitem [{\citenamefont {Barret}\ \emph {et~al.}(2005)\citenamefont {Barret},
  \citenamefont {Hurtmans}, \citenamefont {Carleer}, \citenamefont
  {Mazi{\"e}re}, \citenamefont {Mahieu},\ and\ \citenamefont
  {Coheur}}]{barret2005}%
  \BibitemOpen
  \bibfield  {author} {\bibinfo {author} {\bibfnamefont {B.}~\bibnamefont
  {Barret}}, \bibinfo {author} {\bibfnamefont {D.}~\bibnamefont {Hurtmans}},
  \bibinfo {author} {\bibfnamefont {M.~R.}\ \bibnamefont {Carleer}}, \bibinfo
  {author} {\bibfnamefont {M.~D.}\ \bibnamefont {Mazi{\"e}re}}, \bibinfo
  {author} {\bibfnamefont {E.}~\bibnamefont {Mahieu}}, \ and\ \bibinfo {author}
  {\bibfnamefont {P.-F.}\ \bibnamefont {Coheur}},\ }\href {\doibase
  10.1016/j.jqsrt.2004.12.005} {\bibfield  {journal} {\bibinfo  {journal} {J.
  Quant. Spectrosc. Radiat. Transf.}\ }\textbf {\bibinfo {volume} {95}},\
  \bibinfo {pages} {499–519} (\bibinfo {year} {2005})}\BibitemShut {NoStop}%
\bibitem [{\citenamefont {Ramachandra}\ and\ \citenamefont
  {Oka}(1987)}]{ramachandra1987}%
  \BibitemOpen
  \bibfield  {author} {\bibinfo {author} {\bibfnamefont {R.~D.}\ \bibnamefont
  {Ramachandra}}\ and\ \bibinfo {author} {\bibfnamefont {T.}~\bibnamefont
  {Oka}},\ }\href {\doibase 10.1016/0022-2852(87)90215-3} {\bibfield  {journal}
  {\bibinfo  {journal} {J. Mol. Spectrosc.}\ }\textbf {\bibinfo {volume}
  {122}},\ \bibinfo {pages} {16} (\bibinfo {year} {1987})}\BibitemShut
  {NoStop}%
\bibitem [{\citenamefont {Rank}\ \emph {et~al.}(1963)\citenamefont {Rank},
  \citenamefont {Eastman}, \citenamefont {Rao},\ and\ \citenamefont
  {Wiggins}}]{rank1963}%
  \BibitemOpen
  \bibfield  {author} {\bibinfo {author} {\bibfnamefont {D.~H.}\ \bibnamefont
  {Rank}}, \bibinfo {author} {\bibfnamefont {D.~P.}\ \bibnamefont {Eastman}},
  \bibinfo {author} {\bibfnamefont {B.~S.}\ \bibnamefont {Rao}}, \ and\
  \bibinfo {author} {\bibfnamefont {T.~A.}\ \bibnamefont {Wiggins}},\ }\href
  {\doibase 10.1016/0022-2852(63)90152-8} {\bibfield  {journal} {\bibinfo
  {journal} {J. Mol. Spectrosc.}\ }\textbf {\bibinfo {volume} {10}},\ \bibinfo
  {pages} {34} (\bibinfo {year} {1963})}\BibitemShut {NoStop}%
\bibitem [{\citenamefont {Pine}\ and\ \citenamefont {Fried}(1985)}]{pine1985}%
  \BibitemOpen
  \bibfield  {author} {\bibinfo {author} {\bibfnamefont {A.~S.}\ \bibnamefont
  {Pine}}\ and\ \bibinfo {author} {\bibfnamefont {A.}~\bibnamefont {Fried}},\
  }\href {\doibase 10.1016/0022-2852(85)90344-3} {\bibfield  {journal}
  {\bibinfo  {journal} {J. Mol. Spectrosc.}\ }\textbf {\bibinfo {volume}
  {144}},\ \bibinfo {pages} {148} (\bibinfo {year} {1985})}\BibitemShut
  {NoStop}%
\bibitem [{\citenamefont {Benedict}\ \emph {et~al.}(1956)\citenamefont
  {Benedict}, \citenamefont {Herman}, \citenamefont {Moore},\ and\
  \citenamefont {Silverman}}]{benedict1956}%
  \BibitemOpen
  \bibfield  {author} {\bibinfo {author} {\bibfnamefont {W.~S.}\ \bibnamefont
  {Benedict}}, \bibinfo {author} {\bibfnamefont {R.}~\bibnamefont {Herman}},
  \bibinfo {author} {\bibfnamefont {G.~E.}\ \bibnamefont {Moore}}, \ and\
  \bibinfo {author} {\bibfnamefont {S.}~\bibnamefont {Silverman}},\ }\href
  {\doibase 10.1139/p56-092} {\bibfield  {journal} {\bibinfo  {journal} {Can.
  J. Phys.}\ }\textbf {\bibinfo {volume} {34}},\ \bibinfo {pages} {850}
  (\bibinfo {year} {1956})}\BibitemShut {NoStop}%
\bibitem [{\citenamefont {Marcus}(1970)}]{Marcus_1970}%
  \BibitemOpen
  \bibfield  {author} {\bibinfo {author} {\bibfnamefont {R.}~\bibnamefont
  {Marcus}},\ }\href {\doibase 10.1016/0009-2614(70)80164-6} {\bibfield
  {journal} {\bibinfo  {journal} {Chem.~Phys.~Lett.}\ }\textbf {\bibinfo
  {volume} {7}},\ \bibinfo {pages} {525} (\bibinfo {year} {1970})}\BibitemShut
  {NoStop}%
\bibitem [{\citenamefont {Miller}(1970)}]{Miller_1970}%
  \BibitemOpen
  \bibfield  {author} {\bibinfo {author} {\bibfnamefont {W.~H.}\ \bibnamefont
  {Miller}},\ }\href {\doibase 10.1063/1.1674275} {\bibfield  {journal}
  {\bibinfo  {journal} {J.~Chem.~Phys.}\ }\textbf {\bibinfo {volume} {53}},\
  \bibinfo {pages} {1949} (\bibinfo {year} {1970})}\BibitemShut {NoStop}%
\bibitem [{\citenamefont {Fitz}\ and\ \citenamefont
  {Marcus}(1973)}]{Fitz_1973}%
  \BibitemOpen
  \bibfield  {author} {\bibinfo {author} {\bibfnamefont {D.~E.}\ \bibnamefont
  {Fitz}}\ and\ \bibinfo {author} {\bibfnamefont {R.~A.}\ \bibnamefont
  {Marcus}},\ }\href {\doibase 10.1063/1.1680636} {\bibfield  {journal}
  {\bibinfo  {journal} {J.~Chem.~Phys.}\ }\textbf {\bibinfo {volume} {59}},\
  \bibinfo {pages} {4380} (\bibinfo {year} {1973})}\BibitemShut {NoStop}%
\bibitem [{\citenamefont {Gordon}\ \emph {et~al.}(2022)\citenamefont {Gordon},
  \citenamefont {Rothman}, \citenamefont {Hargreaves}, \citenamefont {Hashemi},
  \citenamefont {Karlovets}, \citenamefont {Skinner}, \citenamefont {Conway},
  \citenamefont {Hill}, \citenamefont {Kochanov}, \citenamefont {Tan},
  \citenamefont {Wcisło}, \citenamefont {Finenko}, \citenamefont {Nelson},
  \citenamefont {Bernath}, \citenamefont {Birk}, \citenamefont {Boudon},
  \citenamefont {Campargue}, \citenamefont {Chance}, \citenamefont {Coustenis},
  \citenamefont {Drouin}, \citenamefont {Flaud}, \citenamefont {Gamache},
  \citenamefont {Hodges}, \citenamefont {Jacquemart}, \citenamefont {Mlawer},
  \citenamefont {Nikitin}, \citenamefont {Perevalov}, \citenamefont {Rotger},
  \citenamefont {Tennyson}, \citenamefont {Toon}, \citenamefont {Tran},
  \citenamefont {Tyuterev}, \citenamefont {Adkins}, \citenamefont {Baker},
  \citenamefont {Barbe}, \citenamefont {Canè}, \citenamefont {Császár},
  \citenamefont {Dudaryonok}, \citenamefont {Egorov}, \citenamefont {Fleisher},
  \citenamefont {Fleurbaey}, \citenamefont {Foltynowicz}, \citenamefont
  {Furtenbacher}, \citenamefont {Harrison}, \citenamefont {Hartmann},
  \citenamefont {Horneman}, \citenamefont {Huang}, \citenamefont {Karman},
  \citenamefont {Karns}, \citenamefont {Kassi}, \citenamefont {Kleiner},
  \citenamefont {Kofman}, \citenamefont {Kwabia–Tchana}, \citenamefont
  {Lavrentieva}, \citenamefont {Lee}, \citenamefont {Long}, \citenamefont
  {Lukashevskaya}, \citenamefont {Lyulin}, \citenamefont {Makhnev},
  \citenamefont {Matt}, \citenamefont {Massie}, \citenamefont {Melosso},
  \citenamefont {Mikhailenko}, \citenamefont {Mondelain}, \citenamefont
  {Müller}, \citenamefont {Naumenko}, \citenamefont {Perrin}, \citenamefont
  {Polyansky}, \citenamefont {Raddaoui}, \citenamefont {Raston}, \citenamefont
  {Reed}, \citenamefont {Rey}, \citenamefont {Richard}, \citenamefont
  {Tóbiás}, \citenamefont {Sadiek}, \citenamefont {Schwenke}, \citenamefont
  {Starikova}, \citenamefont {Sung}, \citenamefont {Tamassia}, \citenamefont
  {Tashkun}, \citenamefont {{Vander Auwera}}, \citenamefont {Vasilenko},
  \citenamefont {Vigasin}, \citenamefont {Villanueva}, \citenamefont {Vispoel},
  \citenamefont {Wagner}, \citenamefont {Yachmenev},\ and\ \citenamefont
  {Yurchenko}}]{hitran2022}%
  \BibitemOpen
  \bibfield  {author} {\bibinfo {author} {\bibfnamefont {I.}~\bibnamefont
  {Gordon}}, \bibinfo {author} {\bibfnamefont {L.}~\bibnamefont {Rothman}},
  \bibinfo {author} {\bibfnamefont {R.}~\bibnamefont {Hargreaves}}, \bibinfo
  {author} {\bibfnamefont {R.}~\bibnamefont {Hashemi}}, \bibinfo {author}
  {\bibfnamefont {E.}~\bibnamefont {Karlovets}}, \bibinfo {author}
  {\bibfnamefont {F.}~\bibnamefont {Skinner}}, \bibinfo {author} {\bibfnamefont
  {E.}~\bibnamefont {Conway}}, \bibinfo {author} {\bibfnamefont
  {C.}~\bibnamefont {Hill}}, \bibinfo {author} {\bibfnamefont {R.}~\bibnamefont
  {Kochanov}}, \bibinfo {author} {\bibfnamefont {Y.}~\bibnamefont {Tan}},
  \bibinfo {author} {\bibfnamefont {P.}~\bibnamefont {Wcisło}}, \bibinfo
  {author} {\bibfnamefont {A.}~\bibnamefont {Finenko}}, \bibinfo {author}
  {\bibfnamefont {K.}~\bibnamefont {Nelson}}, \bibinfo {author} {\bibfnamefont
  {P.}~\bibnamefont {Bernath}}, \bibinfo {author} {\bibfnamefont
  {M.}~\bibnamefont {Birk}}, \bibinfo {author} {\bibfnamefont {V.}~\bibnamefont
  {Boudon}}, \bibinfo {author} {\bibfnamefont {A.}~\bibnamefont {Campargue}},
  \bibinfo {author} {\bibfnamefont {K.}~\bibnamefont {Chance}}, \bibinfo
  {author} {\bibfnamefont {A.}~\bibnamefont {Coustenis}}, \bibinfo {author}
  {\bibfnamefont {B.}~\bibnamefont {Drouin}}, \bibinfo {author} {\bibfnamefont
  {J.}~\bibnamefont {Flaud}}, \bibinfo {author} {\bibfnamefont
  {R.}~\bibnamefont {Gamache}}, \bibinfo {author} {\bibfnamefont
  {J.}~\bibnamefont {Hodges}}, \bibinfo {author} {\bibfnamefont
  {D.}~\bibnamefont {Jacquemart}}, \bibinfo {author} {\bibfnamefont
  {E.}~\bibnamefont {Mlawer}}, \bibinfo {author} {\bibfnamefont
  {A.}~\bibnamefont {Nikitin}}, \bibinfo {author} {\bibfnamefont
  {V.}~\bibnamefont {Perevalov}}, \bibinfo {author} {\bibfnamefont
  {M.}~\bibnamefont {Rotger}}, \bibinfo {author} {\bibfnamefont
  {J.}~\bibnamefont {Tennyson}}, \bibinfo {author} {\bibfnamefont
  {G.}~\bibnamefont {Toon}}, \bibinfo {author} {\bibfnamefont {H.}~\bibnamefont
  {Tran}}, \bibinfo {author} {\bibfnamefont {V.}~\bibnamefont {Tyuterev}},
  \bibinfo {author} {\bibfnamefont {E.}~\bibnamefont {Adkins}}, \bibinfo
  {author} {\bibfnamefont {A.}~\bibnamefont {Baker}}, \bibinfo {author}
  {\bibfnamefont {A.}~\bibnamefont {Barbe}}, \bibinfo {author} {\bibfnamefont
  {E.}~\bibnamefont {Canè}}, \bibinfo {author} {\bibfnamefont
  {A.}~\bibnamefont {Császár}}, \bibinfo {author} {\bibfnamefont
  {A.}~\bibnamefont {Dudaryonok}}, \bibinfo {author} {\bibfnamefont
  {O.}~\bibnamefont {Egorov}}, \bibinfo {author} {\bibfnamefont
  {A.}~\bibnamefont {Fleisher}}, \bibinfo {author} {\bibfnamefont
  {H.}~\bibnamefont {Fleurbaey}}, \bibinfo {author} {\bibfnamefont
  {A.}~\bibnamefont {Foltynowicz}}, \bibinfo {author} {\bibfnamefont
  {T.}~\bibnamefont {Furtenbacher}}, \bibinfo {author} {\bibfnamefont
  {J.}~\bibnamefont {Harrison}}, \bibinfo {author} {\bibfnamefont
  {J.}~\bibnamefont {Hartmann}}, \bibinfo {author} {\bibfnamefont
  {V.}~\bibnamefont {Horneman}}, \bibinfo {author} {\bibfnamefont
  {X.}~\bibnamefont {Huang}}, \bibinfo {author} {\bibfnamefont
  {T.}~\bibnamefont {Karman}}, \bibinfo {author} {\bibfnamefont
  {J.}~\bibnamefont {Karns}}, \bibinfo {author} {\bibfnamefont
  {S.}~\bibnamefont {Kassi}}, \bibinfo {author} {\bibfnamefont
  {I.}~\bibnamefont {Kleiner}}, \bibinfo {author} {\bibfnamefont
  {V.}~\bibnamefont {Kofman}}, \bibinfo {author} {\bibfnamefont
  {F.}~\bibnamefont {Kwabia–Tchana}}, \bibinfo {author} {\bibfnamefont
  {N.}~\bibnamefont {Lavrentieva}}, \bibinfo {author} {\bibfnamefont
  {T.}~\bibnamefont {Lee}}, \bibinfo {author} {\bibfnamefont {D.}~\bibnamefont
  {Long}}, \bibinfo {author} {\bibfnamefont {A.}~\bibnamefont {Lukashevskaya}},
  \bibinfo {author} {\bibfnamefont {O.}~\bibnamefont {Lyulin}}, \bibinfo
  {author} {\bibfnamefont {V.}~\bibnamefont {Makhnev}}, \bibinfo {author}
  {\bibfnamefont {W.}~\bibnamefont {Matt}}, \bibinfo {author} {\bibfnamefont
  {S.}~\bibnamefont {Massie}}, \bibinfo {author} {\bibfnamefont
  {M.}~\bibnamefont {Melosso}}, \bibinfo {author} {\bibfnamefont
  {S.}~\bibnamefont {Mikhailenko}}, \bibinfo {author} {\bibfnamefont
  {D.}~\bibnamefont {Mondelain}}, \bibinfo {author} {\bibfnamefont
  {H.}~\bibnamefont {Müller}}, \bibinfo {author} {\bibfnamefont
  {O.}~\bibnamefont {Naumenko}}, \bibinfo {author} {\bibfnamefont
  {A.}~\bibnamefont {Perrin}}, \bibinfo {author} {\bibfnamefont
  {O.}~\bibnamefont {Polyansky}}, \bibinfo {author} {\bibfnamefont
  {E.}~\bibnamefont {Raddaoui}}, \bibinfo {author} {\bibfnamefont
  {P.}~\bibnamefont {Raston}}, \bibinfo {author} {\bibfnamefont
  {Z.}~\bibnamefont {Reed}}, \bibinfo {author} {\bibfnamefont {M.}~\bibnamefont
  {Rey}}, \bibinfo {author} {\bibfnamefont {C.}~\bibnamefont {Richard}},
  \bibinfo {author} {\bibfnamefont {R.}~\bibnamefont {Tóbiás}}, \bibinfo
  {author} {\bibfnamefont {I.}~\bibnamefont {Sadiek}}, \bibinfo {author}
  {\bibfnamefont {D.}~\bibnamefont {Schwenke}}, \bibinfo {author}
  {\bibfnamefont {E.}~\bibnamefont {Starikova}}, \bibinfo {author}
  {\bibfnamefont {K.}~\bibnamefont {Sung}}, \bibinfo {author} {\bibfnamefont
  {F.}~\bibnamefont {Tamassia}}, \bibinfo {author} {\bibfnamefont
  {S.}~\bibnamefont {Tashkun}}, \bibinfo {author} {\bibfnamefont
  {J.}~\bibnamefont {{Vander Auwera}}}, \bibinfo {author} {\bibfnamefont
  {I.}~\bibnamefont {Vasilenko}}, \bibinfo {author} {\bibfnamefont
  {A.}~\bibnamefont {Vigasin}}, \bibinfo {author} {\bibfnamefont
  {G.}~\bibnamefont {Villanueva}}, \bibinfo {author} {\bibfnamefont
  {B.}~\bibnamefont {Vispoel}}, \bibinfo {author} {\bibfnamefont
  {G.}~\bibnamefont {Wagner}}, \bibinfo {author} {\bibfnamefont
  {A.}~\bibnamefont {Yachmenev}}, \ and\ \bibinfo {author} {\bibfnamefont
  {S.}~\bibnamefont {Yurchenko}},\ }\href {\doibase
  h10.1016/j.jqsrt.2021.107949} {\bibfield  {journal} {\bibinfo  {journal} {J.
  Quant. Spectrosc. Radiat. Transf.}\ }\textbf {\bibinfo {volume} {277}},\
  \bibinfo {pages} {107949} (\bibinfo {year} {2022})}\BibitemShut {NoStop}%
\bibitem [{\citenamefont {Delahaye}\ \emph {et~al.}(2021)\citenamefont
  {Delahaye}, \citenamefont {Armante}, \citenamefont {Scott}, \citenamefont
  {Jacquinet-Husson}, \citenamefont {Chédin}, \citenamefont {Crépeau},
  \citenamefont {Crevoisier}, \citenamefont {Douet}, \citenamefont {Perrin},
  \citenamefont {Barbe}, \citenamefont {Boudon}, \citenamefont {Campargue},
  \citenamefont {Coudert}, \citenamefont {Ebert}, \citenamefont {Flaud},
  \citenamefont {Gamache}, \citenamefont {Jacquemart}, \citenamefont {Jolly},
  \citenamefont {{Kwabia Tchana}}, \citenamefont {Kyuberis}, \citenamefont
  {Li}, \citenamefont {Lyulin}, \citenamefont {Manceron}, \citenamefont
  {Mikhailenko}, \citenamefont {Moazzen-Ahmadi}, \citenamefont {Müller},
  \citenamefont {Naumenko}, \citenamefont {Nikitin}, \citenamefont {Perevalov},
  \citenamefont {Richard}, \citenamefont {Starikova}, \citenamefont {Tashkun},
  \citenamefont {Tyuterev}, \citenamefont {{Vander Auwera}}, \citenamefont
  {Vispoel}, \citenamefont {Yachmenev},\ and\ \citenamefont
  {Yurchenko}}]{geisa2021}%
  \BibitemOpen
  \bibfield  {author} {\bibinfo {author} {\bibfnamefont {T.}~\bibnamefont
  {Delahaye}}, \bibinfo {author} {\bibfnamefont {R.}~\bibnamefont {Armante}},
  \bibinfo {author} {\bibfnamefont {N.}~\bibnamefont {Scott}}, \bibinfo
  {author} {\bibfnamefont {N.}~\bibnamefont {Jacquinet-Husson}}, \bibinfo
  {author} {\bibfnamefont {A.}~\bibnamefont {Chédin}}, \bibinfo {author}
  {\bibfnamefont {L.}~\bibnamefont {Crépeau}}, \bibinfo {author}
  {\bibfnamefont {C.}~\bibnamefont {Crevoisier}}, \bibinfo {author}
  {\bibfnamefont {V.}~\bibnamefont {Douet}}, \bibinfo {author} {\bibfnamefont
  {A.}~\bibnamefont {Perrin}}, \bibinfo {author} {\bibfnamefont
  {A.}~\bibnamefont {Barbe}}, \bibinfo {author} {\bibfnamefont
  {V.}~\bibnamefont {Boudon}}, \bibinfo {author} {\bibfnamefont
  {A.}~\bibnamefont {Campargue}}, \bibinfo {author} {\bibfnamefont
  {L.}~\bibnamefont {Coudert}}, \bibinfo {author} {\bibfnamefont
  {V.}~\bibnamefont {Ebert}}, \bibinfo {author} {\bibfnamefont {J.-M.}\
  \bibnamefont {Flaud}}, \bibinfo {author} {\bibfnamefont {R.}~\bibnamefont
  {Gamache}}, \bibinfo {author} {\bibfnamefont {D.}~\bibnamefont {Jacquemart}},
  \bibinfo {author} {\bibfnamefont {A.}~\bibnamefont {Jolly}}, \bibinfo
  {author} {\bibfnamefont {F.}~\bibnamefont {{Kwabia Tchana}}}, \bibinfo
  {author} {\bibfnamefont {A.}~\bibnamefont {Kyuberis}}, \bibinfo {author}
  {\bibfnamefont {G.}~\bibnamefont {Li}}, \bibinfo {author} {\bibfnamefont
  {O.}~\bibnamefont {Lyulin}}, \bibinfo {author} {\bibfnamefont
  {L.}~\bibnamefont {Manceron}}, \bibinfo {author} {\bibfnamefont
  {S.}~\bibnamefont {Mikhailenko}}, \bibinfo {author} {\bibfnamefont
  {N.}~\bibnamefont {Moazzen-Ahmadi}}, \bibinfo {author} {\bibfnamefont
  {H.}~\bibnamefont {Müller}}, \bibinfo {author} {\bibfnamefont
  {O.}~\bibnamefont {Naumenko}}, \bibinfo {author} {\bibfnamefont
  {A.}~\bibnamefont {Nikitin}}, \bibinfo {author} {\bibfnamefont
  {V.}~\bibnamefont {Perevalov}}, \bibinfo {author} {\bibfnamefont
  {C.}~\bibnamefont {Richard}}, \bibinfo {author} {\bibfnamefont
  {E.}~\bibnamefont {Starikova}}, \bibinfo {author} {\bibfnamefont
  {S.}~\bibnamefont {Tashkun}}, \bibinfo {author} {\bibfnamefont
  {V.}~\bibnamefont {Tyuterev}}, \bibinfo {author} {\bibfnamefont
  {J.}~\bibnamefont {{Vander Auwera}}}, \bibinfo {author} {\bibfnamefont
  {B.}~\bibnamefont {Vispoel}}, \bibinfo {author} {\bibfnamefont
  {A.}~\bibnamefont {Yachmenev}}, \ and\ \bibinfo {author} {\bibfnamefont
  {S.}~\bibnamefont {Yurchenko}},\ }\href {\doibase 10.1016/j.jms.2021.111510}
  {\bibfield  {journal} {\bibinfo  {journal} {J. Mol. Spectrosc.}\ }\textbf
  {\bibinfo {volume} {380}},\ \bibinfo {pages} {111510} (\bibinfo {year}
  {2021})}\BibitemShut {NoStop}%
\bibitem [{\citenamefont {Thibault}\ \emph {et~al.}(2017)\citenamefont
  {Thibault}, \citenamefont {Patkowski}, \citenamefont {Żuchowski},
  \citenamefont {Jóźwiak}, \citenamefont {Ciuryło},\ and\ \citenamefont
  {Wcisło}}]{thibault2017}%
  \BibitemOpen
  \bibfield  {author} {\bibinfo {author} {\bibfnamefont {F.}~\bibnamefont
  {Thibault}}, \bibinfo {author} {\bibfnamefont {K.}~\bibnamefont {Patkowski}},
  \bibinfo {author} {\bibfnamefont {P.~S.}\ \bibnamefont {Żuchowski}},
  \bibinfo {author} {\bibfnamefont {H.}~\bibnamefont {Jóźwiak}}, \bibinfo
  {author} {\bibfnamefont {R.}~\bibnamefont {Ciuryło}}, \ and\ \bibinfo
  {author} {\bibfnamefont {P.}~\bibnamefont {Wcisło}},\ }\href {\doibase
  10.1016/j.jqsrt.2017.08.014} {\bibfield  {journal} {\bibinfo  {journal} {J.
  Quant. Spectrosc. Radiat. Transf.}\ }\textbf {\bibinfo {volume} {202}},\
  \bibinfo {pages} {308} (\bibinfo {year} {2017})}\BibitemShut {NoStop}%
\bibitem [{\citenamefont {J{\'{o}}{\'{z}}wiak}\ \emph
  {et~al.}(2018)\citenamefont {J{\'{o}}{\'{z}}wiak}, \citenamefont {Thibault},
  \citenamefont {Stolarczyk},\ and\ \citenamefont {Wcis{\l}o}}]{Jozwiak_2018}%
  \BibitemOpen
  \bibfield  {author} {\bibinfo {author} {\bibfnamefont {H.}~\bibnamefont
  {J{\'{o}}{\'{z}}wiak}}, \bibinfo {author} {\bibfnamefont {F.}~\bibnamefont
  {Thibault}}, \bibinfo {author} {\bibfnamefont {N.}~\bibnamefont
  {Stolarczyk}}, \ and\ \bibinfo {author} {\bibfnamefont {P.}~\bibnamefont
  {Wcis{\l}o}},\ }\href {\doibase 10.1016/j.jqsrt.2018.08.023} {\bibfield
  {journal} {\bibinfo  {journal} {J. Quant. Spectrosc. Radiat. Transf.}\
  }\textbf {\bibinfo {volume} {219}},\ \bibinfo {pages} {313} (\bibinfo {year}
  {2018})}\BibitemShut {NoStop}%
\bibitem [{\citenamefont {Mart{\'{\i}}nez}\ \emph {et~al.}(2018)\citenamefont
  {Mart{\'{\i}}nez}, \citenamefont {Bermejo}, \citenamefont {Thibault},\ and\
  \citenamefont {Wcis{\l}o}}]{Martinez_2018}%
  \BibitemOpen
  \bibfield  {author} {\bibinfo {author} {\bibfnamefont {R.~Z.}\ \bibnamefont
  {Mart{\'{\i}}nez}}, \bibinfo {author} {\bibfnamefont {D.}~\bibnamefont
  {Bermejo}}, \bibinfo {author} {\bibfnamefont {F.}~\bibnamefont {Thibault}}, \
  and\ \bibinfo {author} {\bibfnamefont {P.}~\bibnamefont {Wcis{\l}o}},\ }\href
  {\doibase 10.1002/jrs.5391} {\bibfield  {journal} {\bibinfo  {journal} {J.
  Raman Spectrosc.}\ }\textbf {\bibinfo {volume} {49}},\ \bibinfo {pages}
  {1339} (\bibinfo {year} {2018})}\BibitemShut {NoStop}%
\bibitem [{\citenamefont {Thibault}\ \emph {et~al.}(2020)\citenamefont
  {Thibault}, \citenamefont {Mart{\'{\i}}nez}, \citenamefont {Bermejo},\ and\
  \citenamefont {Wcis{\l}o}}]{thibault_2020}%
  \BibitemOpen
  \bibfield  {author} {\bibinfo {author} {\bibfnamefont {F.}~\bibnamefont
  {Thibault}}, \bibinfo {author} {\bibfnamefont {R.~Z.}\ \bibnamefont
  {Mart{\'{\i}}nez}}, \bibinfo {author} {\bibfnamefont {D.}~\bibnamefont
  {Bermejo}}, \ and\ \bibinfo {author} {\bibfnamefont {P.}~\bibnamefont
  {Wcis{\l}o}},\ }\href {\doibase 10.1016/j.molap.2020.100063} {\bibfield
  {journal} {\bibinfo  {journal} {Mol. Astroph.}\ }\textbf {\bibinfo {volume}
  {19}},\ \bibinfo {pages} {100063} (\bibinfo {year} {2020})}\BibitemShut
  {NoStop}%
\bibitem [{\citenamefont {Kowzan}\ \emph
  {et~al.}(2020{\natexlab{a}})\citenamefont {Kowzan}, \citenamefont
  {Wcis{\l}o}, \citenamefont {S{\l}owi{\'{n}}ski}, \citenamefont
  {Mas{\l}owski}, \citenamefont {Viel},\ and\ \citenamefont
  {Thibault}}]{Kowzan_2020a}%
  \BibitemOpen
  \bibfield  {author} {\bibinfo {author} {\bibfnamefont {G.}~\bibnamefont
  {Kowzan}}, \bibinfo {author} {\bibfnamefont {P.}~\bibnamefont {Wcis{\l}o}},
  \bibinfo {author} {\bibfnamefont {M.}~\bibnamefont {S{\l}owi{\'{n}}ski}},
  \bibinfo {author} {\bibfnamefont {P.}~\bibnamefont {Mas{\l}owski}}, \bibinfo
  {author} {\bibfnamefont {A.}~\bibnamefont {Viel}}, \ and\ \bibinfo {author}
  {\bibfnamefont {F.}~\bibnamefont {Thibault}},\ }\href {\doibase
  10.1016/j.jqsrt.2019.106803} {\bibfield  {journal} {\bibinfo  {journal} {J.
  Quant. Spectrosc. Radiat. Transf.}\ }\textbf {\bibinfo {volume} {243}},\
  \bibinfo {pages} {106803} (\bibinfo {year} {2020}{\natexlab{a}})}\BibitemShut
  {NoStop}%
\bibitem [{\citenamefont {Kowzan}\ \emph
  {et~al.}(2020{\natexlab{b}})\citenamefont {Kowzan}, \citenamefont {Cybulski},
  \citenamefont {Wcis\l{}o}, \citenamefont {S\l{}owi\ifmmode~\acute{n}\else
  \'{n}\fi{}ski}, \citenamefont {Viel}, \citenamefont {Mas\l{}owski},\ and\
  \citenamefont {Thibault}}]{Kowzan_2020b}%
  \BibitemOpen
  \bibfield  {author} {\bibinfo {author} {\bibfnamefont {G.}~\bibnamefont
  {Kowzan}}, \bibinfo {author} {\bibfnamefont {H.}~\bibnamefont {Cybulski}},
  \bibinfo {author} {\bibfnamefont {P.}~\bibnamefont {Wcis\l{}o}}, \bibinfo
  {author} {\bibfnamefont {M.}~\bibnamefont {S\l{}owi\ifmmode~\acute{n}\else
  \'{n}\fi{}ski}}, \bibinfo {author} {\bibfnamefont {A.}~\bibnamefont {Viel}},
  \bibinfo {author} {\bibfnamefont {P.}~\bibnamefont {Mas\l{}owski}}, \ and\
  \bibinfo {author} {\bibfnamefont {F.}~\bibnamefont {Thibault}},\ }\href
  {\doibase 10.1103/PhysRevA.102.012821} {\bibfield  {journal} {\bibinfo
  {journal} {Phys. Rev. A}\ }\textbf {\bibinfo {volume} {102}},\ \bibinfo
  {pages} {012821} (\bibinfo {year} {2020}{\natexlab{b}})}\BibitemShut
  {NoStop}%
\bibitem [{\citenamefont {Słowiński}\ \emph {et~al.}(2020)\citenamefont
  {Słowiński}, \citenamefont {F}, \citenamefont {Tan}, \citenamefont {Wang},
  \citenamefont {Liu}, \citenamefont {Hu}, \citenamefont {Kassi}, \citenamefont
  {Campargue}, \citenamefont {Konefał}, \citenamefont {Jóźwiak},
  \citenamefont {Patkowski}, \citenamefont {Żuchowski}, \citenamefont
  {Ciuryło}, \citenamefont {Lisak},\ and\ \citenamefont
  {Wcisło}}]{słowiński2020}%
  \BibitemOpen
  \bibfield  {author} {\bibinfo {author} {\bibfnamefont {M.}~\bibnamefont
  {Słowiński}}, \bibinfo {author} {\bibfnamefont {T.}~\bibnamefont {F}},
  \bibinfo {author} {\bibfnamefont {Y.}~\bibnamefont {Tan}}, \bibinfo {author}
  {\bibfnamefont {J.}~\bibnamefont {Wang}}, \bibinfo {author} {\bibfnamefont
  {A.-W.}\ \bibnamefont {Liu}}, \bibinfo {author} {\bibfnamefont {S.-M.}\
  \bibnamefont {Hu}}, \bibinfo {author} {\bibfnamefont {S.}~\bibnamefont
  {Kassi}}, \bibinfo {author} {\bibfnamefont {A.}~\bibnamefont {Campargue}},
  \bibinfo {author} {\bibfnamefont {M.}~\bibnamefont {Konefał}}, \bibinfo
  {author} {\bibfnamefont {H.}~\bibnamefont {Jóźwiak}}, \bibinfo {author}
  {\bibfnamefont {K.}~\bibnamefont {Patkowski}}, \bibinfo {author}
  {\bibfnamefont {P.}~\bibnamefont {Żuchowski}}, \bibinfo {author}
  {\bibfnamefont {R.}~\bibnamefont {Ciuryło}}, \bibinfo {author}
  {\bibfnamefont {D.}~\bibnamefont {Lisak}}, \ and\ \bibinfo {author}
  {\bibfnamefont {P.}~\bibnamefont {Wcisło}},\ }\href {\doibase
  10.1103/PhysRevA.101.052705} {\bibfield  {journal} {\bibinfo  {journal}
  {Phys. Rev. A}\ }\textbf {\bibinfo {volume} {101}},\ \bibinfo {pages}
  {052705} (\bibinfo {year} {2020})}\BibitemShut {NoStop}%
\bibitem [{\citenamefont {Stankiewicz}\ \emph {et~al.}(2020)\citenamefont
  {Stankiewicz}, \citenamefont {J{\'{o}}{\'{z}}wiak}, \citenamefont
  {Gancewski}, \citenamefont {Stolarczyk}, \citenamefont {Thibault},\ and\
  \citenamefont {Wcis{\l}o}}]{Stankiewicz_2020}%
  \BibitemOpen
  \bibfield  {author} {\bibinfo {author} {\bibfnamefont {K.}~\bibnamefont
  {Stankiewicz}}, \bibinfo {author} {\bibfnamefont {H.}~\bibnamefont
  {J{\'{o}}{\'{z}}wiak}}, \bibinfo {author} {\bibfnamefont {M.}~\bibnamefont
  {Gancewski}}, \bibinfo {author} {\bibfnamefont {N.}~\bibnamefont
  {Stolarczyk}}, \bibinfo {author} {\bibfnamefont {F.}~\bibnamefont
  {Thibault}}, \ and\ \bibinfo {author} {\bibfnamefont {P.}~\bibnamefont
  {Wcis{\l}o}},\ }\href {\doibase 10.1016/j.jqsrt.2020.107194} {\bibfield
  {journal} {\bibinfo  {journal} {J. Quant. Spectrosc. Radiat. Transfer}\
  }\textbf {\bibinfo {volume} {254}},\ \bibinfo {pages} {107194} (\bibinfo
  {year} {2020})}\BibitemShut {NoStop}%
\bibitem [{\citenamefont {Serov}\ \emph {et~al.}(2021)\citenamefont {Serov},
  \citenamefont {Stolarczyk}, \citenamefont {Makarov}, \citenamefont {Vilkov},
  \citenamefont {Golubiatnikov}, \citenamefont {Balashov}, \citenamefont
  {Koshelev}, \citenamefont {Wcis{\l}o}, \citenamefont {Thibault},\ and\
  \citenamefont {Tretyakov}}]{Serov_2021}%
  \BibitemOpen
  \bibfield  {author} {\bibinfo {author} {\bibfnamefont {E.}~\bibnamefont
  {Serov}}, \bibinfo {author} {\bibfnamefont {N.}~\bibnamefont {Stolarczyk}},
  \bibinfo {author} {\bibfnamefont {D.}~\bibnamefont {Makarov}}, \bibinfo
  {author} {\bibfnamefont {I.}~\bibnamefont {Vilkov}}, \bibinfo {author}
  {\bibfnamefont {G.~Y.}\ \bibnamefont {Golubiatnikov}}, \bibinfo {author}
  {\bibfnamefont {A.}~\bibnamefont {Balashov}}, \bibinfo {author}
  {\bibfnamefont {M.}~\bibnamefont {Koshelev}}, \bibinfo {author}
  {\bibfnamefont {P.}~\bibnamefont {Wcis{\l}o}}, \bibinfo {author}
  {\bibfnamefont {F.}~\bibnamefont {Thibault}}, \ and\ \bibinfo {author}
  {\bibfnamefont {M.~Y.}\ \bibnamefont {Tretyakov}},\ }\href {\doibase
  10.1016/j.jqsrt.2021.107807} {\bibfield  {journal} {\bibinfo  {journal} {J.
  Quant. Spectrosc. Radiat. Transfer}\ }\textbf {\bibinfo {volume} {272}},\
  \bibinfo {pages} {107807} (\bibinfo {year} {2021})}\BibitemShut {NoStop}%
\bibitem [{\citenamefont {Słowiński}\ \emph {et~al.}(2022)\citenamefont
  {Słowiński}, \citenamefont {Jóźwiak}, \citenamefont {Gancewski},
  \citenamefont {Stankiewicz}, \citenamefont {Stolarczyk}, \citenamefont {Tan},
  \citenamefont {Wang}, \citenamefont {Liu}, \citenamefont {Hu}, \citenamefont
  {Kassi}, \citenamefont {Campargue}, \citenamefont {Patkowski}, \citenamefont
  {Żuchowski}, \citenamefont {Ciuryło}, \citenamefont {Thibault},\ and\
  \citenamefont {Wcisło}}]{słowiński2022}%
  \BibitemOpen
  \bibfield  {author} {\bibinfo {author} {\bibfnamefont {M.}~\bibnamefont
  {Słowiński}}, \bibinfo {author} {\bibfnamefont {H.}~\bibnamefont
  {Jóźwiak}}, \bibinfo {author} {\bibfnamefont {M.}~\bibnamefont
  {Gancewski}}, \bibinfo {author} {\bibfnamefont {K.}~\bibnamefont
  {Stankiewicz}}, \bibinfo {author} {\bibfnamefont {N.}~\bibnamefont
  {Stolarczyk}}, \bibinfo {author} {\bibfnamefont {Y.}~\bibnamefont {Tan}},
  \bibinfo {author} {\bibfnamefont {J.}~\bibnamefont {Wang}}, \bibinfo {author}
  {\bibfnamefont {A.-W.}\ \bibnamefont {Liu}}, \bibinfo {author} {\bibfnamefont
  {S.-M.}\ \bibnamefont {Hu}}, \bibinfo {author} {\bibfnamefont
  {S.}~\bibnamefont {Kassi}}, \bibinfo {author} {\bibfnamefont
  {A.}~\bibnamefont {Campargue}}, \bibinfo {author} {\bibfnamefont
  {K.}~\bibnamefont {Patkowski}}, \bibinfo {author} {\bibfnamefont
  {P.}~\bibnamefont {Żuchowski}}, \bibinfo {author} {\bibfnamefont
  {R.}~\bibnamefont {Ciuryło}}, \bibinfo {author} {\bibfnamefont
  {F.}~\bibnamefont {Thibault}}, \ and\ \bibinfo {author} {\bibfnamefont
  {P.}~\bibnamefont {Wcisło}},\ }\href {\doibase 10.1016/j.jqsrt.2021.107951}
  {\bibfield  {journal} {\bibinfo  {journal} {J. Quant. Spectrosc. Radiat.
  Transf.}\ }\textbf {\bibinfo {volume} {277}},\ \bibinfo {pages} {107951}
  (\bibinfo {year} {2022})}\BibitemShut {NoStop}%
\bibitem [{\citenamefont {Stolarczyk}\ \emph {et~al.}(2023)\citenamefont
  {Stolarczyk}, \citenamefont {Kowzan}, \citenamefont {Thibault}, \citenamefont
  {Cybulski}, \citenamefont {Słowiński}, \citenamefont {Tan}, \citenamefont
  {Wang}, \citenamefont {Liu}, \citenamefont {Hu},\ and\ \citenamefont
  {Wcisło}}]{stolarczyk2023}%
  \BibitemOpen
  \bibfield  {author} {\bibinfo {author} {\bibfnamefont {N.}~\bibnamefont
  {Stolarczyk}}, \bibinfo {author} {\bibfnamefont {G.}~\bibnamefont {Kowzan}},
  \bibinfo {author} {\bibfnamefont {F.}~\bibnamefont {Thibault}}, \bibinfo
  {author} {\bibfnamefont {H.}~\bibnamefont {Cybulski}}, \bibinfo {author}
  {\bibfnamefont {M.}~\bibnamefont {Słowiński}}, \bibinfo {author}
  {\bibfnamefont {Y.}~\bibnamefont {Tan}}, \bibinfo {author} {\bibfnamefont
  {J.}~\bibnamefont {Wang}}, \bibinfo {author} {\bibfnamefont {A.-W.}\
  \bibnamefont {Liu}}, \bibinfo {author} {\bibfnamefont {S.-M.}\ \bibnamefont
  {Hu}}, \ and\ \bibinfo {author} {\bibfnamefont {P.}~\bibnamefont {Wcisło}},\
  }\href {\doibase 10.1063/5.0139229} {\bibfield  {journal} {\bibinfo
  {journal} {J. Chem. Phys.}\ }\textbf {\bibinfo {volume} {158}},\ \bibinfo
  {pages} {094303} (\bibinfo {year} {2023})}\BibitemShut {NoStop}%
\bibitem [{\citenamefont {Wcisło}\ \emph {et~al.}(2018)\citenamefont
  {Wcisło}, \citenamefont {Thibault}, \citenamefont {Zaborowski},
  \citenamefont {Wójtewicz}, \citenamefont {Cygan}, \citenamefont {Kowzan},
  \citenamefont {Masłowski}, \citenamefont {Komasa}, \citenamefont
  {Puchalski}, \citenamefont {Pachucki}, \citenamefont {Ciuryło},\ and\
  \citenamefont {Lisak}}]{wcisło2018}%
  \BibitemOpen
  \bibfield  {author} {\bibinfo {author} {\bibfnamefont {P.}~\bibnamefont
  {Wcisło}}, \bibinfo {author} {\bibfnamefont {F.}~\bibnamefont {Thibault}},
  \bibinfo {author} {\bibfnamefont {M.}~\bibnamefont {Zaborowski}}, \bibinfo
  {author} {\bibfnamefont {S.}~\bibnamefont {Wójtewicz}}, \bibinfo {author}
  {\bibfnamefont {A.}~\bibnamefont {Cygan}}, \bibinfo {author} {\bibfnamefont
  {G.}~\bibnamefont {Kowzan}}, \bibinfo {author} {\bibfnamefont
  {P.}~\bibnamefont {Masłowski}}, \bibinfo {author} {\bibfnamefont
  {J.}~\bibnamefont {Komasa}}, \bibinfo {author} {\bibfnamefont
  {M.}~\bibnamefont {Puchalski}}, \bibinfo {author} {\bibfnamefont
  {K.}~\bibnamefont {Pachucki}}, \bibinfo {author} {\bibfnamefont
  {R.}~\bibnamefont {Ciuryło}}, \ and\ \bibinfo {author} {\bibfnamefont
  {D.}~\bibnamefont {Lisak}},\ }\href {\doibase 10.1016/j.jqsrt.2018.04.011}
  {\bibfield  {journal} {\bibinfo  {journal} {J. Quant. Spectrosc. Radiat.
  Transf.}\ }\textbf {\bibinfo {volume} {213}},\ \bibinfo {pages} {41}
  (\bibinfo {year} {2018})}\BibitemShut {NoStop}%
\bibitem [{\citenamefont {Lamperti}\ \emph {et~al.}(2023)\citenamefont
  {Lamperti}, \citenamefont {Rutkowski}, \citenamefont {Ronchetti},
  \citenamefont {Gatti}, \citenamefont {Gotti}, \citenamefont {Cerullo},
  \citenamefont {Thibault}, \citenamefont {J{\'{o}}{\'{z}}wiak}, \citenamefont
  {W{\'{o}}jtewicz}, \citenamefont {Mas{\l}owski}, \citenamefont {Wcis{\l}o},
  \citenamefont {Polli},\ and\ \citenamefont {Marangoni}}]{Lamperti_2023}%
  \BibitemOpen
  \bibfield  {author} {\bibinfo {author} {\bibfnamefont {M.}~\bibnamefont
  {Lamperti}}, \bibinfo {author} {\bibfnamefont {L.}~\bibnamefont {Rutkowski}},
  \bibinfo {author} {\bibfnamefont {D.}~\bibnamefont {Ronchetti}}, \bibinfo
  {author} {\bibfnamefont {D.}~\bibnamefont {Gatti}}, \bibinfo {author}
  {\bibfnamefont {R.}~\bibnamefont {Gotti}}, \bibinfo {author} {\bibfnamefont
  {G.}~\bibnamefont {Cerullo}}, \bibinfo {author} {\bibfnamefont
  {F.}~\bibnamefont {Thibault}}, \bibinfo {author} {\bibfnamefont
  {H.}~\bibnamefont {J{\'{o}}{\'{z}}wiak}}, \bibinfo {author} {\bibfnamefont
  {S.}~\bibnamefont {W{\'{o}}jtewicz}}, \bibinfo {author} {\bibfnamefont
  {P.}~\bibnamefont {Mas{\l}owski}}, \bibinfo {author} {\bibfnamefont
  {P.}~\bibnamefont {Wcis{\l}o}}, \bibinfo {author} {\bibfnamefont
  {D.}~\bibnamefont {Polli}}, \ and\ \bibinfo {author} {\bibfnamefont
  {M.}~\bibnamefont {Marangoni}},\ }\href {\doibase 10.1038/s42005-023-01187-z}
  {\bibfield  {journal} {\bibinfo  {journal} {Comm.~Phys.}\ }\textbf {\bibinfo
  {volume} {6}} (\bibinfo {year} {2023}),\
  10.1038/s42005-023-01187-z}\BibitemShut {NoStop}%
\bibitem [{\citenamefont {Paredes-Roib{\'{a}}s}\ \emph
  {et~al.}(2021)\citenamefont {Paredes-Roib{\'{a}}s}, \citenamefont
  {Mart{\'{\i}}nez}, \citenamefont {J{\'{o}}{\'{z}}wiak},\ and\ \citenamefont
  {Thibault}}]{Paredes_Roibas_2021}%
  \BibitemOpen
  \bibfield  {author} {\bibinfo {author} {\bibfnamefont {D.}~\bibnamefont
  {Paredes-Roib{\'{a}}s}}, \bibinfo {author} {\bibfnamefont {R.~Z.}\
  \bibnamefont {Mart{\'{\i}}nez}}, \bibinfo {author} {\bibfnamefont
  {H.}~\bibnamefont {J{\'{o}}{\'{z}}wiak}}, \ and\ \bibinfo {author}
  {\bibfnamefont {F.}~\bibnamefont {Thibault}},\ }\href {\doibase
  10.1016/j.jqsrt.2021.107868} {\bibfield  {journal} {\bibinfo  {journal} {J.
  Quant. Spectrosc. Radiat. Transf.}\ }\textbf {\bibinfo {volume} {275}},\
  \bibinfo {pages} {107868} (\bibinfo {year} {2021})}\BibitemShut {NoStop}%
\bibitem [{\citenamefont {Knizia}\ \emph {et~al.}(2009)\citenamefont {Knizia},
  \citenamefont {Adler},\ and\ \citenamefont {Werner}}]{knizia2009simplified}%
  \BibitemOpen
  \bibfield  {author} {\bibinfo {author} {\bibfnamefont {G.}~\bibnamefont
  {Knizia}}, \bibinfo {author} {\bibfnamefont {T.~B.}\ \bibnamefont {Adler}}, \
  and\ \bibinfo {author} {\bibfnamefont {H.-J.}\ \bibnamefont {Werner}},\
  }\href@noop {} {\bibfield  {journal} {\bibinfo  {journal} {J. Chem. Phys.}\
  }\textbf {\bibinfo {volume} {130}},\ \bibinfo {pages} {054104} (\bibinfo
  {year} {2009})}\BibitemShut {NoStop}%
\bibitem [{\citenamefont {Peterson}\ \emph {et~al.}(2008)\citenamefont
  {Peterson}, \citenamefont {Adler},\ and\ \citenamefont
  {Werner}}]{peterson2008systematically}%
  \BibitemOpen
  \bibfield  {author} {\bibinfo {author} {\bibfnamefont {K.~A.}\ \bibnamefont
  {Peterson}}, \bibinfo {author} {\bibfnamefont {T.~B.}\ \bibnamefont {Adler}},
  \ and\ \bibinfo {author} {\bibfnamefont {H.-J.}\ \bibnamefont {Werner}},\
  }\href@noop {} {\bibfield  {journal} {\bibinfo  {journal} {J. Chem. Phys.}\
  }\textbf {\bibinfo {volume} {128}},\ \bibinfo {pages} {084102} (\bibinfo
  {year} {2008})}\BibitemShut {NoStop}%
\bibitem [{\citenamefont {Feller}\ \emph {et~al.}(2006)\citenamefont {Feller},
  \citenamefont {Peterson},\ and\ \citenamefont
  {Crawford}}]{feller2006sources}%
  \BibitemOpen
  \bibfield  {author} {\bibinfo {author} {\bibfnamefont {D.}~\bibnamefont
  {Feller}}, \bibinfo {author} {\bibfnamefont {K.~A.}\ \bibnamefont
  {Peterson}}, \ and\ \bibinfo {author} {\bibfnamefont {T.~D.}\ \bibnamefont
  {Crawford}},\ }\href@noop {} {\bibfield  {journal} {\bibinfo  {journal} {The
  Journal of chemical physics}\ }\textbf {\bibinfo {volume} {124}},\ \bibinfo
  {pages} {054107} (\bibinfo {year} {2006})}\BibitemShut {NoStop}%
\bibitem [{\citenamefont {Werner}\ \emph {et~al.}(2012)\citenamefont {Werner},
  \citenamefont {Knowles}, \citenamefont {Knizia}, \citenamefont {Manby},\ and\
  \citenamefont {Sch{\"{u}}tz}}]{Werner2012-molpro}%
  \BibitemOpen
  \bibfield  {author} {\bibinfo {author} {\bibfnamefont {H.-J.}\ \bibnamefont
  {Werner}}, \bibinfo {author} {\bibfnamefont {P.~J.}\ \bibnamefont {Knowles}},
  \bibinfo {author} {\bibfnamefont {G.}~\bibnamefont {Knizia}}, \bibinfo
  {author} {\bibfnamefont {F.~R.}\ \bibnamefont {Manby}}, \ and\ \bibinfo
  {author} {\bibfnamefont {M.}~\bibnamefont {Sch{\"{u}}tz}},\ }\href {\doibase
  10.1002/wcms.82} {\bibfield  {journal} {\bibinfo  {journal} {Wiley
  Interdisciplinary Reviews: Computational Molecular Science}\ }\textbf
  {\bibinfo {volume} {2}},\ \bibinfo {pages} {242} (\bibinfo {year}
  {2012})}\BibitemShut {NoStop}%
\bibitem [{\citenamefont {Dumouchel}\ \emph {et~al.}(2023)\citenamefont
  {Dumouchel}, \citenamefont {Quintas-S{\'a}nchez}, \citenamefont
  {Balan{\c{c}}a}, \citenamefont {Dawes}, \citenamefont {Lique},\ and\
  \citenamefont {Feautrier}}]{dumouchel2023collisional}%
  \BibitemOpen
  \bibfield  {author} {\bibinfo {author} {\bibfnamefont {F.}~\bibnamefont
  {Dumouchel}}, \bibinfo {author} {\bibfnamefont {E.}~\bibnamefont
  {Quintas-S{\'a}nchez}}, \bibinfo {author} {\bibfnamefont {C.}~\bibnamefont
  {Balan{\c{c}}a}}, \bibinfo {author} {\bibfnamefont {R.}~\bibnamefont
  {Dawes}}, \bibinfo {author} {\bibfnamefont {F.}~\bibnamefont {Lique}}, \ and\
  \bibinfo {author} {\bibfnamefont {N.}~\bibnamefont {Feautrier}},\ }\href@noop
  {} {\bibfield  {journal} {\bibinfo  {journal} {The Journal of Chemical
  Physics}\ }\textbf {\bibinfo {volume} {158}} (\bibinfo {year}
  {2023})}\BibitemShut {NoStop}%
\bibitem [{\citenamefont {Zadrożny}\ \emph {et~al.}(2022)\citenamefont
  {Zadrożny}, \citenamefont {Jóźwiak}, \citenamefont {Ernesto},
  \citenamefont {Dawes},\ and\ \citenamefont {Wcisło}}]{zadrożny2022}%
  \BibitemOpen
  \bibfield  {author} {\bibinfo {author} {\bibfnamefont {A.}~\bibnamefont
  {Zadrożny}}, \bibinfo {author} {\bibfnamefont {H.}~\bibnamefont
  {Jóźwiak}}, \bibinfo {author} {\bibfnamefont {E.~Q.-S.}\ \bibnamefont
  {Ernesto}}, \bibinfo {author} {\bibfnamefont {R.}~\bibnamefont {Dawes}}, \
  and\ \bibinfo {author} {\bibfnamefont {P.}~\bibnamefont {Wcisło}},\ }\href
  {\doibase 10.1063/5.0115654} {\bibfield  {journal} {\bibinfo  {journal} {J.
  Chem. Phys.}\ }\textbf {\bibinfo {volume} {157}},\ \bibinfo {pages} {174310}
  (\bibinfo {year} {2022})}\BibitemShut {NoStop}%
\bibitem [{\citenamefont {Denis-Alpizar}\ \emph {et~al.}(2022)\citenamefont
  {Denis-Alpizar}, \citenamefont {Quintas-S{\'a}nchez},\ and\ \citenamefont
  {Dawes}}]{denis2022state}%
  \BibitemOpen
  \bibfield  {author} {\bibinfo {author} {\bibfnamefont {O.}~\bibnamefont
  {Denis-Alpizar}}, \bibinfo {author} {\bibfnamefont {E.}~\bibnamefont
  {Quintas-S{\'a}nchez}}, \ and\ \bibinfo {author} {\bibfnamefont
  {R.}~\bibnamefont {Dawes}},\ }\href@noop {} {\bibfield  {journal} {\bibinfo
  {journal} {Monthly Notices of the Royal Astronomical Society}\ }\textbf
  {\bibinfo {volume} {512}},\ \bibinfo {pages} {5546} (\bibinfo {year}
  {2022})}\BibitemShut {NoStop}%
\bibitem [{\citenamefont {Ajili}\ \emph {et~al.}(2022)\citenamefont {Ajili},
  \citenamefont {Quintas-S{\'a}nchez}, \citenamefont {Mehnen}, \citenamefont
  {{\.Z}uchowski}, \citenamefont {Brz{\k{e}}k}, \citenamefont {El-Kork},
  \citenamefont {Gacesa}, \citenamefont {Dawes},\ and\ \citenamefont
  {Hochlaf}}]{ajili2022theoretical}%
  \BibitemOpen
  \bibfield  {author} {\bibinfo {author} {\bibfnamefont {Y.}~\bibnamefont
  {Ajili}}, \bibinfo {author} {\bibfnamefont {E.}~\bibnamefont
  {Quintas-S{\'a}nchez}}, \bibinfo {author} {\bibfnamefont {B.}~\bibnamefont
  {Mehnen}}, \bibinfo {author} {\bibfnamefont {P.~S.}\ \bibnamefont
  {{\.Z}uchowski}}, \bibinfo {author} {\bibfnamefont {F.}~\bibnamefont
  {Brz{\k{e}}k}}, \bibinfo {author} {\bibfnamefont {N.}~\bibnamefont
  {El-Kork}}, \bibinfo {author} {\bibfnamefont {M.}~\bibnamefont {Gacesa}},
  \bibinfo {author} {\bibfnamefont {R.}~\bibnamefont {Dawes}}, \ and\ \bibinfo
  {author} {\bibfnamefont {M.}~\bibnamefont {Hochlaf}},\ }\href@noop {}
  {\bibfield  {journal} {\bibinfo  {journal} {Physical Chemistry Chemical
  Physics}\ }\textbf {\bibinfo {volume} {24}},\ \bibinfo {pages} {28984}
  (\bibinfo {year} {2022})}\BibitemShut {NoStop}%
\bibitem [{\citenamefont {Quintas-S{\'a}nchez}\ and\ \citenamefont
  {Dawes}(2019)}]{quintas2018autosurf}%
  \BibitemOpen
  \bibfield  {author} {\bibinfo {author} {\bibfnamefont {E.}~\bibnamefont
  {Quintas-S{\'a}nchez}}\ and\ \bibinfo {author} {\bibfnamefont
  {R.}~\bibnamefont {Dawes}},\ }\href@noop {} {\bibfield  {journal} {\bibinfo
  {journal} {J. Chem. Inf. Model.}\ }\textbf {\bibinfo {volume} {59}},\
  \bibinfo {pages} {262} (\bibinfo {year} {2019})}\BibitemShut {NoStop}%
\bibitem [{\citenamefont {Dawes}\ and\ \citenamefont
  {Quintas-S\'anchez}(2018)}]{Dawes2018}%
  \BibitemOpen
  \bibfield  {author} {\bibinfo {author} {\bibfnamefont {R.}~\bibnamefont
  {Dawes}}\ and\ \bibinfo {author} {\bibfnamefont {E.}~\bibnamefont
  {Quintas-S\'anchez}},\ }\href {\doibase 10.1002/9781119518068.ch5} {\emph
  {\bibinfo {title} {Reviews in Computational Chemistry vol. 31}}}\ (\bibinfo
  {publisher} {John Wiley \& Sons, Inc.},\ \bibinfo {year} {2018})\
  Chap.~\bibinfo {chapter} {5}, pp.\ \bibinfo {pages} {199--264}\BibitemShut
  {NoStop}%
\bibitem [{\citenamefont {Majumder}\ \emph {et~al.}(2016)\citenamefont
  {Majumder}, \citenamefont {Ndengu\'e},\ and\ \citenamefont
  {Dawes}}]{majumder2016automated}%
  \BibitemOpen
  \bibfield  {author} {\bibinfo {author} {\bibfnamefont {M.}~\bibnamefont
  {Majumder}}, \bibinfo {author} {\bibfnamefont {S.~A.}\ \bibnamefont
  {Ndengu\'e}}, \ and\ \bibinfo {author} {\bibfnamefont {R.}~\bibnamefont
  {Dawes}},\ }\href {\doibase 10.1080/00268976.2015.1096974} {\bibfield
  {journal} {\bibinfo  {journal} {Mol. Phys.}\ }\textbf {\bibinfo {volume}
  {114}},\ \bibinfo {pages} {1} (\bibinfo {year} {2016})}\BibitemShut {NoStop}%
\bibitem [{\citenamefont {Dawes}\ \emph {et~al.}(2010)\citenamefont {Dawes},
  \citenamefont {Wang}, \citenamefont {Jasper},\ and\ \citenamefont
  {Carrington~Jr}}]{dawes2010nitrous}%
  \BibitemOpen
  \bibfield  {author} {\bibinfo {author} {\bibfnamefont {R.}~\bibnamefont
  {Dawes}}, \bibinfo {author} {\bibfnamefont {X.-G.}\ \bibnamefont {Wang}},
  \bibinfo {author} {\bibfnamefont {A.~W.}\ \bibnamefont {Jasper}}, \ and\
  \bibinfo {author} {\bibfnamefont {T.}~\bibnamefont {Carrington~Jr}},\
  }\href@noop {} {\bibfield  {journal} {\bibinfo  {journal} {The Journal of
  chemical physics}\ }\textbf {\bibinfo {volume} {133}},\ \bibinfo {pages}
  {134304} (\bibinfo {year} {2010})}\BibitemShut {NoStop}%
\bibitem [{\citenamefont {Sobol}(1976)}]{sobol1976uniformly}%
  \BibitemOpen
  \bibfield  {author} {\bibinfo {author} {\bibfnamefont {I.~M.}\ \bibnamefont
  {Sobol}},\ }\href@noop {} {\bibfield  {journal} {\bibinfo  {journal} {USSR
  Computational Mathematics and Mathematical Physics}\ }\textbf {\bibinfo
  {volume} {16}},\ \bibinfo {pages} {236} (\bibinfo {year} {1976})}\BibitemShut
  {NoStop}%
\bibitem [{\citenamefont {Dawes}\ \emph {et~al.}(2013)\citenamefont {Dawes},
  \citenamefont {Wang},\ and\ \citenamefont {Carrington}}]{Dawes2013}%
  \BibitemOpen
  \bibfield  {author} {\bibinfo {author} {\bibfnamefont {R.}~\bibnamefont
  {Dawes}}, \bibinfo {author} {\bibfnamefont {X.~G.}\ \bibnamefont {Wang}}, \
  and\ \bibinfo {author} {\bibfnamefont {T.}~\bibnamefont {Carrington}},\
  }\href {\doibase 10.1021/jp404888d} {\bibfield  {journal} {\bibinfo
  {journal} {Journal of Physical Chemistry A}\ }\textbf {\bibinfo {volume}
  {117}},\ \bibinfo {pages} {7612} (\bibinfo {year} {2013})}\BibitemShut
  {NoStop}%
\bibitem [{\citenamefont {Jóźwiak}\ \emph {et~al.}(2021)\citenamefont
  {Jóźwiak}, \citenamefont {Thibault}, \citenamefont {Cybulski},\ and\
  \citenamefont {Wcisło}}]{jóźwiak2021}%
  \BibitemOpen
  \bibfield  {author} {\bibinfo {author} {\bibfnamefont {H.}~\bibnamefont
  {Jóźwiak}}, \bibinfo {author} {\bibfnamefont {F.}~\bibnamefont {Thibault}},
  \bibinfo {author} {\bibfnamefont {H.}~\bibnamefont {Cybulski}}, \ and\
  \bibinfo {author} {\bibfnamefont {P.}~\bibnamefont {Wcisło}},\ }\href
  {\doibase 10.1063/5.0040438} {\bibfield  {journal} {\bibinfo  {journal} {J.
  Chem. Phys.}\ }\textbf {\bibinfo {volume} {154}},\ \bibinfo {pages} {054314}
  (\bibinfo {year} {2021})}\BibitemShut {NoStop}%
\bibitem [{\citenamefont {Green}(1975)}]{green1975}%
  \BibitemOpen
  \bibfield  {author} {\bibinfo {author} {\bibfnamefont {S.}~\bibnamefont
  {Green}},\ }\href {\doibase 10.1063/1.430752} {\bibfield  {journal} {\bibinfo
   {journal} {J. Chem. Phys.}\ }\textbf {\bibinfo {volume} {62}},\ \bibinfo
  {pages} {2271} (\bibinfo {year} {1975})}\BibitemShut {NoStop}%
\bibitem [{\citenamefont {Gancewski}\ \emph {et~al.}(2021)\citenamefont
  {Gancewski}, \citenamefont {Jóźwiak}, \citenamefont {Ernesto},
  \citenamefont {Dawes}, \citenamefont {Thibault},\ and\ \citenamefont
  {Wcisło}}]{gancewski2021}%
  \BibitemOpen
  \bibfield  {author} {\bibinfo {author} {\bibfnamefont {M.}~\bibnamefont
  {Gancewski}}, \bibinfo {author} {\bibfnamefont {H.}~\bibnamefont
  {Jóźwiak}}, \bibinfo {author} {\bibfnamefont {E.~Q.-S.}\ \bibnamefont
  {Ernesto}}, \bibinfo {author} {\bibfnamefont {R.}~\bibnamefont {Dawes}},
  \bibinfo {author} {\bibfnamefont {F.}~\bibnamefont {Thibault}}, \ and\
  \bibinfo {author} {\bibfnamefont {P.}~\bibnamefont {Wcisło}},\ }\href
  {\doibase 10.1063/5.0063006} {\bibfield  {journal} {\bibinfo  {journal} {J.
  Chem. Phys.}\ }\textbf {\bibinfo {volume} {155}},\ \bibinfo {pages} {124307}
  (\bibinfo {year} {2021})}\BibitemShut {NoStop}%
\bibitem [{\citenamefont {Launay}(1977)}]{Launay_1977}%
  \BibitemOpen
  \bibfield  {author} {\bibinfo {author} {\bibfnamefont {J.~M.}\ \bibnamefont
  {Launay}},\ }\href {\doibase 10.1088/0022-3700/10/18/023} {\bibfield
  {journal} {\bibinfo  {journal} {J.~Phys.~B-At.~Mol.~Opt.}\ }\textbf {\bibinfo
  {volume} {10}},\ \bibinfo {pages} {3665} (\bibinfo {year}
  {1977})}\BibitemShut {NoStop}%
\bibitem [{\citenamefont {Alexander}\ and\ \citenamefont
  {DePristo}(1977)}]{Alexander_1977}%
  \BibitemOpen
  \bibfield  {author} {\bibinfo {author} {\bibfnamefont {M.~H.}\ \bibnamefont
  {Alexander}}\ and\ \bibinfo {author} {\bibfnamefont {A.~E.}\ \bibnamefont
  {DePristo}},\ }\href {\doibase 10.1063/1.434132} {\bibfield  {journal}
  {\bibinfo  {journal} {J.~Chem.~Phys.}\ }\textbf {\bibinfo {volume} {66}},\
  \bibinfo {pages} {2166} (\bibinfo {year} {1977})}\BibitemShut {NoStop}%
\bibitem [{\citenamefont {Lill}\ \emph {et~al.}(1982)\citenamefont {Lill},
  \citenamefont {Parker},\ and\ \citenamefont {Light}}]{Lill_1982}%
  \BibitemOpen
  \bibfield  {author} {\bibinfo {author} {\bibfnamefont {J.~V.}\ \bibnamefont
  {Lill}}, \bibinfo {author} {\bibfnamefont {G.~A.}\ \bibnamefont {Parker}}, \
  and\ \bibinfo {author} {\bibfnamefont {J.~C.}\ \bibnamefont {Light}},\ }\href
  {\doibase 10.1016/0009-2614(82)83051-0} {\bibfield  {journal} {\bibinfo
  {journal} {Chemical Physics Letters}\ }\textbf {\bibinfo {volume} {89}},\
  \bibinfo {pages} {483} (\bibinfo {year} {1982})}\BibitemShut {NoStop}%
\bibitem [{\citenamefont {Coxon}\ and\ \citenamefont
  {Hajigeorgiou}(2015)}]{coxon2015}%
  \BibitemOpen
  \bibfield  {author} {\bibinfo {author} {\bibfnamefont {J.~A.}\ \bibnamefont
  {Coxon}}\ and\ \bibinfo {author} {\bibfnamefont {P.~G.}\ \bibnamefont
  {Hajigeorgiou}},\ }\href {\doibase 10.1016/j.jqsrt.2014.08.028} {\bibfield
  {journal} {\bibinfo  {journal} {J. Quant. Spectrosc. Radiat. Transf.}\
  }\textbf {\bibinfo {volume} {151}},\ \bibinfo {pages} {133} (\bibinfo {year}
  {2015})}\BibitemShut {NoStop}%
\bibitem [{\citenamefont {Cazzoli}\ and\ \citenamefont
  {Puzzarini}(2004)}]{Cazzoli_2004}%
  \BibitemOpen
  \bibfield  {author} {\bibinfo {author} {\bibfnamefont {G.}~\bibnamefont
  {Cazzoli}}\ and\ \bibinfo {author} {\bibfnamefont {C.}~\bibnamefont
  {Puzzarini}},\ }\href {\doibase 10.1016/j.jms.2004.03.020} {\bibfield
  {journal} {\bibinfo  {journal} {J. Mol. Spectrosc.}\ }\textbf {\bibinfo
  {volume} {226}},\ \bibinfo {pages} {161} (\bibinfo {year}
  {2004})}\BibitemShut {NoStop}%
\bibitem [{\citenamefont {Hajigeorgiou}(2013)}]{photos2013}%
  \BibitemOpen
  \bibfield  {author} {\bibinfo {author} {\bibfnamefont {P.~G.}\ \bibnamefont
  {Hajigeorgiou}},\ }\href {\doibase 10.1063/1.4773285} {\bibfield  {journal}
  {\bibinfo  {journal} {J. Chem. Phys.}\ }\textbf {\bibinfo {volume} {138}},\
  \bibinfo {pages} {014309} (\bibinfo {year} {2013})}\BibitemShut {NoStop}%
\bibitem [{\citenamefont {Rabitz}(1975)}]{Rabitz_1975}%
  \BibitemOpen
  \bibfield  {author} {\bibinfo {author} {\bibfnamefont {H.}~\bibnamefont
  {Rabitz}},\ }\href {\doibase 10.1063/1.431304} {\bibfield  {journal}
  {\bibinfo  {journal} {J.~Chem.~Phys.}\ }\textbf {\bibinfo {volume} {63}},\
  \bibinfo {pages} {5208} (\bibinfo {year} {1975})}\BibitemShut {NoStop}%
\bibitem [{\citenamefont {Johnson}(1978)}]{Johnson_1978}%
  \BibitemOpen
  \bibfield  {author} {\bibinfo {author} {\bibfnamefont {B.~R.}\ \bibnamefont
  {Johnson}},\ }\href {\doibase 10.1063/1.436421} {\bibfield  {journal}
  {\bibinfo  {journal} {J. Chem. Phys.}\ }\textbf {\bibinfo {volume} {69}},\
  \bibinfo {pages} {4678} (\bibinfo {year} {1978})}\BibitemShut {NoStop}%
\bibitem [{\citenamefont {Jóźwiak}\ \emph {et~al.}()\citenamefont
  {Jóźwiak}, \citenamefont {Gancewski}, \citenamefont {Grabowski},
  \citenamefont {Stankiewicz}, \citenamefont {Zadrożny}, \citenamefont
  {Olejnik},\ and\ \citenamefont {Wcisło}}]{BIGOS}%
  \BibitemOpen
  \bibfield  {author} {\bibinfo {author} {\bibfnamefont {H.}~\bibnamefont
  {Jóźwiak}}, \bibinfo {author} {\bibfnamefont {M.}~\bibnamefont
  {Gancewski}}, \bibinfo {author} {\bibfnamefont {A.}~\bibnamefont
  {Grabowski}}, \bibinfo {author} {\bibfnamefont {K.}~\bibnamefont
  {Stankiewicz}}, \bibinfo {author} {\bibfnamefont {A.}~\bibnamefont
  {Zadrożny}}, \bibinfo {author} {\bibfnamefont {A.}~\bibnamefont {Olejnik}},
  \ and\ \bibinfo {author} {\bibfnamefont {P.}~\bibnamefont {Wcisło}},\
  }\href@noop {} {\enquote {\bibinfo {title} {{BIGOS Computer Code}},}\
  }\BibitemShut {NoStop}%
\bibitem [{\citenamefont {Monchick}\ and\ \citenamefont
  {Hunter}(1986)}]{monchick1986diatomic}%
  \BibitemOpen
  \bibfield  {author} {\bibinfo {author} {\bibfnamefont {L.}~\bibnamefont
  {Monchick}}\ and\ \bibinfo {author} {\bibfnamefont {L.}~\bibnamefont
  {Hunter}},\ }\href {\doibase 10.1063/1.451277} {\bibfield  {journal}
  {\bibinfo  {journal} {J. Chem. Phys.}\ }\textbf {\bibinfo {volume} {85}},\
  \bibinfo {pages} {713} (\bibinfo {year} {1986})}\BibitemShut {NoStop}%
\bibitem [{\citenamefont {Sch{\"a}fer}\ and\ \citenamefont
  {Monchick}(1992)}]{schafer1992line}%
  \BibitemOpen
  \bibfield  {author} {\bibinfo {author} {\bibfnamefont {J.}~\bibnamefont
  {Sch{\"a}fer}}\ and\ \bibinfo {author} {\bibfnamefont {L.}~\bibnamefont
  {Monchick}},\ }\href {\doibase 10.1063/1.453612} {\bibfield  {journal}
  {\bibinfo  {journal} {Astronomy and Astrophysics}\ }\textbf {\bibinfo
  {volume} {265}},\ \bibinfo {pages} {859} (\bibinfo {year}
  {1992})}\BibitemShut {NoStop}%
\bibitem [{\citenamefont {Yutsis}\ \emph {et~al.}(1962)\citenamefont {Yutsis},
  \citenamefont {Levinson},\ and\ \citenamefont {Vangas}}]{Yutsis}%
  \BibitemOpen
  \bibfield  {author} {\bibinfo {author} {\bibfnamefont {A.~P.}\ \bibnamefont
  {Yutsis}}, \bibinfo {author} {\bibfnamefont {I.~B.}\ \bibnamefont
  {Levinson}}, \ and\ \bibinfo {author} {\bibfnamefont {V.~V.}\ \bibnamefont
  {Vangas}},\ }\href@noop {} {\emph {\bibinfo {title} {Theory of angular
  momentum}}}\ (\bibinfo  {publisher} {Israel Program for Scientific
  Translations, Jerusalem},\ \bibinfo {year} {1962})\BibitemShut {NoStop}%
\bibitem [{\citenamefont {Ben-Reuven}(1966{\natexlab{a}})}]{Ben_Reuven_1966a}%
  \BibitemOpen
  \bibfield  {author} {\bibinfo {author} {\bibfnamefont {A.}~\bibnamefont
  {Ben-Reuven}},\ }\href {\doibase 10.1103/physrev.141.34} {\bibfield
  {journal} {\bibinfo  {journal} {Physical Review}\ }\textbf {\bibinfo {volume}
  {141}},\ \bibinfo {pages} {34} (\bibinfo {year}
  {1966}{\natexlab{a}})}\BibitemShut {NoStop}%
\bibitem [{\citenamefont {Ben-Reuven}(1966{\natexlab{b}})}]{Ben_Reuven_1966b}%
  \BibitemOpen
  \bibfield  {author} {\bibinfo {author} {\bibfnamefont {A.}~\bibnamefont
  {Ben-Reuven}},\ }\href {\doibase 10.1103/physrev.145.7} {\bibfield  {journal}
  {\bibinfo  {journal} {Physical Review}\ }\textbf {\bibinfo {volume} {145}},\
  \bibinfo {pages} {7} (\bibinfo {year} {1966}{\natexlab{b}})}\BibitemShut
  {NoStop}%
\bibitem [{\citenamefont {Corey}\ and\ \citenamefont
  {McCourt}(1984)}]{Corey_1984}%
  \BibitemOpen
  \bibfield  {author} {\bibinfo {author} {\bibfnamefont {G.~C.}\ \bibnamefont
  {Corey}}\ and\ \bibinfo {author} {\bibfnamefont {F.~R.}\ \bibnamefont
  {McCourt}},\ }\href {\doibase 10.1063/1.447930} {\bibfield  {journal}
  {\bibinfo  {journal} {J. Chem. Phys.}\ }\textbf {\bibinfo {volume} {81}},\
  \bibinfo {pages} {2318} (\bibinfo {year} {1984})}\BibitemShut {NoStop}%
\bibitem [{\citenamefont {Wcisło}\ \emph {et~al.}(2015)\citenamefont
  {Wcisło}, \citenamefont {Thibault}, \citenamefont {Cybulski}, \citenamefont
  {Ciuryło}, \citenamefont {Cygan}, \citenamefont {Kowzan}, \citenamefont
  {Masłowski}, \citenamefont {Komasa}, \citenamefont {Puchalski},
  \citenamefont {Pachucki}, \citenamefont {Ciuryło},\ and\ \citenamefont
  {Lisak}}]{wcisło2015}%
  \BibitemOpen
  \bibfield  {author} {\bibinfo {author} {\bibfnamefont {P.}~\bibnamefont
  {Wcisło}}, \bibinfo {author} {\bibfnamefont {F.}~\bibnamefont {Thibault}},
  \bibinfo {author} {\bibfnamefont {H.}~\bibnamefont {Cybulski}}, \bibinfo
  {author} {\bibfnamefont {R.}~\bibnamefont {Ciuryło}}, \bibinfo {author}
  {\bibfnamefont {A.}~\bibnamefont {Cygan}}, \bibinfo {author} {\bibfnamefont
  {G.}~\bibnamefont {Kowzan}}, \bibinfo {author} {\bibfnamefont
  {P.}~\bibnamefont {Masłowski}}, \bibinfo {author} {\bibfnamefont
  {J.}~\bibnamefont {Komasa}}, \bibinfo {author} {\bibfnamefont
  {M.}~\bibnamefont {Puchalski}}, \bibinfo {author} {\bibfnamefont
  {K.}~\bibnamefont {Pachucki}}, \bibinfo {author} {\bibfnamefont
  {R.}~\bibnamefont {Ciuryło}}, \ and\ \bibinfo {author} {\bibfnamefont
  {D.}~\bibnamefont {Lisak}},\ }\href {\doibase 10.1103/PhysRevA.91.052505}
  {\bibfield  {journal} {\bibinfo  {journal} {Phys. Rev. A.}\ }\textbf
  {\bibinfo {volume} {91}},\ \bibinfo {pages} {052505} (\bibinfo {year}
  {2015})}\BibitemShut {NoStop}%
\bibitem [{\citenamefont {Gomez}\ \emph {et~al.}(2011)\citenamefont {Gomez},
  \citenamefont {Ivanov}, \citenamefont {Buzykin},\ and\ \citenamefont
  {Thibault}}]{gomez2011}%
  \BibitemOpen
  \bibfield  {author} {\bibinfo {author} {\bibfnamefont {L.}~\bibnamefont
  {Gomez}}, \bibinfo {author} {\bibfnamefont {S.~V.}\ \bibnamefont {Ivanov}},
  \bibinfo {author} {\bibfnamefont {O.~G.}\ \bibnamefont {Buzykin}}, \ and\
  \bibinfo {author} {\bibfnamefont {F.}~\bibnamefont {Thibault}},\ }\href
  {\doibase 10.1016/j.jqsrt.2011.04.005} {\bibfield  {journal} {\bibinfo
  {journal} {J. Quant. Spectrosc. Radiat. Transf.}\ }\textbf {\bibinfo {volume}
  {112}},\ \bibinfo {pages} {1942–1949} (\bibinfo {year} {2011})}\BibitemShut
  {NoStop}%
\bibitem [{\citenamefont {Thibault}\ \emph {et~al.}(2011)\citenamefont
  {Thibault}, \citenamefont {Martínez}, \citenamefont {Bermejo},\ and\
  \citenamefont {Gómez}}]{thibault2011}%
  \BibitemOpen
  \bibfield  {author} {\bibinfo {author} {\bibfnamefont {F.}~\bibnamefont
  {Thibault}}, \bibinfo {author} {\bibfnamefont {R.~Z.}\ \bibnamefont
  {Martínez}}, \bibinfo {author} {\bibfnamefont {D.}~\bibnamefont {Bermejo}},
  \ and\ \bibinfo {author} {\bibfnamefont {L.}~\bibnamefont {Gómez}},\ }\href
  {\doibase 10.1016/j.jqsrt.2011.07.006} {\bibfield  {journal} {\bibinfo
  {journal} {J. Quant. Spectrosc. Radiat. Transf.}\ }\textbf {\bibinfo {volume}
  {112}},\ \bibinfo {pages} {2542} (\bibinfo {year} {2011})}\BibitemShut
  {NoStop}%
\bibitem [{\citenamefont {Stolarczyk}\ \emph {et~al.}(2020)\citenamefont
  {Stolarczyk}, \citenamefont {Thibault}, \citenamefont {Cybulski},
  \citenamefont {Jóźwiak}, \citenamefont {Kowzan}, \citenamefont {Vispoel},
  \citenamefont {Gordon}, \citenamefont {Rothman}, \citenamefont {Gamache},\
  and\ \citenamefont {Wcisło}}]{stolarczyk2020}%
  \BibitemOpen
  \bibfield  {author} {\bibinfo {author} {\bibfnamefont {N.}~\bibnamefont
  {Stolarczyk}}, \bibinfo {author} {\bibfnamefont {F.}~\bibnamefont
  {Thibault}}, \bibinfo {author} {\bibfnamefont {H.}~\bibnamefont {Cybulski}},
  \bibinfo {author} {\bibfnamefont {H.}~\bibnamefont {Jóźwiak}}, \bibinfo
  {author} {\bibfnamefont {G.}~\bibnamefont {Kowzan}}, \bibinfo {author}
  {\bibfnamefont {B.}~\bibnamefont {Vispoel}}, \bibinfo {author} {\bibfnamefont
  {I.}~\bibnamefont {Gordon}}, \bibinfo {author} {\bibfnamefont
  {L.}~\bibnamefont {Rothman}}, \bibinfo {author} {\bibfnamefont
  {R.}~\bibnamefont {Gamache}}, \ and\ \bibinfo {author} {\bibfnamefont
  {P.}~\bibnamefont {Wcisło}},\ }\href {\doibase 10.1016/j.jqsrt.2019.106676}
  {\bibfield  {journal} {\bibinfo  {journal} {J. Quant. Spectrosc. Radiat.
  Transf.}\ }\textbf {\bibinfo {volume} {240}},\ \bibinfo {pages} {106676}
  (\bibinfo {year} {2020})}\BibitemShut {NoStop}%
\bibitem [{\citenamefont {Wcisło}\ \emph {et~al.}(2021)\citenamefont
  {Wcisło}, \citenamefont {Thibault}, \citenamefont {Stolarczyk},
  \citenamefont {Jóźwiak}, \citenamefont {Słowiński}, \citenamefont
  {Gancewski}, \citenamefont {Stankiewicz}, \citenamefont {Konefał},
  \citenamefont {Kassi}, \citenamefont {Campargue}, \citenamefont {Tan},
  \citenamefont {Wang}, \citenamefont {Patkowski}, \citenamefont {Ciuryło},
  \citenamefont {Lisak}, \citenamefont {Kochanov}, \citenamefont {Rothman},\
  and\ \citenamefont {Gordon}}]{wcisło2021}%
  \BibitemOpen
  \bibfield  {author} {\bibinfo {author} {\bibfnamefont {P.}~\bibnamefont
  {Wcisło}}, \bibinfo {author} {\bibfnamefont {F.}~\bibnamefont {Thibault}},
  \bibinfo {author} {\bibfnamefont {N.}~\bibnamefont {Stolarczyk}}, \bibinfo
  {author} {\bibfnamefont {H.}~\bibnamefont {Jóźwiak}}, \bibinfo {author}
  {\bibfnamefont {M.}~\bibnamefont {Słowiński}}, \bibinfo {author}
  {\bibfnamefont {M.}~\bibnamefont {Gancewski}}, \bibinfo {author}
  {\bibfnamefont {K.}~\bibnamefont {Stankiewicz}}, \bibinfo {author}
  {\bibfnamefont {M.}~\bibnamefont {Konefał}}, \bibinfo {author}
  {\bibfnamefont {S.}~\bibnamefont {Kassi}}, \bibinfo {author} {\bibfnamefont
  {A.}~\bibnamefont {Campargue}}, \bibinfo {author} {\bibfnamefont
  {Y.}~\bibnamefont {Tan}}, \bibinfo {author} {\bibfnamefont {J.}~\bibnamefont
  {Wang}}, \bibinfo {author} {\bibfnamefont {K.}~\bibnamefont {Patkowski}},
  \bibinfo {author} {\bibfnamefont {R.}~\bibnamefont {Ciuryło}}, \bibinfo
  {author} {\bibfnamefont {D.}~\bibnamefont {Lisak}}, \bibinfo {author}
  {\bibfnamefont {R.}~\bibnamefont {Kochanov}}, \bibinfo {author}
  {\bibfnamefont {L.}~\bibnamefont {Rothman}}, \ and\ \bibinfo {author}
  {\bibfnamefont {I.}~\bibnamefont {Gordon}},\ }\href {\doibase
  10.1016/j.jqsrt.2020.107477} {\bibfield  {journal} {\bibinfo  {journal} {J.
  Quant. Spectrosc. Radiat. Transf.}\ }\textbf {\bibinfo {volume} {260}},\
  \bibinfo {pages} {107477} (\bibinfo {year} {2021})}\BibitemShut {NoStop}%
\bibitem [{\citenamefont {Stankiewicz}\ \emph {et~al.}(2021)\citenamefont
  {Stankiewicz}, \citenamefont {Stolarczyk}, \citenamefont
  {J{\'{o}}{\'{z}}wiak}, \citenamefont {Thibault},\ and\ \citenamefont
  {Wcis{\l}o}}]{Stankiewicz_2021}%
  \BibitemOpen
  \bibfield  {author} {\bibinfo {author} {\bibfnamefont {K.}~\bibnamefont
  {Stankiewicz}}, \bibinfo {author} {\bibfnamefont {N.}~\bibnamefont
  {Stolarczyk}}, \bibinfo {author} {\bibfnamefont {H.}~\bibnamefont
  {J{\'{o}}{\'{z}}wiak}}, \bibinfo {author} {\bibfnamefont {F.}~\bibnamefont
  {Thibault}}, \ and\ \bibinfo {author} {\bibfnamefont {P.}~\bibnamefont
  {Wcis{\l}o}},\ }\href {\doibase 10.1016/j.jqsrt.2021.107911} {\bibfield
  {journal} {\bibinfo  {journal} {J. Quant. Spectrosc. Radiat. Transfer}\
  }\textbf {\bibinfo {volume} {276}},\ \bibinfo {pages} {107911} (\bibinfo
  {year} {2021})}\BibitemShut {NoStop}%
\bibitem [{\citenamefont {Rohart}\ \emph {et~al.}(1994)\citenamefont {Rohart},
  \citenamefont {Mäder},\ and\ \citenamefont {Nicolaisen}}]{rohart1994}%
  \BibitemOpen
  \bibfield  {author} {\bibinfo {author} {\bibfnamefont {F.}~\bibnamefont
  {Rohart}}, \bibinfo {author} {\bibfnamefont {H.}~\bibnamefont {Mäder}}, \
  and\ \bibinfo {author} {\bibfnamefont {H.}~\bibnamefont {Nicolaisen}},\
  }\href {\doibase 10.1063/1.468342} {\bibfield  {journal} {\bibinfo  {journal}
  {J. Chem. Phys.}\ }\textbf {\bibinfo {volume} {101}},\ \bibinfo {pages}
  {6475} (\bibinfo {year} {1994})}\BibitemShut {NoStop}%
\bibitem [{\citenamefont {Pickett}(1980)}]{pickett1980}%
  \BibitemOpen
  \bibfield  {author} {\bibinfo {author} {\bibfnamefont {H.~M.}\ \bibnamefont
  {Pickett}},\ }\href {\doibase 10.1063/1.440145} {\bibfield  {journal}
  {\bibinfo  {journal} {J. Chem. Phys.}\ }\textbf {\bibinfo {volume} {73}},\
  \bibinfo {pages} {6090} (\bibinfo {year} {1980})}\BibitemShut {NoStop}%
\bibitem [{\citenamefont {Wcisło}\ \emph {et~al.}(2016)\citenamefont
  {Wcisło}, \citenamefont {Gordon}, \citenamefont {Tran}, \citenamefont {Tan},
  \citenamefont {Hu}, \citenamefont {Campargue}, \citenamefont {Kassi},
  \citenamefont {Romanini}, \citenamefont {Hill}, \citenamefont {Kochanov},\
  and\ \citenamefont {Rothman}}]{wcisło2016}%
  \BibitemOpen
  \bibfield  {author} {\bibinfo {author} {\bibfnamefont {P.}~\bibnamefont
  {Wcisło}}, \bibinfo {author} {\bibfnamefont {I.~E.}\ \bibnamefont {Gordon}},
  \bibinfo {author} {\bibfnamefont {H.}~\bibnamefont {Tran}}, \bibinfo {author}
  {\bibfnamefont {Y.}~\bibnamefont {Tan}}, \bibinfo {author} {\bibfnamefont
  {S.~M.}\ \bibnamefont {Hu}}, \bibinfo {author} {\bibfnamefont
  {A.}~\bibnamefont {Campargue}}, \bibinfo {author} {\bibfnamefont
  {S.}~\bibnamefont {Kassi}}, \bibinfo {author} {\bibfnamefont
  {D.}~\bibnamefont {Romanini}}, \bibinfo {author} {\bibfnamefont
  {C.}~\bibnamefont {Hill}}, \bibinfo {author} {\bibfnamefont {R.~V.}\
  \bibnamefont {Kochanov}}, \ and\ \bibinfo {author} {\bibfnamefont
  {L.}~\bibnamefont {Rothman}},\ }\href {\doibase 10.1016/j.jqsrt.2016.01.024}
  {\bibfield  {journal} {\bibinfo  {journal} {J. Quant. Spectrosc. Radiat.
  Transf.}\ }\textbf {\bibinfo {volume} {177}},\ \bibinfo {pages} {75}
  (\bibinfo {year} {2016})}\BibitemShut {NoStop}%
\bibitem [{\citenamefont {Hirschfelder}\ \emph {et~al.}(1954)\citenamefont
  {Hirschfelder}, \citenamefont {Curtiss},\ and\ \citenamefont
  {Bird}}]{hirschfelder1954}%
  \BibitemOpen
  \bibfield  {author} {\bibinfo {author} {\bibfnamefont {J.~O.}\ \bibnamefont
  {Hirschfelder}}, \bibinfo {author} {\bibfnamefont {C.~F.}\ \bibnamefont
  {Curtiss}}, \ and\ \bibinfo {author} {\bibfnamefont {R.~B.}\ \bibnamefont
  {Bird}},\ }\href@noop {} {\emph {\bibinfo {title} {{The Molecular Theory of
  Gases and Liquids}}}}\ (\bibinfo  {publisher} {Wiley},\ \bibinfo {year}
  {1954})\BibitemShut {NoStop}%
\bibitem [{\citenamefont {Chapman}\ and\ \citenamefont
  {Cowling}(1939)}]{Chapman_Cowling}%
  \BibitemOpen
  \bibfield  {author} {\bibinfo {author} {\bibfnamefont {S.}~\bibnamefont
  {Chapman}}\ and\ \bibinfo {author} {\bibfnamefont {T.~G.}\ \bibnamefont
  {Cowling}},\ }\href@noop {} {\emph {\bibinfo {title} {The Mathematical Theory
  of Non-uniform Gases}}}\ (\bibinfo  {publisher} {Cambridge University
  Press},\ \bibinfo {year} {1939})\BibitemShut {NoStop}%
\end{thebibliography}%

\end{document}